\documentclass[aps,prx,twocolumn,showpacs,superscriptaddress,longbibliography]{revtex4-2}
\usepackage{amsmath,amssymb,graphicx}
\usepackage{graphicx,epstopdf}
\usepackage{gensymb}
\usepackage[dvipsnames]{xcolor}
\usepackage[T1]{fontenc}
\usepackage[utf8]{inputenc}
\usepackage{lipsum}
\usepackage{adjustbox}
\newlength\mylength
\newcommand{\be}{\begin{equation}}
\newcommand{\ee}{\end{equation}}
\newcommand{\bea}{\begin{eqnarray}}
\newcommand{\eea}{\end{eqnarray}}
\newcommand{\bse}{\begin{subequations}}
\newcommand{\ese}{\end{subequations}}

\usepackage{multibib}
\usepackage{color}
\usepackage[colorlinks,bookmarks=false,citecolor=darkblue,linkcolor=red,urlcolor=blue]{hyperref}

\definecolor{darkred}{rgb}{0.7,0.0,0.0}

\definecolor{darkblue}{rgb}{0,0.02,0.45}

\definecolor{darkgreen}{rgb}{0.02,0.45,0.0}

\definecolor{violet}{rgb}{0.8,0.2,0.6}

\begin{document}

\title{Frustration driven magnetic correlations in the spin-$5/2$ triangular lattice antiferromagnet RbFe(HPO$_{3}$)$_{2}$}
\author{V. Nagpal}
\affiliation{School of Physics, Indian Institute of Science Education and Research Thiruvananthapuram-695551, India}
\author{Sebin J. Sebastian}
\affiliation{School of Physics, Indian Institute of Science Education and Research Thiruvananthapuram-695551, India}
\affiliation{Ames National Laboratory, U.S. DOE, Iowa State University, Ames, Iowa 50011, USA}
\author{Surya P. Patra}
\affiliation{School of Physics, Indian Institute of Science Education and Research Thiruvananthapuram-695551, India}
\author{S. Shibash}
\affiliation{School of Physics, Indian Institute of Science Education and Research Thiruvananthapuram-695551, India}
\author{Q.-P. Ding}
\affiliation{Ames National Laboratory, U.S. DOE, Iowa State University, Ames, Iowa 50011, USA}
\author{Y. Furukawa}
\affiliation{Ames National Laboratory, U.S. DOE, Iowa State University, Ames, Iowa 50011, USA}
\affiliation{Department of Physics and Astronomy, Iowa State University, Ames, Iowa 50011, USA}
\author{R. Nath}
\email{rameshchandra.nath@gmail.com}
\affiliation{School of Physics, Indian Institute of Science Education and Research Thiruvananthapuram-695551, India}
\date{\today}

\begin{abstract}
A detailed study of the structural and magnetic properties of a spin-$5/2$ triangular lattice antiferromagnet RbFe(HPO$_{3}$)$_{2}$ is presented using x-ray diffraction, magnetization, heat capacity, and $^{31}$P nuclear magnetic resonance (NMR) experiments on a polycrystalline sample. The crystal structure features an equilateral triangular lattice of Fe$^{3+}$ ions. The thermodynamic measurements reveal the onset of a magnetic long-range order at $T_{\rm N1} \simeq 7.8$~K in zero-field, followed by another low temperature field induced ordering at $T_{\rm N2}$ in higher fields. The transition at $T_{\rm N1}$ is further confirmed from the NMR spin lattice relaxation measurements. The value of the frustration ratio ($f \simeq 7$) implies moderate spin frustration in the compound. The $^{31}$P NMR spectra exhibit two distinct spectral lines corresponding to two inequivalent phosphorus sites (P1 and P2), consistent with the crystal structure. The P1 site is strongly coupled with an isotropic hyperfine coupling of $A_{\rm hf}^{\rm iso} = 0.55(2)$~T/$\mu_{\rm B}$ while the P2 site is weakly coupled with $A_{\rm hf}^{\rm iso} = 0.25(3)$~T/$\mu_{\rm B}$ with the Fe$^{3+}$ ions. The magnetic susceptibility and NMR shift data are described well assuming a spin-$5/2$ isotropic triangular lattice antiferromagnetic model with an average exchange coupling of $J/k_{\rm B} = 2.8(2)$~K. Below $T_{\rm N1}$, the spectra evolve into a nearly rectangular powder pattern, indicating a commensurate antiferromagnetic type order. The $^{31}$P spin-lattice relaxation rate well below $T_{\rm N1}$ follows a $T^3$ temperature dependence, implying a two-magnon Raman scattering mechanism in the ordered state. Three well-defined phase regimes are clearly ascertained in the $H-T$ phase diagram, reflecting a weak magnetic anisotropy in the compound.
\end{abstract}

\maketitle

\section{Introduction}
The triangular-lattice antiferromagnet (TLAF) has long been regarded as a cornerstone in the study of frustrated magnetism~\cite{Diep2013}. The incompatibility of antiferromagnetic (AFM) exchange interactions on a triangular geometry suppresses conventional N\'eel order and produces a highly degenerate manifold of states. This intrinsic degeneracy makes TLAFs exceptionally sensitive to fluctuations. However, the quantum and thermal fluctuations lift this degeneracy through an order-by-disorder mechanism, resulting in unconventional magnetic ground states. One of the most celebrated possibilities is the quantum spin liquid (QSL), a highly entangled state without magnetic long-range order (LRO), first envisioned by Anderson in the form of a resonating-valence-bond state for spin-$1/2$ TLAFs~\cite{Anderson153,Balents199}. Later, both theoretical and experimental studies established that the ideal isotropic TLAF instead stabilizes in a chiral three-sublattice $120^{\circ}$ spin configuration, for any spin value~\cite{Jolicoeur2727,Chubukov8891}. Importantly, this canonical state is not immutable: perturbations such as exchange anisotropy, interlayer coupling, and external magnetic field can destabilize the $120^{\circ}$ order and yield a variety of exotic phases in both quantum ($S = 1/2$) and classical ($S > 1/2$) limits~\cite{Smirnov037202,Gallegos196702,Melchy064411,Seabra214418}.
%Zhou15720,Coldea134424,Kimura140401,Heidarian012404,Park7264

These theoretical considerations are vividly realized in a broad range of materials. Prototypical examples include $ABX_3$ halides ($A$ = Cs, Rb; $B$ = V, Mn, Ni; $X$ = Cl, Br, I)~\cite{Collins605,Kadowaki751} and Ba$_3MZ_2$O$_9$ ($M$ = Mn, Co, Ni; $Z$ = Sb, Nb, Ta)~\cite{Doi8923,Lee224402,Ranjith115804}, along with other compounds hosting triangular layers~\cite{Lal014429,Sebastian104425,MajumderL060403,Somesh104422}. More recently, members of the \emph{yavapaiite} family $AM(XO_4)_2$ ($A$ = K, Rb, Cs, Ba; $M$ = Fe, Ti, Mo; $X$ = S, Mo, P) have emerged as exemplary quasi-2D TLAFs displaying unusual quantum phases and complex field-dependent properties~\cite{Smirnov037202,Serrano6314,Inami2374,Nilsen113035,Svistov204,Svistov094434,Mitamura147202,Abdel214427}. Within this family, the degree of lattice symmetry crucially determines the ground state. For instance, equilateral TLAFs such as $A$Fe(SO$_4$)$_2$ ($A$ = Rb, Cs) and $A$Fe(MoO$_4$)$_2$ ($A$ = K, Rb, Cs) stabilize three-sublattice $120^{\circ}$ order in the basal plane~\cite{Serrano6314,Inami2374,Svistov204,Svistov094434,Svistov024412}, whereas monoclinic compounds including K$T$(SO$_4$)$_2$ ($T$ = Fe, Ti) and BaMo(PO$_4$)$_2$ favor isosceles triangular motifs with collinear AFM order~\cite{Nilsen113035,Abdel214427}.

An intriguing extension is provided by the \emph{oxofluorophosphate} family $A$Fe(PO$_3$F)$_2$ [$A$ = K, (NH$_4$)$_2$Cl, NH$_4$, Rb, Cs], in which the Fe(PO$_3$F)$_2$ layers are structurally identical to $M(XO_4)_2$ layers of the \emph{yavapaiite} family~\cite{Stefanie7982}. While all the members order antiferromagnetically, their ground states and field responses are diverse. For example, trigonal KFe(PO$_3$F)$_2$ hosts a coplanar $120^{\circ}$ order coexisting with commensurate non-centrosymmetric stacking along the $c$ axis, whereas NH$_4$Fe(PO$_3$F)$_2$ exhibits a weak XY anisotropy manifested as a field-induced spin-flop transition~\cite{Mohanty184435}. These discoveries highlight the richness of the \emph{yavapaiite} and \emph{oxofluorophosphate} based TLAFs, pondering the fact that subtle perturbations may destabilize the canonical $120^{\circ}$ state and give rise to unconventional field-induced phases.

\begin{figure}
\includegraphics[width=\columnwidth]{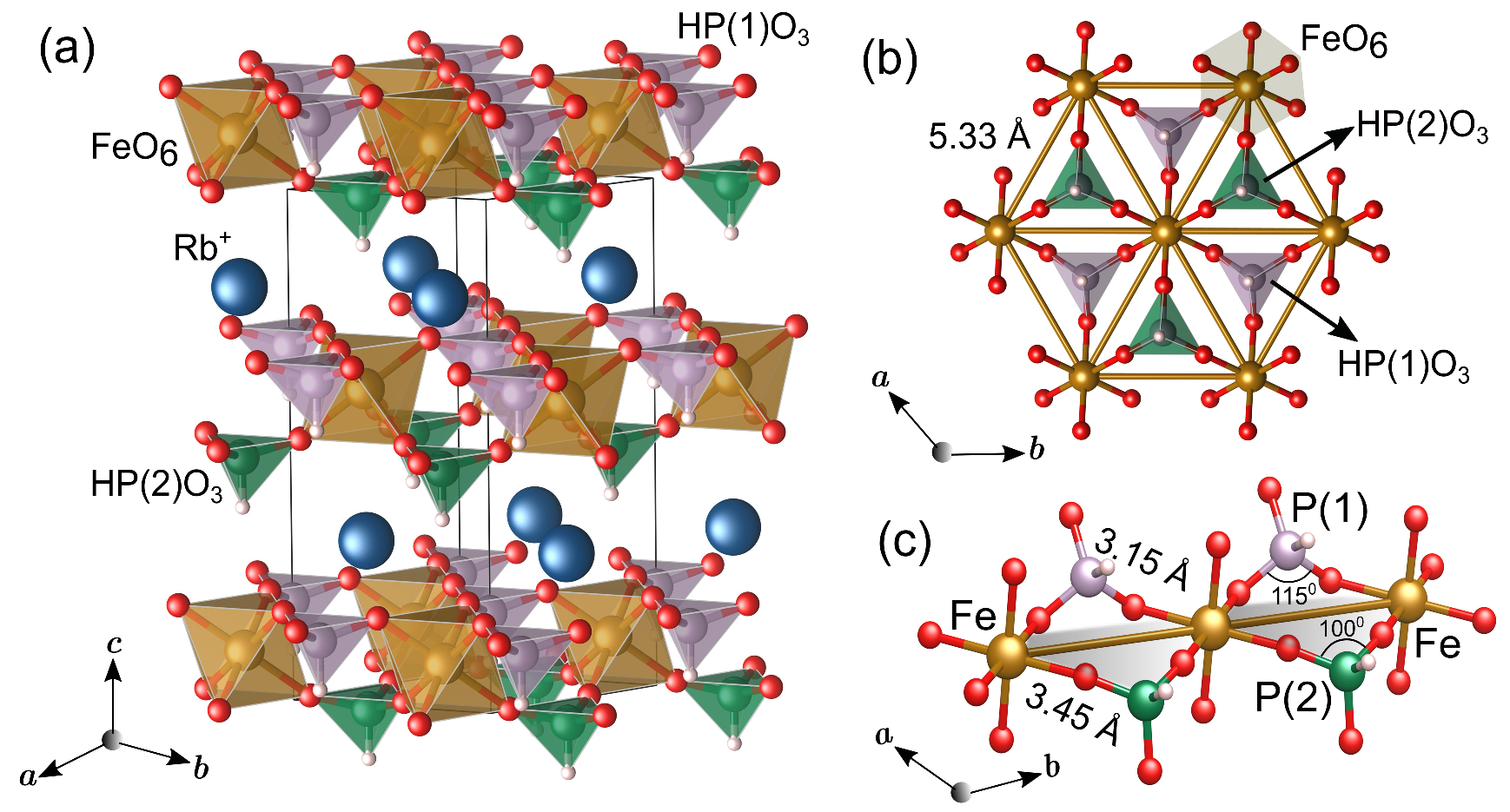}
\caption{\label{Fig1} (a) Crystal structure of RbFe(HPO$_3$)$_2$ showing triangular layers of Fe atoms formed by the corner-shared FeO$_6$ octahedra and HPO$_3$ pseudo-tetrahedra. Rb$^+$ ions separate the two adjacent triangular layers. (b) A section of the layer showing Fe$^{3+}$ ions connected through HPO$_3$ pseudo-tetrahedra in a triangular lattice. (c) Schematic picture showing different environments for P(1) and P(2) sites with respect to Fe atoms.}
\end{figure}
In this work, we present a detailed study of the spin-$5/2$ frustrated \emph{yavapaiite}-type phosphate compound RbFe(HPO$_{3}$)$_{2}$ (RFHPO) by means of dc and ac magnetic, thermodynamic, and nuclear magnetic resonance (NMR) measurements. RFHPO belongs to the family of compounds $A^{1+}M^{3+}$(HPO$_3$)$_2$ [$A$ = K, NH$_4$, Rb and $M$ = V, Fe] and crystallizes in a hexagonal structure with space group $P6_{3}mc$ (No.~186) at room temperature~\cite{Hamchaoui295}. The crystal structure of RFHPO is presented in Fig.~\ref{Fig1}(a) in which Fe$^{3+}$ ions form an FeO$_6$ octahedra and corner-share with pseudo-tetrahedra phosphite group HPO$_3$ in order to form the triangular Fe$^{3+}$ layers in the $ab$-plane. These planes are stacked parallel to each other along the $c$-axis, separated by the non-magnetic Rb$^+$ ions. The Fe$^{3+}$ ions from the adjacent layers are connected via weak hydrogen bonds formed by the Fe-O-H-O-Fe pathway, which leads to negligible interlayer coupling. Further, it is found that Fe$^{3+}$-Fe$^{3+}$ bond lengths in each triangular unit are equal, making equilateral triangles as shown in Fig.~\ref{Fig1}(b). No detailed magnetic measurements on RFHPO are available till date, apart from an earlier report that identified an AFM transition at around $T_{\rm N} \simeq 9$~K from the bulk susceptibility data~\cite{Hamchaoui295}.
%with a large negative Curie--Weiss temperature ($\theta_{\rm CW} \simeq -52$~K) suggesting strong AFM exchange and moderate frustration~\cite{Hamchaoui295}.
However, the microscopic nature of the ordered state and its field evolution remained unexplored. Our magnetic measurements reveal that RFHPO is a frustrated spin-$5/2$ TLAF that undergoes a magnetic LRO at $T_{\rm N1}\simeq 7.8$~K despite a large negative Curie–Weiss temperature ($\theta_{\rm CW} \simeq -52$~K), which is of commensurate AFM type. The $H$–$T$ phase diagram divulges two magnetic phases, reflecting spin reorientation driven by a weak magnetic anisotropy and/or frustration.

\section{Methods}
\begin{figure}
\includegraphics[width=\columnwidth]{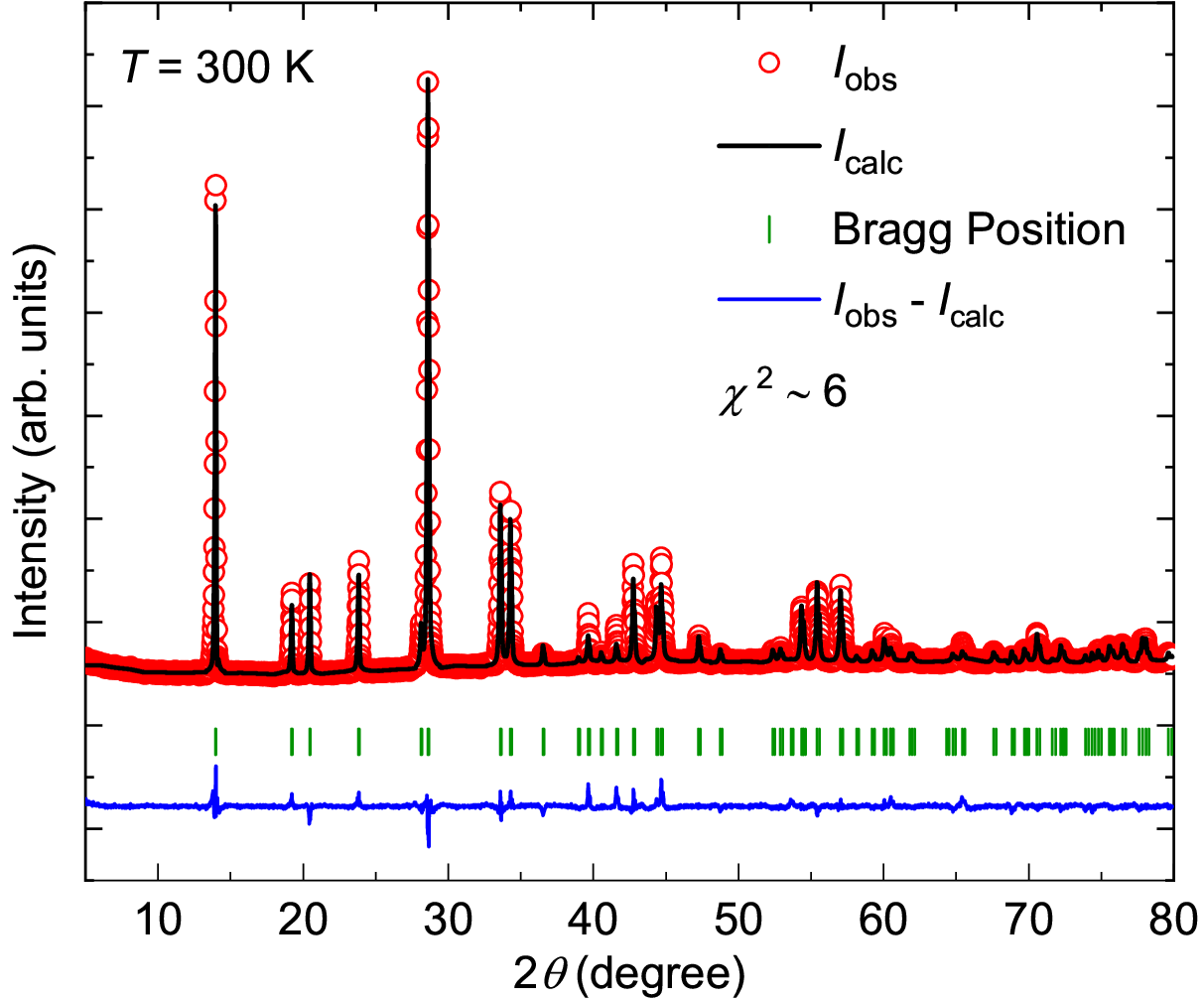}
\caption{\label{Fig2} Powder XRD data collected at room temperature. The red open circles are the experimental data, the black solid line is the Rietveld fit to the data, the vertical bars are the Bragg-peak positions, and the blue line at the bottom is the difference between the experimentally observed and calculated intensities.}
\end{figure}

A polycrystalline sample of RFHPO was synthesized using the hydrothermal reaction method. The initial precursors $0.913$~g Rb$_{2}$CO$_{3}$ (Aldrich, 99.9$\%$), $0.534$~g FeCl$_{3}.6$H$_{2}$O (Loba Chemie, 97$\%$), and $0.608$~g H$_3$PO$_3$ (Alfa Aesar, 99.8$\%$) were taken and mixed in a 10 mL deionized water. The obtained mixture was immediately placed in a 23 mL Teflon-lined stainless steel autoclave and then heated to 180$\degree$ C in an oven for 3 days followed by a slow cooling in 3 days. The resultant product was then washed with water/ethanol several times in order to remove any unwanted impurities and dried overnight at 100$\degree$~C. Finally, a white-purplish coloured polycrystalline sample was obtained. Powder x-ray diffraction (XRD) of the resultant product was carried out at room temperature to confirm the phase purity using a PANalytical x-ray diffractometer (Cu\textit{K$_{\alpha}$} radiation, $\lambda_{\rm avg}\simeq 1.5418$~\AA). Figure~\ref{Fig2} shows the powder XRD pattern of RFHPO along with the Rietveld fit. All the diffracted peaks could be indexed with the hexagonal unit cell [$P6_{3}mc$], confirming single phase of the polycrystalline product. The obtained lattice parameters at room temperature after the refinement are $a = b = 5.33(1)$~\AA, $c= 12.68(3)$~\AA,~and unit-cell volume $V_{\rm cell} = 313.62(2)$~\AA$^3$, consistent with the previous report~\cite{Hamchaoui295}.

The dc magnetization ($M$) was measured as a function of temperature ($2$~K~$\leq T \leq$~380~K) and magnetic field (0~$\leq \mu_0H \leq$~9~T) using the VSM option of a Physical Property Measurement System (PPMS, Quantum Design). The ac magnetization was measured in the temperature range ($2$~K~$\leq T \leq$~30~K) using the ACMS option of PPMS. The ac-measurements were done by varying the frequency ($100$~Hz~$\leq f \leq$~10~KHz) in an applied ac field of $H_{\rm ac}=5$~Oe and then by varying the dc magnetic field (0.01~$\leq H_{\rm dc} \leq$~9~T) keeping $H_{\rm ac}=5$~Oe and $f=499$~Hz fixed. Similarly, heat capacity ($C_{\rm p}$) as a function of $T$ ($2$~K~$\leq T \leq$~200~K) and $H$ (0~$\leq \mu_0H \leq 9$~T) was measured on a small piece of sintered pellet using the standard relaxation technique in PPMS.

Nuclear magnetic resonance (NMR) measurements were performed using a laboratory-built phase-coherent spin-echo pulse spectrometer over a temperature range 1.6~K~$\leq T \leq 300$~K on the $^{31}$P nucleus ($I = 1/2$) with a gyromagnetic ratio of $\gamma_{\rm N}/2\pi = 17.2356$ MHz/T. The NMR data were collected at two different radio frequencies, $\nu$ = 77.1 and 120.6~MHz, which correspond to magnetic fields of $\mu_0 H = 4.47$ and 7~T, respectively. The NMR spectra were obtained by sweeping the magnetic field at a fixed resonance frequency, employing a standard $\pi/2 - \tau - \pi$ pulse sequence with $\tau = 20~\mu$s. The NMR shift $K(T)$ was extracted from the spectral peak position as $K = (2\pi\nu/\gamma_{\rm N} - H_{\rm p})/ H_{\rm p}$, where $H_{\rm p}$ represents the magnetic field of the peak position at each temperature. The nuclear spin-lattice relaxation rate ($1/T_1$) was extracted by measuring the longitudinal magnetization as a function of waiting time by employing the saturation pulse sequence $\pi/2-\tau_1-\pi/2-\tau_2-\pi$, at the spectral peak positions. 

Full diagonalization was performed using the \verb|fulldiag| algorithm~\cite{Todo047203} of the \verb|ALPS| package~\cite{Albuquerque1187}.

\begin{table}[ptb]
	\caption{Refined structural parameters of RFHPO obtained from the Rietveld refinement of the powder-XRD data collected at room temperature. Listed are the Wyckoff positions, refined atomic coordinates, and occupancy of each atom.}
	\label{TableRefinementRFHPO}
	\begin{ruledtabular}
		\begin{tabular}{cccccc}
			Atom & Wyckoff & $x$ & $y$ & $z$ & Occ. \\ \hline
			Rb1 & $2$b & $0.667$ & $0.333$ & $0.790(2)$ & $1.000$ \\
		    Fe & $2$b & $0.667$ & $0.333$ & $1.104(2)$ & $1.000$ \\
			P1 & $2$b & $0.333$ & $0.667$ & $0.967(1)$ & $1.000$ \\
			P2 & $2$a & $0$ & $0$ & $0.634(1)$ & $1.000$\\
            O1 & $6$c & $0.520(1)$ & $1.040(2)$ & $0.978(2)$ & $1.000$ \\
			O2 & $6$c & $0.141(1)$ & $0.282(2)$ & $0.699(2)$ & $1.000$\\
            H1 & $2$b & $0.333$ & $0.667$ & $0.783(2)$ & $1.000$ \\
			H2 & $2$a & $0$ & $0$ & $0.615(2)$ & $1.000$ \\
		\end{tabular}
	\end{ruledtabular}
\end{table}

\section{Results and Discussion}
\subsection{Magnetization}
\begin{figure}
\includegraphics[width=\columnwidth]{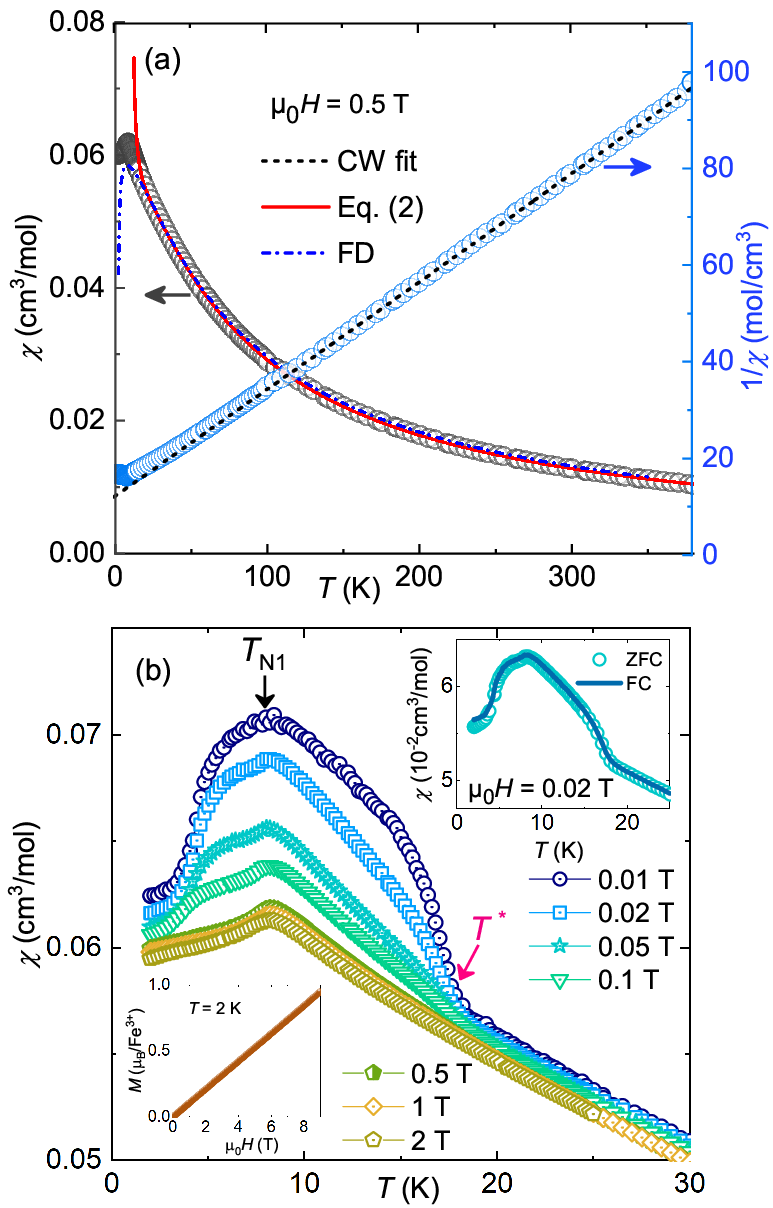} 
\caption{\label{Fig3} (a) The dc magnetic susceptibility $\chi$ and its inverse ($1/\chi$) as a function of temperature for $\mu_0 H = 0.5$~T in the left and right axes, respectively. The solid line represents the fit using the isotropic triangular lattice model to $\chi(T)$. The dash-dotted line is the simulation of $\chi(T)$ of a $S=5/2$ isotropic TLAF using FD method. The dashed line represents the CW fit to $1/\chi(T)$. (b) $\chi$ vs $T$ in various applied magnetic fields for $\mu_0 H \leq 2$~T. The arrows indicate the transitions at $T_{\rm N1}$ and $T^*$. Top inset: FC and ZFC $\chi(T)$ data measured at $\mu_0H=0.02$~T. Bottom inset: $M$ vs $H$ measured at $T=2$~K.}
\end{figure}
Temperature-dependent dc magnetic susceptibility $\chi~[\equiv M/H]$ measured in an applied magnetic field of $\mu_0 H = 0.5$~T is shown in the left $y$-axis of Fig.~\ref{Fig3}(a). In the high temperature region, $\chi(T)$ increases with lowering temperature following the Curie-Weiss (CW) behaviour. At low temperatures, it exhibits a broad peak at around $T_{\rm N1} \simeq 8$~K, implying the transition to a magnetic LRO state. This transition temperature is slightly lower than the previous report~\cite{Hamchaoui295}. To analyze the data, the inverse susceptibility $1/\chi$ for $T > 150$~K was fitted using the CW equation,
\begin{equation}
\chi(T)=\chi_0+\frac{C}{(T-\theta_{\rm CW})}.
\label{Eq1}
\end{equation}
Here, $\chi_0$ is the temperature-independent susceptibility, $C$ is the Curie constant, and $\theta_{\rm CW}$ is the characteristic CW temperature. The fit yields $\chi_0 = -7.79(5) \times 10^{-5}$~cm$^3$/mol, $C\ = 4.50(4)$~cm$^3$K/mol, and $\theta_{\rm CW} = -52.7(3)$~K. The large negative value of $\theta_{\rm CW}$ indicates dominant AFM interaction between Fe$^{3+}$ ions. $\chi_0$ comprises temperature-independent core diamagnetism and Van-Vleck paramagnetism. The core diamagnetic susceptibility ($\chi_{\rm dia}$) of RFHPO is calculated to be $-1.08\times10^{-4}$~cm$^3$/mol by adding the core diamagnetic susceptibilities of elemental ions Rb$^+$, Fe$^{3+}$, P$^{5+}$, O$^{2-}$, and H$^{+}$~\cite{Bain532}. The Van-Vleck paramagnetic susceptibility ($\chi_{\rm VV}$) is deduced to be $\sim 3\times10^{-5}$~cm$^3$/mol by the subtraction of $\chi_{\rm dia}$ from $\chi_0$. Using the value of $C$, the effective magnetic moment is calculated to be $\mu_{\rm eff} = 6.0(1)~\mu_{\rm B}$ ($\mu_{\rm eff}=\sqrt{3k_{\rm B}C/N_{\rm A}}$, where $k_{\rm B}$ is the Boltzmann constant, $N_{\rm A}$ is the Avogadro's number, and $\mu_{\rm B}$ is the Bohr magneton). The obtained $\mu_{\rm eff}$ is close to the expected spin-only effective moment for a spin-$5/2$ system with $g \simeq 2.03$.

While $\theta_{\rm CW}$ represents the overall energy scale of the exchange interactions among the localized spins, the inherent geometrical frustration leads to the suppression of $T_{\rm N}$ as compared to $\theta_{\rm CW}$. Such a suppression of $T_{\rm N}$ is often used as an empirical measure of the magnetic frustration, quantified by the frustration index, $f(= |{\theta_{\rm CW}}|/T_{\rm N1})$. The value of $f$ for RFHPO is calculated to be $\sim 7$, indicating a moderate frustration. Similarly, the magnitude of $\theta_{\rm CW}$ is defined as the sum of all possible exchange interactions $|\theta_{\rm CW}| = JzS(S+1)/3k_{\rm B}$, where $z$ is the number of nearest neighbors of Fe$^{3+}$ ions~\cite{Domb296}. For RFHPO, taking $z = 6$ and $S=5/2$, the average intralayer exchange coupling is calculated to be $J/k_{\rm B} \simeq 3$~K. In the Heisenberg TLAF, the saturation field ($H_{\rm sat}$) can be written in terms of $J/k_{\rm B}$ as $\mu_0H_{\rm sat} = 9JS/g\mu_B$~\cite{Sebastian104425,Kawamura4530}. In our case, with $J/k_{\rm B} = 2.8$~K and $g = 2.03$, this value is calculated to be $\mu_0 H_{\rm sat}\simeq 50$~T.

Further, to model $\chi(T)$ and estimate the exchange coupling between the Fe$^{3+}$ ions, $\chi(T)$ is divided into two components
\begin{equation}
\chi(T)=\chi_0 + \chi_{\rm spin}(T).
\label{Eq2}
\end{equation}
Here, $\chi_{\rm spin}(T)$ is the high temperature series expansion (HTSE) for a spin-$5/2$ Heisenberg isotropic TLAF model given by~\cite{Sebastian104425,Delmas55}
\begin{equation}
\frac{N_{\rm A}{\mu_{\rm B}^2}{g^2}}{3|J|\chi_{\rm spin}}= x + 4 + \frac{3.20}{x} -\frac{2.186}{x^2} + \frac{0.03}{x^3} + \frac{3.45}{x^4} -\frac{3.99}{x^5},
\label{Eq3}
\end{equation}
where $x = {k_{\rm B}}T/|J|S(S + 1)$. This expression holds for $T \geq JS(S + 1)$~\cite{Schmidt104443}. As shown in Fig.~\ref{Fig3}(a), $\chi(T)$ above $30$~K is fitted by Eq.~\eqref{Eq2} that yields $\chi_0 = -1.29(2) \times 10^{-4}$~cm$^3$/mol, $g = 2.01(1)$, and the average AFM exchange coupling $J/k_{\rm B} = 2.8(2)$~K. Indeed, this value of $J/k_{\rm B}$ is in good agreement with the one obtained from $\theta_{\rm CW}$.
%With this value of $J/k_{\rm B}$, the saturation field is estimated to be about $\mu_0H_{\rm sat} \simeq 50$~T.
To further establish the 2D TLAF model, we also simulated $\chi(T)$ using full diagonalization method for a $S=5/2$ TLAF taking $J/k_{\rm B} = 2.8$~K. As seen in Fig.~\ref{Fig3}(a), it reproduces the overall shape of $\chi(T)$ data very well.
% in the entire temperature range.

To assess the magnetic ground state of the system, we measured $\chi(T)$ at low fields ($\mu_0H \leq 2$~T), as shown in Fig.~\ref{Fig3}(b). At $\mu_0H = 0.01$~T, it features two distinct anomalies, labeled $T^* \simeq 17$~K and $T_{\rm N1} \simeq 8.2$~K. With increasing field, the broad hump at $T^*$ is gradually reduced and is completely suppressed for $\mu_0H > 0.5$~T. This feature possibly indicates the development of a weak FM correlation among the magnetic ions, which is a short-range type~\cite{Nalbandyan1705,Ghosh15779}.
Such a weak ferromagnetism could be arising due a small spin canting triggered by magnetic anisotropy~\cite{Nath014407}. However, the anomaly at $T_{\rm N1}$ almost remains unchanged upto $\mu_0H = 2$~T. Further, we measured the field-cooled (FC) and zero-field-cooled (ZFC) $\chi(T)$ data at $\mu_0H = 0.02$~T [top inset of Fig.~\ref{Fig3}(b)], which show no splitting at the transition temperatures, suggesting the absence of spin-freezing.
%In addition, below $T_{\rm N1}$, the low-field $\chi(T)$ data display a cusp, possibly indicating another magnetic phase transition~\cite{Muthu174430}. With increasing magnetic field, both the $T^*$ anomaly and the low-temperature cusp are progressively suppressed and vanish for $\mu_0H \geq 0.5$~T. This suppression, together with the overall decrease of $\chi(T)$ with field, supports the presence of weak FM correlations superimposed on an AFM background.
The magnetization isotherm $M(H)$ measured at $T = 2$~K [bottom inset of Fig.~\ref{Fig3}(b)] shows an almost linear increase without reaching saturation upto 9~T, consistent with the dominant AFM exchange interaction in RFHPO.
%A full-quadrant $M(H)$ measurement at low fields (not shown here) also reveals no hysteresis.

\begin{figure}
\includegraphics[width=\columnwidth]{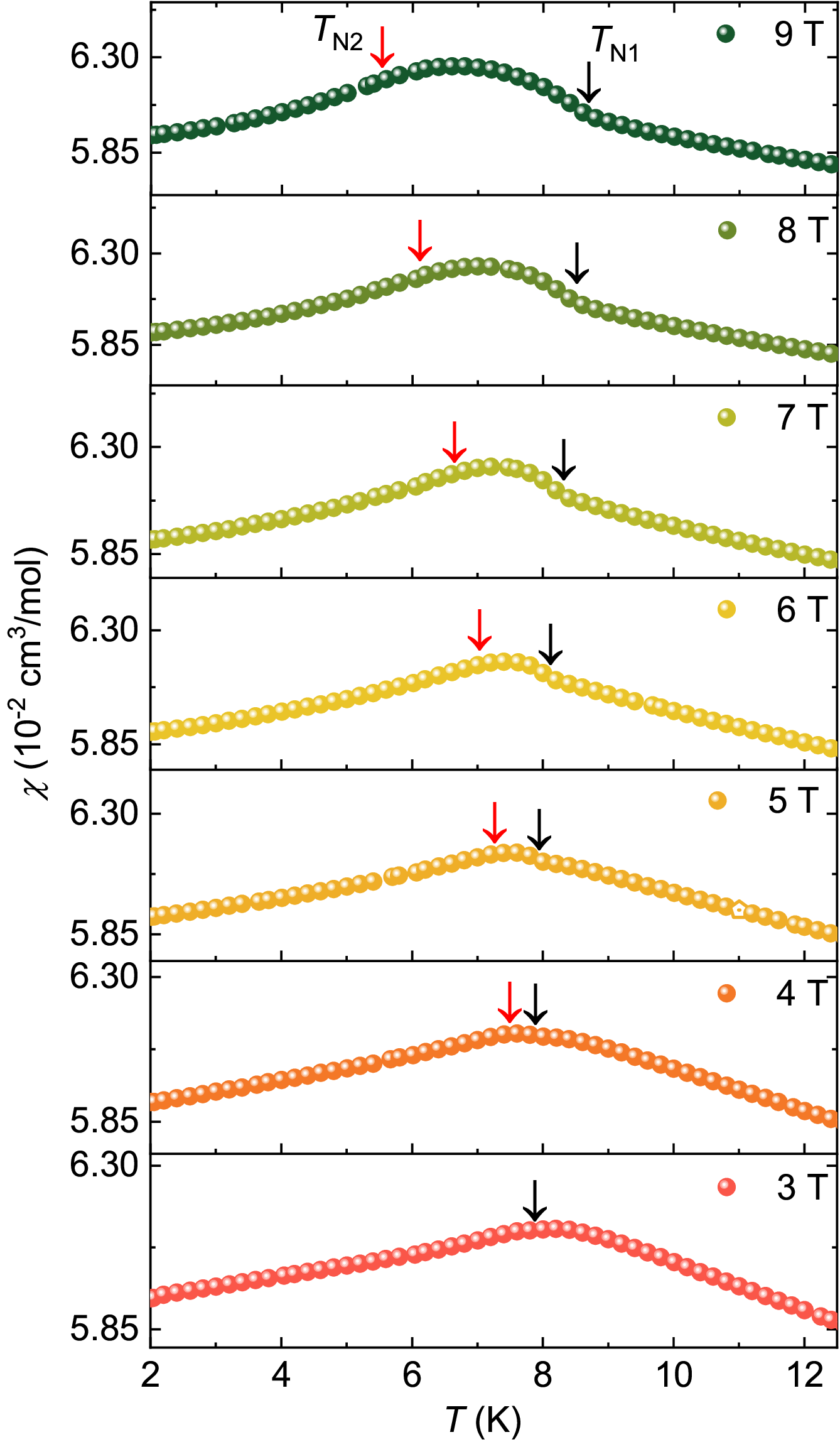}
\caption{\label{Fig4} Low temperature $\chi(T)$ in various applied fields 3~T~$\leq \mu_0H \leq 9$~T. The vertical arrows indicate the two transitions, $T_{\rm N1}$ and $T_{\rm N2}$, respectively.}
\end{figure}
Figure~\ref{Fig4} presents $\chi(T)$ measured in different applied magnetic fields, $3$~T $\leq \mu_{0}H \leq 9$~T. For $\mu_0H \leq 3$~T, the anomaly at $T_{\rm N1}$ remains almost unchanged. However, for $\mu_0H > 3$~T, $T_{\rm N1}$ moves towards higher temperatures and another field induced kink ($T_{\rm N2}$) appears which moves to lower temperature with increasing field. These two transitions are distinctly visible in the $d\chi/dT$ vs $T$ plots (not shown).
%With increasing field, this broad maximum gradually splits into two distinct anomalies, labeled $T_{\rm N1}$ and $T_{\rm N2}$. The transition $T_{\rm N1}$, identified by the sharp upturn in $\chi(T)$ above the broad maximum, shifts to higher temperatures with field. In contrast, $T_{\rm N2}$, which appears below the broad maximum, moves to lower temperatures as the field increases.Zvereva1550
Such a field-induced transition has also been reported in other high-spin, low-dimensional TLAFs~\cite{Svistov024412,Stefanie7982,Ishii17001,Yue214430}. These two transitions are further corroborated by our ac susceptibility and heat capacity measurements, discussed in the following sections.

\subsection{AC susceptibility}
\begin{figure}
\includegraphics[width=\columnwidth]{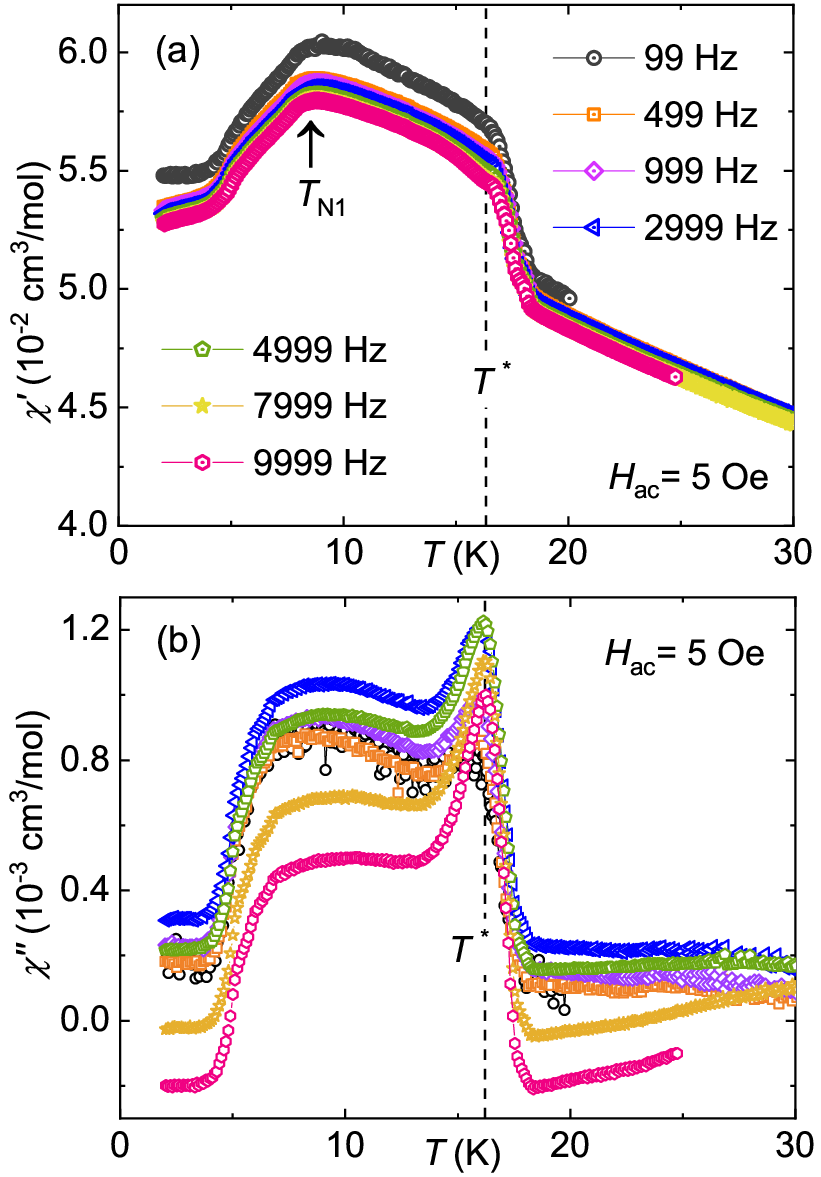}	
\caption{\label{Fig5} (a) Temperature dependence of $\chi'$ measured in different frequencies. The transitions at $T_{\rm N1}$ and $T^*$ are highlighted. (b) $\chi''$ vs $T$ measured in different frequencies.}
\end{figure}
To further investigate the magnetic transitions in zero dc field, we performed ac susceptibility measurements in an applied ac field of $H_{\rm ac} = 5$~Oe at various frequencies. As shown in Fig.~\ref{Fig5}(a), the real part of ac susceptibility $\chi'(T)$ exhibits behavior similar to the low-field dc $\chi(T)$ data. It shows an asymmetric profile with a kink at $T_{\rm N1}\simeq 7.8$~K. In addition, a hump appears near $T^*\simeq 16$~K, consistent with the anomaly observed in low field dc $\chi(T)$. The imaginary part $\chi''(T)$ [see Fig.~\ref{Fig5}(b)] also displays a sharp peak at $T^*$ along with a broad shoulder near $T_{\rm N1}$. Both the features in $\chi'(T)$ and $\chi''(T)$ are found to be frequency independent.

\begin{figure}
\includegraphics[width=\columnwidth]{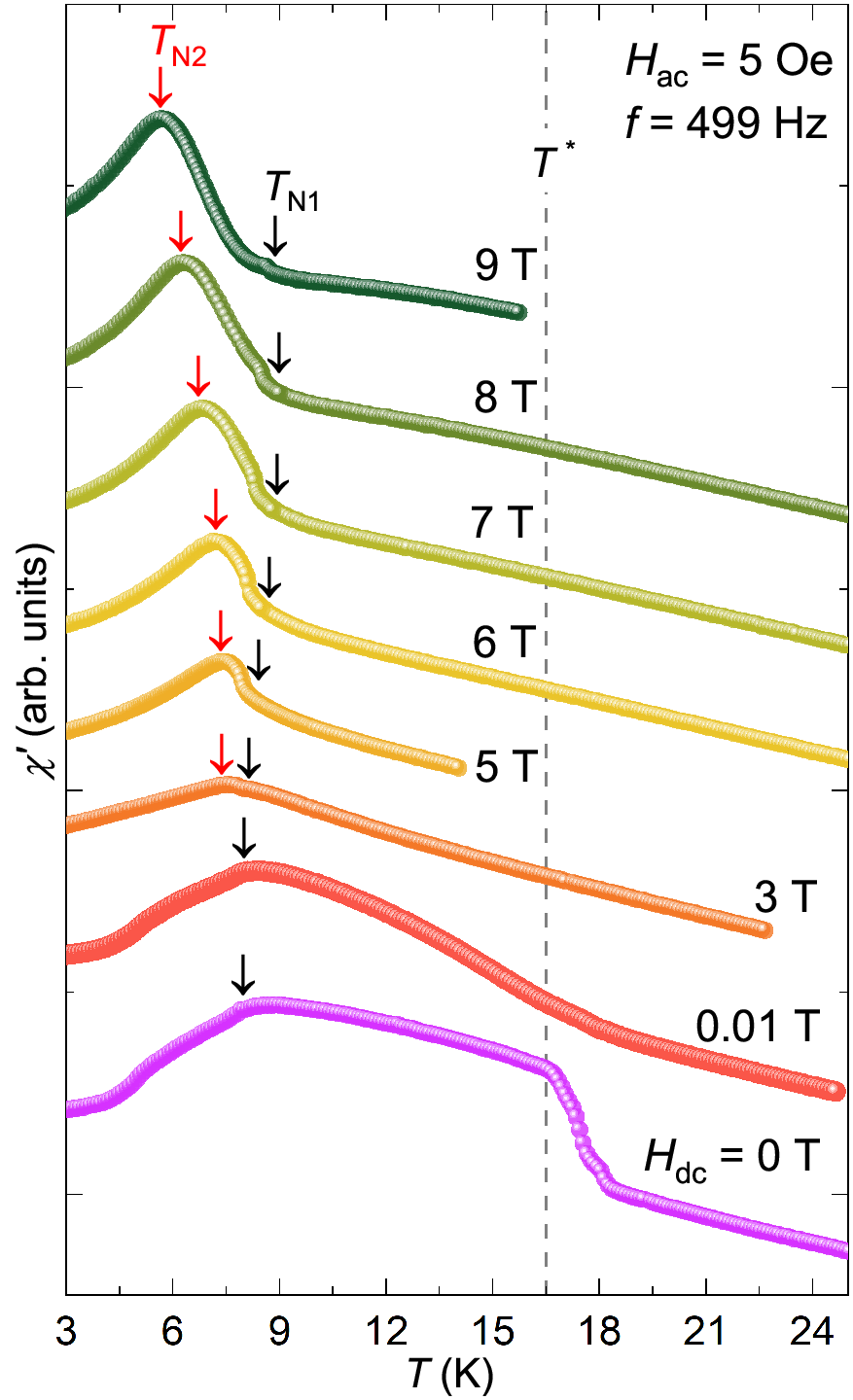}	
\caption{\label{Fig6} $\chi'$ vs $T$ at various applied dc fields and at a fixed frequency $f = 499$~Hz. The data are vertically translated for clarity.}
\end{figure}
To probe the field-induced transition, we measured $\chi'(T)$ in different dc fields (see Fig.~\ref{Fig6}) at a fixed frequency $f = 499$~Hz and $H_{\rm ac} = 5$~Oe. In zero dc field, both $T_{\rm N1}$ and $T^*$ are clearly visible. With increasing dc field, $T^*$ is completely suppressed, confirming its origin as short-range FM correlation. For $H_{\rm dc} \geq 3$~T, $\chi'(T)$ shows a kink associated with $T_{\rm N1}$ followed by another peak at lower temperatures, corresponding to the second transition $T_{\rm N2}$. With further increase in dc field, $T_{\rm N1}$ shifts to higher temperatures, while $T_{\rm N2}$ moves to lower temperatures, similar to the dc $\chi(T)$ data.

\subsection{Heat capacity}
\begin{figure}
\includegraphics[width=\columnwidth]{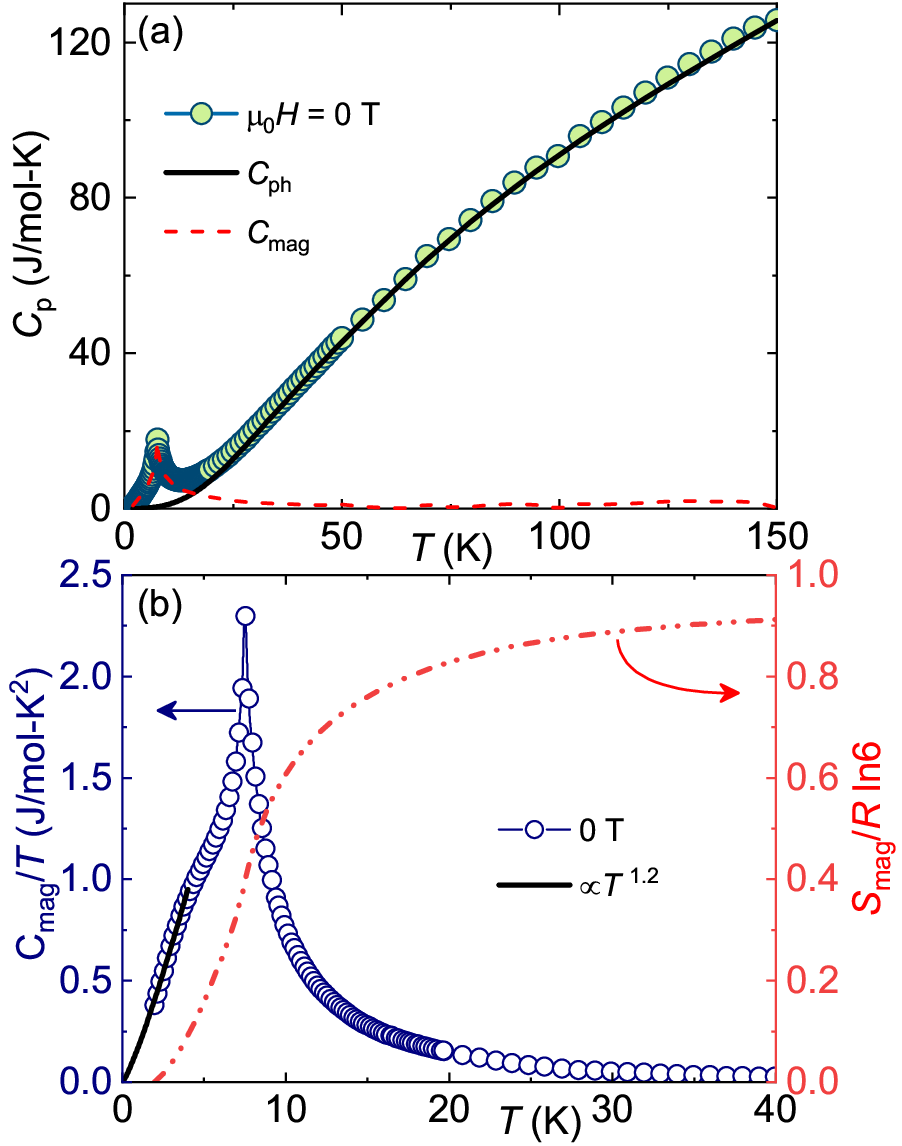} 
\caption{\label{Fig7} (a) $C_{\rm p}$ vs $T$ in zero field. The solid line denotes $C_{\rm ph}$, while the dashed line represents $C_{\rm mag}$. (b) $C_{\rm mag}/T$ and $S_{\rm mag}$ vs $T$ in the left and right $y$-axes, respectively. The solid line is the power law fit to the $C_{\rm mag}$ data at low temperatures, below 3~K.}
\end{figure}
Figure~\ref{Fig7}(a) shows the temperature dependence of heat capacity [$C_{\rm p}(T)$] measured in zero-field. With decreasing temperature, $C_{\rm p}(T)$ exhibits a pronounced $\lambda$-type anomaly at $T_{\rm N1} \simeq 7.6$~K, signaling the onset of a magnetic LRO. No anomaly associated with $T^*$ is observed in zero-field data.

In magnetic insulators, the total heat capacity can be expressed as $C_{\rm p}(T) = C_{\rm ph}(T) + C_{\rm mag}(T)$, where the phonon contribution $C_{\rm ph}(T)$ dominates at high temperatures, while the magnetic contribution $C_{\rm mag}(T)$ dominates at low temperatures. To separate the two, we fitted $C_{\rm p}(T)$ data using a Debye–Einstein model consisting of one Debye term and four Einstein terms~\cite{Gopal2012,Sebastian064413}:
\begin{equation}
	C_{\rm ph}(T) = f_{\rm D}C_{\rm D}(\theta_{\rm D},T) + \sum_{i=1}^{4} g_i C_{{\rm E}_i}(\theta_{{\rm E}_i},T),
	\label{Eq4}
\end{equation}
where $f_{\rm D}$ and $g_i$ are weight factors that satisfy $f_{\rm D} + \sum g_i = 1$, consistent with the Dulong–Petit limit ($\sim 3nR$) at high temperatures~\cite{Fitzgerel545}. The Debye term that accounts for the low-energy vibrations (acoustic modes) of heavy atoms can be written as 
\begin{equation}
	C_{\rm D}(\theta_{\rm D},T) = 9nR\left(\frac{T}{\theta_{\rm D}}\right)^3 \int_0^{\theta_{\rm D}/T} \frac{x^4 e^x}{(e^x - 1)^2} dx.
	\label{Eq5}
\end{equation}
The Einstein term describes the higher-energy optical modes and can be expressed as  
\begin{equation}
	C_{\rm E}(\theta_{\rm E},T) = 3nR\left(\frac{\theta_{\rm E}}{T}\right)^2 \frac{e^{\theta_{\rm E}/T}}{(e^{\theta_{\rm E}/T}-1)^2}.
	\label{Eq6}
\end{equation}
Here, $\theta_{\rm D}$ and $\theta_{\rm E}$ are the characteristic Debye and Einstein temperatures, respectively. The fit [solid line in Fig.~\ref{Fig7}(a)] in the high temperatures yields $f_{\rm D} \simeq 0.079$, $g_1 \simeq 0.147$, $g_2 \simeq 0.255$, $g_3 \simeq 0.378$, $g_4 \simeq 0.141$, $\theta_{\rm D}=120 (2)$~K, $\theta_{{\rm E}_1} = 154(3)$~K, $\theta_{{\rm E}_2} = 341(3)$~K, $\theta_{{\rm E}_3} = 865(4)$~K, and $\theta_{{\rm E}_4} = 1600(3)$~K. Subtracting $C_{\rm ph}(T)$ from the measured $C_{\rm p}(T)$ gives the magnetic part $C_{\rm mag}(T)$, shown as the dashed line in Fig.~\ref{Fig7}(a). 

Figure~\ref{Fig7}(b) presents $C_{\rm mag}(T)/T$ vs $T$, which exhibits a sharp peak at $T_{\rm N1}$ along with a broad hump below it, typical of high-spin systems~\cite{Sebastian104425,Nath024431}. Furthermore, the magnetic entropy was estimated as $S_{\rm mag}(T) = \int_{1.95\,{\rm K}}^{T} \frac{C_{\rm mag}(T')}{T'} dT'$. The resulting entropy, plotted on the right $y$-axis of Fig.~\ref{Fig7}(b), saturates to a value $S_{\rm mag} \simeq 13.85$~J/mol-K above 50~K, which is close to the expected value $R\ln(2S+1) = 14.89$~J/mol-K for $S=5/2$. At very low temperatures ($T<T_{\rm N1}$), $C_{\rm mag}(T)$ follows a power law with an exponent $\alpha \simeq 2.2$ which possibly reflects dominant 2D magnon excitation in the AFM ordered state~\cite{Takatsu104424,Somesh104422}.
%The inset of Fig~\ref{Fig7}(b) highlights the $T^2$ dependence of $C_{\rm mag}(T)$ at low temperatures, consistent with the 2D AFM order and gapless spin excitations~\cite{Zvereva1550,Takatsu104424}.

\begin{figure}
\includegraphics[width=\columnwidth]{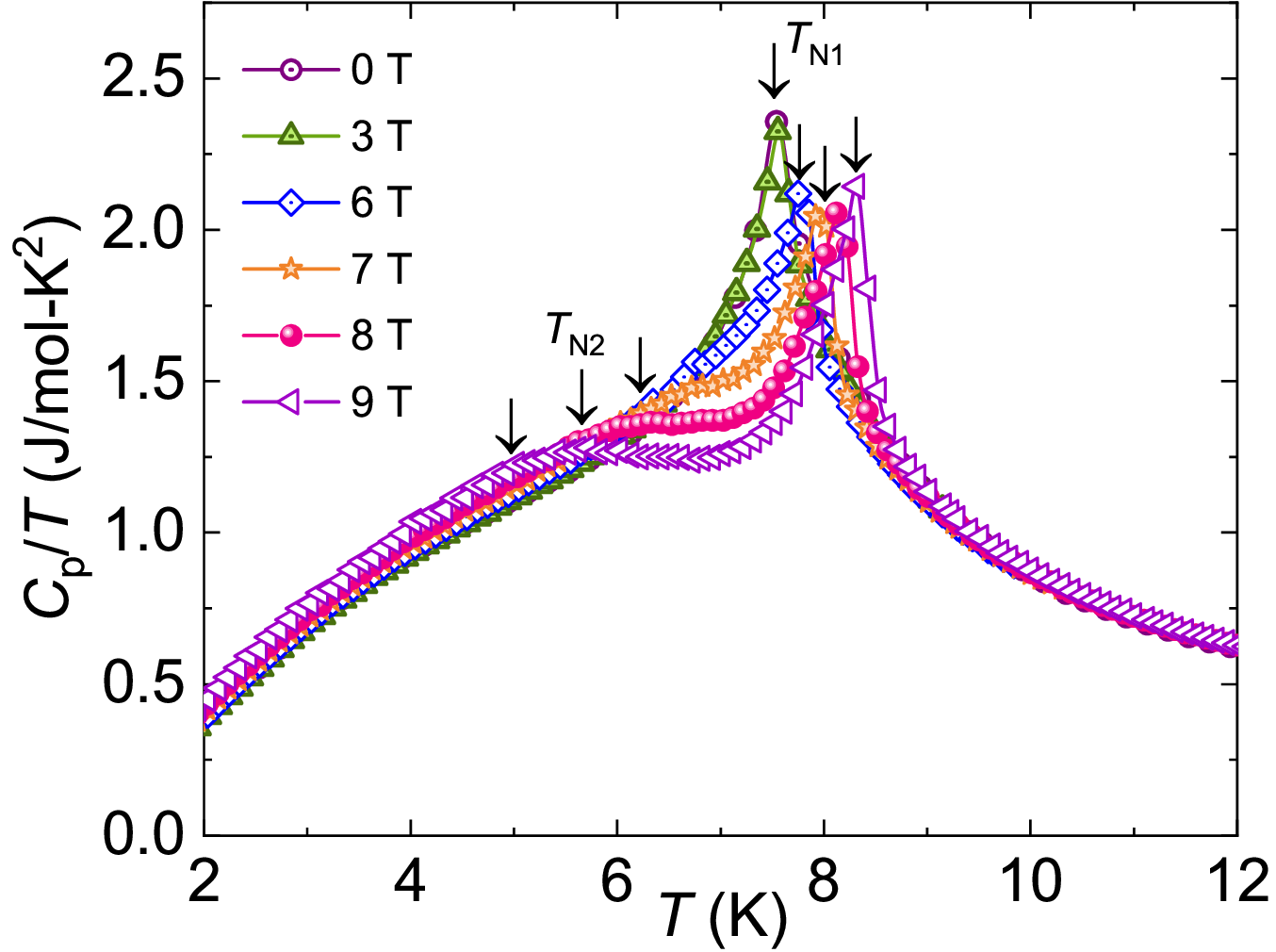}	
\caption{\label{Fig8} $C_{\rm p}/T$ vs $T$ in the low-$T$ regime measured in different fields. The respective vertical arrows point to the transitions at $T_{\rm N1}$ and $T_{\rm N2}$.}
\end{figure}
To access the field-induced transition, $C_{\rm p}(T)$ in the low-$T$ region measured in different fields is shown in Fig.~\ref{Fig8}. For $\mu_0H \leq 6$~T, the transition at $T_{\rm N1}$ remains nearly unchanged. At higher fields ($\mu_0H \geq 7$~T), an additional shoulder-like anomaly emerges at lower temperatures, which becomes more pronounced with increasing field and is assigned to the field-induced transition $T_{\rm N2}$. These findings are consistent with the $\chi(T)$ data. Similar behavior demonstrating field-induced transitions in specific heat is reported in other frustrated magnets and is generally attributed to the magnetic anisotropy~\cite{Ishii17001,Yue214430}.

\subsection{$^{31}$P NMR}
Nuclear magnetic resonance (NMR) is a powerful local probe for investigating both static and dynamic properties of correlated spin systems. Figure~\ref{Fig1}(a) illustrates the unit cell of RFHPO containing two crystallographically inequivalent HP(1)O$_3$ and HP(2)O$_3$ tetrahedra connected to the Fe$^{3+}$ions. As shown in Fig.~\ref{Fig1}(c), the Fe--Fe connectivity involves the pathway Fe--O--P--O--Fe. The structural analysis reveals the bond angles $\angle \text{Fe--P(1)--Fe} \simeq 115\degree$ and $\angle \text{Fe--P(2)--Fe} \simeq 100\degree$ and bond distances P(1)--Fe~$\simeq 3.15$~\AA~and P(2)--Fe~$\simeq 3.45$~\AA. Because of shorter bond distance and larger angle, one expects stronger hyperfine coupling for the P(1) site as compared to the P(2) site. Since the $^{31}$P nuclei have a nuclear spin $I = 1/2$, each site is expected to produce a single resonance line in the paramagnetic regime.

\subsubsection{$^{31}$P NMR spectra ($T > T_{\rm N1}$)}
\begin{figure}[h]
\centering
\includegraphics[width=\linewidth]{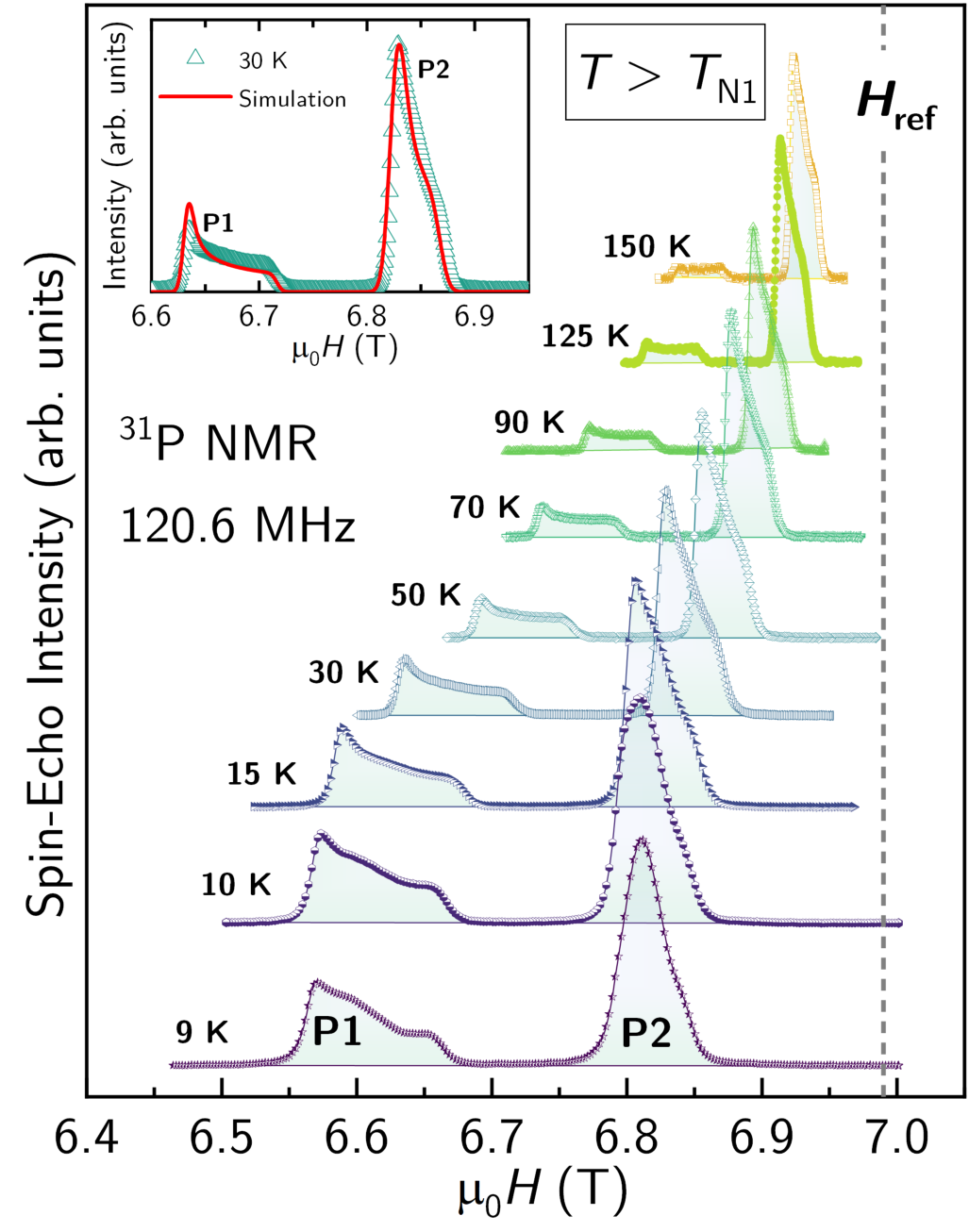}
\caption{Temperature-dependent field swept $^{31}$P NMR spectra measured at 120.6~MHz ($\mu_0H \approx 7$~T), down to 9~K ($T > T_{\rm N_1}$). The vertical dashed line marks the zero-shift position or reference field for $^{31}$P nucleus. Inset: spectral simulation at $T = 30$~K, accounting for the contributions from two inequivalent phosphorus sites (P1 and P2).}
\label{Fig9}
\end{figure}
\begin{figure}
\centering
\includegraphics[width=\linewidth]{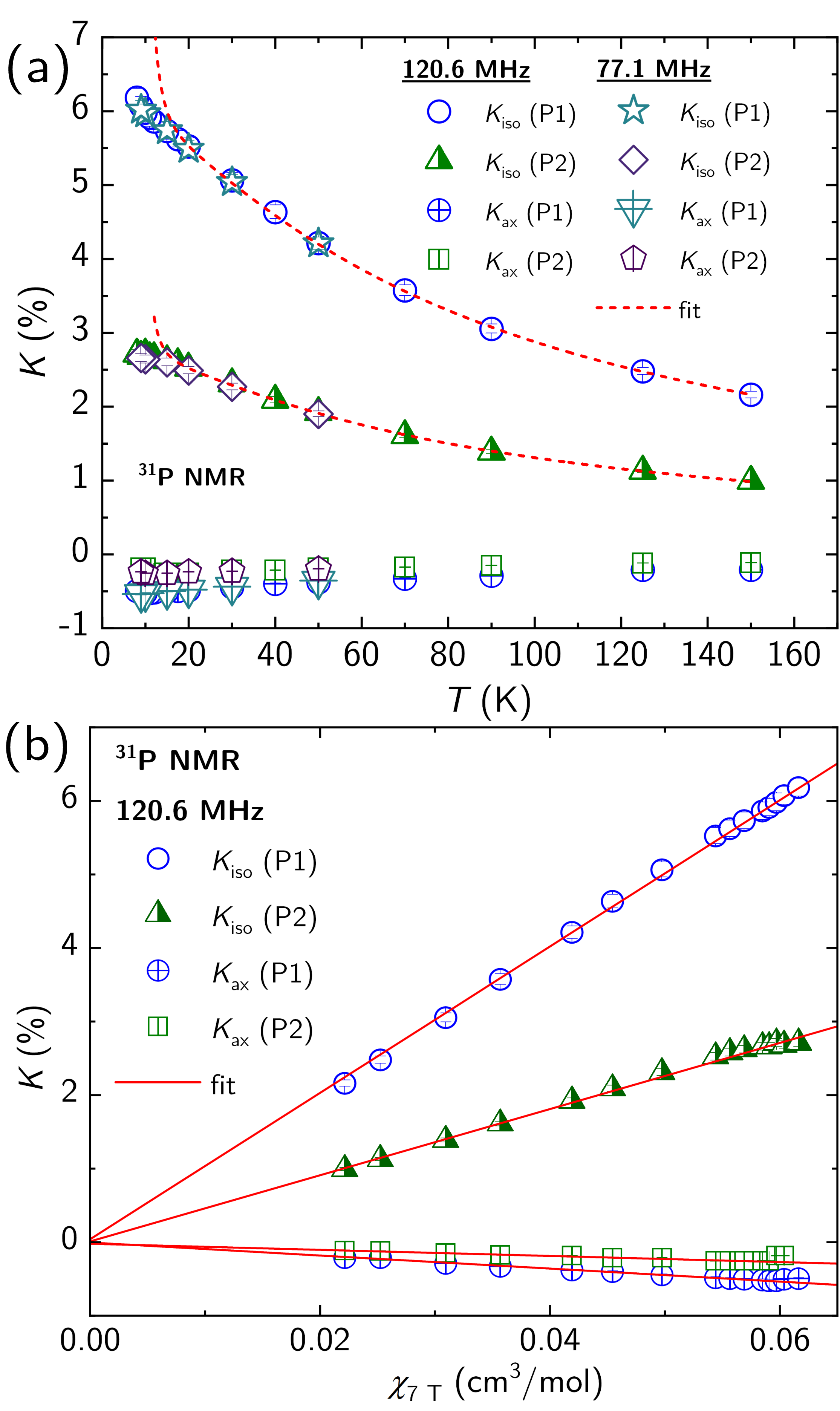}
\caption{(a) Temperature dependence of $K_{\rm iso}$ and $K_{\rm ax}$ for both P1 and P2 sites, measured at two different frequencies. The dashed curves represent the fits using Eq.~\eqref{K} to $K_{\rm iso}(T)$. (b) Clogston-Jaccarino plot of $K_{\rm iso}$ and $K_{\rm ax}$ vs $\chi$ with temperature as an implicit parameter. Solid lines represent linear fits.}
\label{Fig10}
\end{figure}
Figure~\ref{Fig9} shows the field-swept $^{31}$P NMR spectra measured at 120.6~MHz ($\mu_0H \simeq 7$~T) at different temperatures down to 9~K ($T > T_{\rm N1}$). In the paramagnetic regime, two distinct resonance lines with finite asymmetry are observed, corresponding to the P1 and P2 sites. The presence of two inequivalent lines reflects different local magnetic environments, originating from site-dependent hyperfine interactions with Fe$^{3+}$ moments. The broad line can be assigned to the strongly coupled P1 site, while the narrow line corresponds to the weakly coupled P2 site. The asymmetric line shape for both the P-sites can be attributed either to the asymmetry in hyperfine coupling or anisotropy in $\chi(T)$~\cite{Islam174432,Yogi024413}. Upon cooling, both lines broaden and shift progressively towards lower magnetic fields.

To simulate the anisotropic powder spectra, we used the standard expression for the NMR shift ($K$) 
\begin{equation}
K = K_{\text{iso}} + K_{\text{ax}}(3\cos^2\theta - 1),
\label{eq:NMR}
\end{equation}
where $K_{\text{iso}}$ and $K_{\text{ax}}$ denote the isotropic and axial components of the NMR shift, respectively, and $\theta$ is the angle between the applied magnetic field and the principal axis of the hyperfine tensor~\cite{slichter2013,Sebastian104428}. A representative simulation of the spectrum at $T = 30$~K based on a two-site model, incorporating finite anisotropic shifts, is shown in the inset of Fig.~\ref{Fig9}. This approach was extended to other temperatures to extract the temperature dependence of $K$.
It should be noted that at low temperatures just above $T_{\rm N1}$, the NMR spectra were not reproduced well by simulation, especially for the P1 site. 
This indicates that the P1 site has more than one environment with slightly different hyperfine coupling constants for unknown reasons, which produces the complicated NMR spectra in the magnetically ordered states.

The extracted $K_{\rm iso}(T)$ and $K_{\rm ax}(T)$ for both the P1 and P2 sites are plotted in Fig.~\ref{Fig10}(a) for two frequencies. Both the components of $K$ are found to be frequency independent, as anticipated from the $\chi(T)$ data. With decreasing temperature, both the components of $K$ increase, consistent with the Curie-Weiss behavior for $T > T_{\rm N1}$. Since $K(T)$ reflects the intrinsic spin susceptibility $\chi_{\rm spin}(T)$, it can be expressed as  
\begin{equation}
K(T) = K_0 + \frac{A_{\rm hf}}{N_{\rm A}\mu_{\rm B}}\,\chi_{\rm spin}(T),
\label{K}
\end{equation}
where $K_0$ is the temperature-independent chemical shift and $A_{\rm hf}$ is the hyperfine coupling constant between the $^{31}$P nucleus and Fe$^{3+}$ spins. In Fig.~\ref{Fig10}(b), $K_{\rm iso}$ and $K_{\rm ax}$ are plotted against $\chi$ measured at 7~T, with temperature as an implicit parameter. The observed linearity confirms that the NMR shift is tracking $\chi_{\rm spin}$, with no contribution from impurities or orphan spins. A linear fit yields the isotropic hyperfine coupling $A_{\rm hf}^{\rm iso} = 0.55(2)$~T/$\mu_{\rm B}$ and $0.25(3)$~T/$\mu_{\rm B}$ for the P1 and P2 sites, respectively. Similarly, the obtained axial hyperfine coupling is $A_{\rm hf}^{\rm ax} = -0.049(2)$~T/$\mu_{\rm B}$ and $-0.023(1)$~T/$\mu_{\rm B}$ for P1 and P2, respectively. These results indicate that P1 is strongly coupled to Fe$^{3+}$ spins than P2, consistent with our site assignment based on the crystal structure. The obtained $A_{\rm hf}$ values are comparable to those reported in other phosphate-based AFMs~\cite{Islam174432,Ambika015803,Mohanty184435}.

Finally, the exchange coupling $J/k_{\rm B}$ between Fe$^{3+}$ spins was estimated by fitting $K_{\rm iso}(T)$ for both sites using Eq.~\eqref{K}, with $\chi_{\rm spin}(T)$ given in Eq.~\eqref{Eq3}. During the fit, $A_{\rm hf}^{\rm iso}$ was fixed to the values obtained from the $K-–\chi$ analysis. As shown in Fig.~\ref{Fig10}(a), the fits above 15~K for both the sites yield $J/k_{\rm B} = 2.6(2)$~K and $g = 2.01(1)$, which are in good agreement with the $\chi(T)$ analysis.

\subsubsection{$^{31}$P NMR spectra ($T < T_{\rm N1}$)}
\begin{figure}
    \centering
 \includegraphics[width=\linewidth]{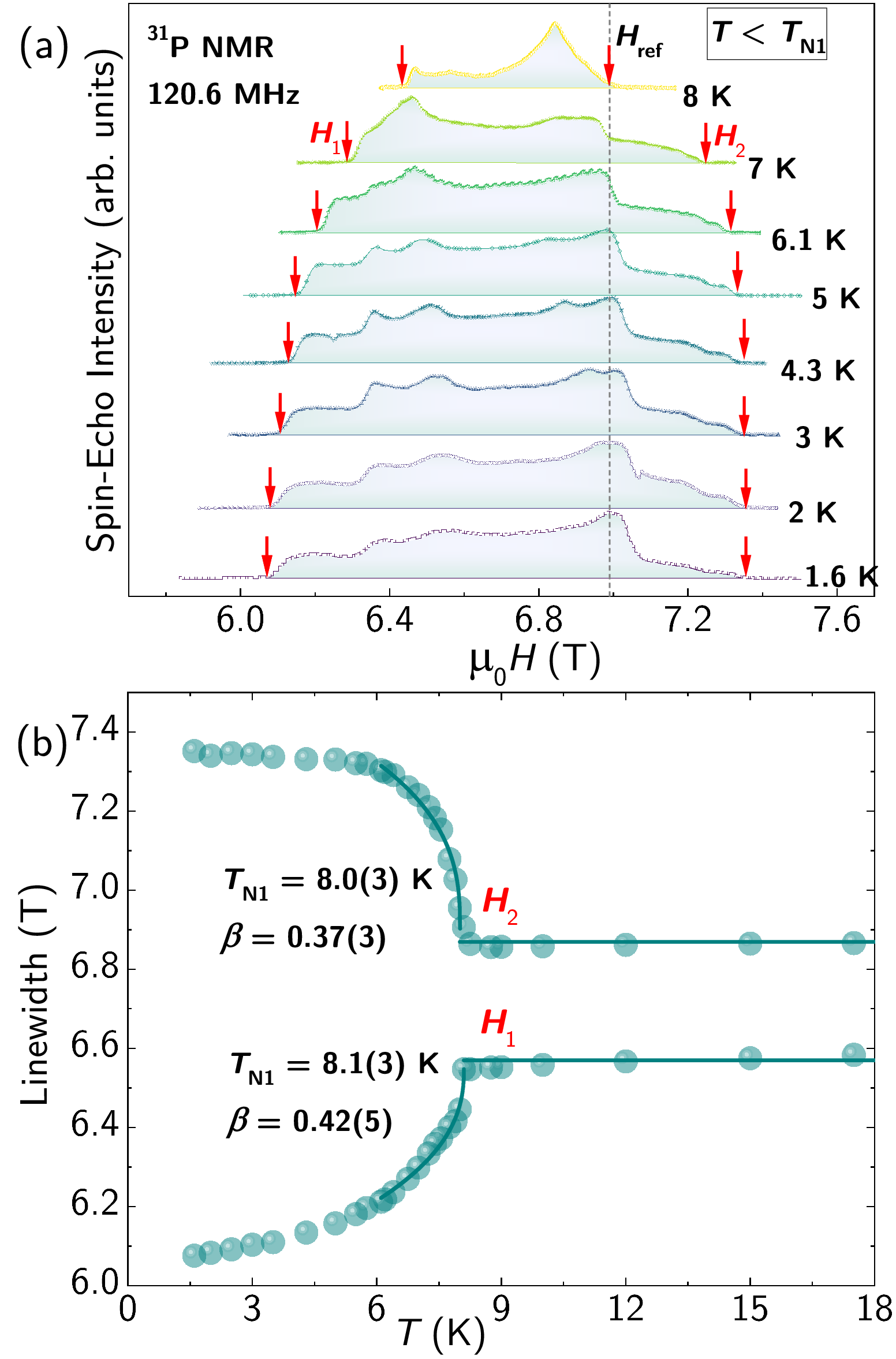}
    \caption{(a) $^{31}$P NMR spectra measured below 9~K at 120.6~MHz. The vertical dashed line corresponds to $^{31}$P zero-shift position. The downward arrows indicate the low-$H$ and high-$H$ edges of the NMR spectra. (b) Temperature dependence of the low- and high-$H$ edge positions ($H_1$ and $H_2$) of the NMR spectra. The dark solid lines are fitting results using Eq.~\eqref{Hint} in the critical regime.}
    \label{Fig11}
\end{figure}
Figure~\ref{Fig11}(a) shows the low temperature $^{31}$P NMR spectra at 120.6~MHz. Below $T_{\rm N1} \simeq 8$~K, the spectra broaden markedly and the two lines observed in the paramagnetic state merge into a single, broad, and anisotropic line. This rapid broadening signals the development of a static internal field at the $^{31}$P sites as the system enters the magnetic ordered state. At low temperatures, the spectra evolve into a nearly rectangular shape with multiple features. Such a rectangular pattern is a characteristic feature of a commensurate AFM state which arises due to the random orientation of internal hyperfine fields with respect to the external field in a polycrystalline sample~\cite{Yamada1751,Ranjith014415,Ranjith024422,Ambika015803}. 
%Such a lineshape arises from the random orientation of internal hyperfine fields relative to the external field in a polycrystalline sample~\cite{Yamada1751}.At 120.6~MHz, which corresponds to a magnetic field of $\sim 7$~T, the field-induced transition in $\chi(T)$ and $C_{\rm p}(T)$ appears at $T_{\rm N2}\simeq 6.8$~K. 
The features in our NMR spectra below $T_{\rm N1}$ evolve systematically with more number of shoulders as the temperature is lowered. No clear features could be assigned solely to $T_{\rm N2}$ expected from $\chi(T)$ and $C_{\rm p}(T)$.
%For a tentative estimation of line width, we simulated the line shape assuming the superposition of two P-sites and the commensurate AFM order. Though, the simulated rectangular powder pattern could not reproduce the entire experimental line shape, yet it gives a reasonable fit [see Fig.~\ref{Fig11}(a)].

In general, as the broadening of the lines is due to the static internal field at the P sites produced by the ordered Fe moments, the temperature dependence of the linewidth reflects that of the sublattice magnetization (i.e., the order parameter) in the magnetically ordered state. 
However, as described above, the observed spectra exhibit a complex shape due to the two inequivalent P sites, as well as the additional P1 site with a slightly different hyperfine coupling. As a result,  it is not straightforward to extract the temperature dependence of the static internal field from the spectra. 
Therefore, we tracked the temperature dependence of the low- and high-magnetic field edge positions of the spectra [$H_1$ and $H_2$, indicated by the arrows in Fig.~\ref{Fig11}(a)]. Assuming the low-field edge originates from  the internal field at the P1 site ($H_{\rm int,P1}$), $H_1$ can be written as $H_1(T)=H_{\rm 0,P1}-H_{\rm int, P1}(T)$, where $H_{\rm 0,P1}$ is the center position for the P1 line. Similarly, since the high-field edge H$_2$ most likely arises from the P2 line,  one can express it as  $H_2(T)=H_{\rm 0,P2}+H_{\rm int, P2}(T)$, where $H_{\rm int,P2}$ is the internal field at P2 and  $H_{\rm 0,P2}$ is the center position of the P2 line. The temperature dependencies of  $H_1$ and $H_2$ are shown in Fig.~\ref{Fig11}(b). Notably, $H_1$ and $H_2$ exhibit different temperature dependence: $H_2$ nearly saturates below $\sim$ 5 K, whereas $H_1$ keeps decreasing down to the lowest temperature measured. This indicates that $H_{\rm 0,P1}$ and $H_{\rm 0,P2}$ are not strictly temperature independent but decrease slightly with decreasing temperature. Nevertheless, we used the data to extract the critical exponent of the order parameter and fitted  the data
by a power-law of the following form   near $T_{\rm N1}$ (critical regime) assuming $H_{\rm 0,P1}$ and $H_{\rm 0,P2}$ are effectively constant in the limited regime:
\begin{equation}
H_i(T)  = H_{\rm 0,Pi} \mp H_{\rm int, Pi}(0)\left(1 - \frac{T}{T_{\rm N1}}\right)^{\beta}.
\label{Hint}
\end{equation}
Here, $\beta$ is the critical exponent of the order parameter and $i$ = 1 and 2 with minus or plus signs in the formula, respectively. The fits  in the range 6.1 - 8.1~K yields $\beta = 0.42(5)$, $H_{\rm int, P1}(0) = 0.59$~T, $T_{\rm N1} = 8.1(3)$~K,  and $H_{\rm  0,P1} \simeq$ 6.55 T for P1, and $\beta = 0.37(3)$, $H_{\rm int, P2}(0) = 0.70$ ~T, $T_{\rm N1} = 8.0(3)$~K,  and $H_{\rm 0,P2} \simeq$ 6.90 T for P2, respectively.
The slight different values of $\beta$ are due to the temperature dependence of $H_{\rm 0,P1}$ and $H_{\rm 0,P2}$ and are considered to be possible upper and lower values.  
Because these values are close to those expected for a 3D AFM~\cite{Nath214430,Ambika015803} and significantly greater than those expected for 2D AFM systems (0.125 for Ising model and 0.231 for XY model)~\cite{Collins1989,Ozeki2007,Bramwell1994}, the results suggest a 3D AFM nature of RFHPO.

\subsubsection{$^{31}$P spin-lattice relaxation rate ($1/T_1$)}
\begin{figure*}
\centering
 \includegraphics[width=\linewidth]{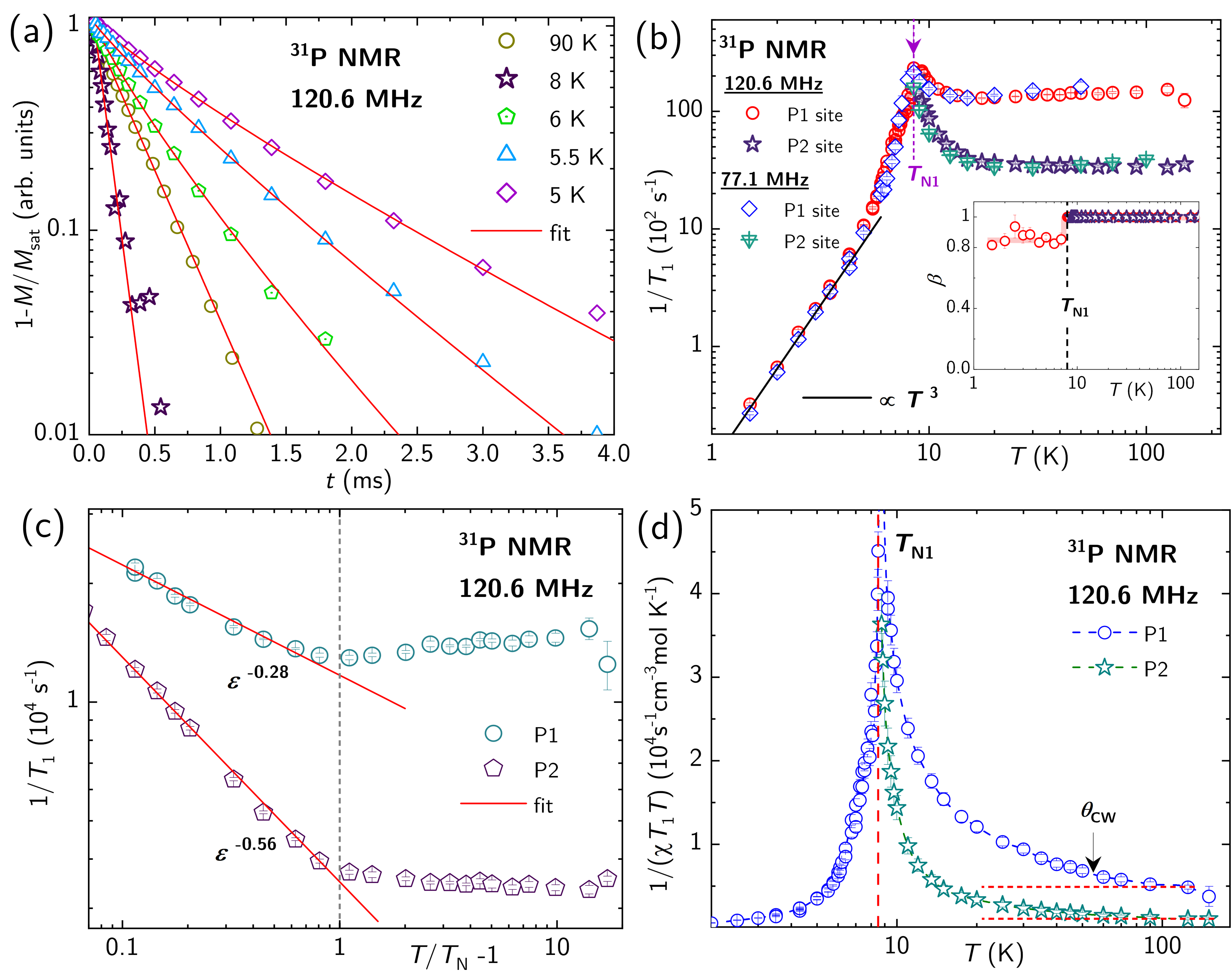}
\caption{(a)  Recovery of the longitudinal nuclear magnetization at 120.6~MHz measured at the P1 site. Solid lines are fits using Eq.~\eqref{exp}. (b) Temperature dependence of the $^{31}$P $1/T_1$ measured at 120.6~MHz and 77.1~MHz for both P1 and P2 sites. The downward arrow indicates $T_{\rm N1}$. Solid line below $T_{\rm N1}$ represent $T^3$ fit. Inset: temperature dependence of exponent ($\beta$) for both P1 and P2 sites. (c) $1/T_1$ plotted as a function of reduced temperature $\epsilon \equiv (T/T_{\rm N1} - 1)$ for both P-sites. Solid lines are the power-law fits in the critical regime. (d) Temperature dependence of $1/\chi T_1T$. The horizontal dotted lines mark the constant behaviour at high temperatures and the vertical dash-dotted line denotes $T_{\rm N}$.
}
\label{Fig12}
\end{figure*}
To probe the local spin dynamics, we measured the temperature-dependent $^{31}$P spin-lattice relaxation rate ($1/T_1$) down to 9~K at the central peak positions of both the P1 and P2 sites and at two frequencies 120.6~MHz ($\mu_0H \approx 7$~T) and 77.1~MHz ($\mu_0H \approx 4.5$~T). However, for $T<9$~K, since the line is very broad and two P-sites are merged, we measured $1/T_1$ at the extreme left and right shoulder positions. The values of $1/T_1$ at both the shoulder positions are found to be the same. For a nucleus with spin $I=1/2$, the recovery of the longitudinal magnetization typically follows a single exponential function. In the present case, the recovery curves are best described by a stretched exponential
\begin{equation}
1-\frac{M(t)}{M(\infty)} = A\, e^{-(t/T_1)^{\beta}},
\label{exp}
\end{equation}
where $M(t)$ and $M(\infty)$ are the nuclear magnetizations at time $t$ and at equilibrium, respectively, and $\beta$ is the stretch exponent. %Representative recovery curves at different temperatures and their fits are presented in Fig.~\ref{Fig12}(a). 
Above $T_{\rm N1}$, the recovery is single-exponential with $\beta = 1$, while at $T_{\rm N1}$ the value of $\beta$ drops suddenly and saturates to $\beta \simeq 0.8$ [see inset of Fig.~\ref{Fig12}(b)]. This suggests a distribution of relaxation time below $T_{\rm N1}$~\cite{Mohanty184435}.

Figure~\ref{Fig12}(b) presents the temperature dependence of $1/T_1$ for both P-sites. In the paramagnetic regime, the two P-sites exhibit distinct $1/T_1$ values due to their different hyperfine couplings. Above $\sim 15$~K, $1/T_1$ remains nearly constant, reflecting localized paramagnetic fluctuations. On approaching $T_{\rm N1}$, $1/T_1$ exhibits a sharp peak at both fields, evidencing critical slowing down of spin fluctuations near the magnetic ordering~\cite{Sebastian064413}. This value of $T_{\rm N1} \simeq 8$~K is consistent with the $\chi(T)$ and $C_{\rm p}(T)$ results. Below $T_{\rm N1}$, $1/T_1$ decreases rapidly, and at low-$T$s it follows a $T^3$ dependence, consistent with the two-magnon Raman process, as reported in other TLAFs~\cite{Ambika015803}. In contrast to the $\chi(T)$ and $C_{\rm p}(T)$ data, the field-induced transition at $T_{\rm N2}$ is not manifested in NMR $1/T_1(T)$. 
%One possible reason could be that %the associated fluctuations near $T_{\rm N2}$ occur on a time scale faster than the NMR window ($\mathcal{O}\approx 10^{-6}$~s), leading to a nearly featureless $1/T_1(T)$. Alternatively,  the ordering vector at $T_{\rm N2}$ may coincide with a form-factor node of the probe nuclei [$A(\vec{q})\approx 0$], which effectively filters out the critical fluctuations.

The divergent behaviour of $1/T_1$ near $T_{\rm N1}$ reflects critical fluctuations associated with a second-order transition. To analyze this behavior, we plotted $1/T_1$ versus reduced temperature $\epsilon = (T-T_{\rm N1})/T_{\rm N1}$ in Fig.~\ref{Fig12}(c). A power-law fit yields the critical exponent $\sim 0.28(3)$ and $\sim 0.58(3)$ for the P1 and P2 sites, respectively. %The value for P1 is close to the expected value ($\sim 0.3$) for a 3D Heisenberg universality class~\cite{Benner1990}, whereas the larger exponent for P2 may possibly arises from its weaker $A_{\rm hf}$. A weaker hyperfine coupling reduces the sensitivity to critical fluctuations, leading to a large uncertainty in the exponent value.
%Thus, the discrepancy between P1 and P2 reflects site-dependent variations in the local magnetic environment.
Compared to the P1 site, the P2 site shows more clear critical behavior.
The critical exponent value $\sim 0.56(5)$ for the P2 site is close to the expected value ($\sim 0.6$) for 3D Ising universality class~\cite{Benner1990}, whereas the smaller exponent for P1 may arise from a smaller form-factor. 
The smaller form-factor for P1 can also be inferred based on the observation that the $H_{\rm int, P1}$ is slightly smaller than $H_{\rm int, P2}$ as shown above.
Since the hyperfine coupling constant $A_{\rm hf}$ for P1 is twice as that for P2, the smaller internal magnetic field must stem from a smaller form-factor in P1.
Similar behaviors have been observed in iron-based superconductors CaK(Fe$_{1-x}$Ni$_x$)$_4$As$_4$ in which distinct critical fluctuations were observed at different As sites due to different form factors  \cite{Cui104512,Ding220510,Ding137204}. 

To probe the nature of spin fluctuations in the paramagnetic regime (above $T_{\rm N}$), we analyzed the temperature dependence of $1/(\chi T_1T)$, as presented in Fig.~\ref{Fig12}(d). At high temperatures, $1/(\chi T_1T)$ remains nearly constant, while it shows a gradual increase below $\sim 60$~K. The nuclear spin-lattice relaxation rate divided by temperature, $\frac{1}{T_1T}$, is related to the imaginary part of the dynamical spin susceptibility $\chi^{\prime\prime}_M(\vec{q},\omega_{\rm N})$ at the NMR frequency $\omega_{\rm N}$ through the relation~\cite{Moriya516}
\begin{equation}
\frac{1}{T_{1}T} = \frac{2\gamma_{\rm N}^{2}k_{\rm B}}{N_{\rm A}^{2}}
\sum\limits_{\vec{q}}\mid A(\vec{q})\mid
^{2}\frac{\chi^{\prime\prime}_{\rm M}(\vec{q},\omega_{\rm N})}{\omega_{\rm N}},
\label{t1form}
\end{equation}
where the summation extends over all wave vectors $\vec{q}$ in the first Brillouin zone, and $A(\vec{q})$ denotes the hyperfine form factor. For $q=0$ and $\omega_{\rm N}=0$, the real part of $\chi_{\rm M}(\vec{q},\omega_{\rm N})$ corresponds to the uniform static susceptibility ($\chi$). Therefore, the temperature-independent behavior of $1/(\chi T_1T)$ in the high-temperature region ($T > \theta_{\rm CW}$) implies that $1/T_1T$ is mainly governed by the uniform susceptibility $\chi$. The mild upturn below $\sim 60$~K signifies the development of spin correlations with $q \neq 0$ or short-range AFM fluctuations, as typically observed in low-dimensional frustrated spin systems~\cite{Nath214430}. 

\section{Discussion and Summary}
\begin{figure}[h]
\includegraphics[width=\columnwidth]{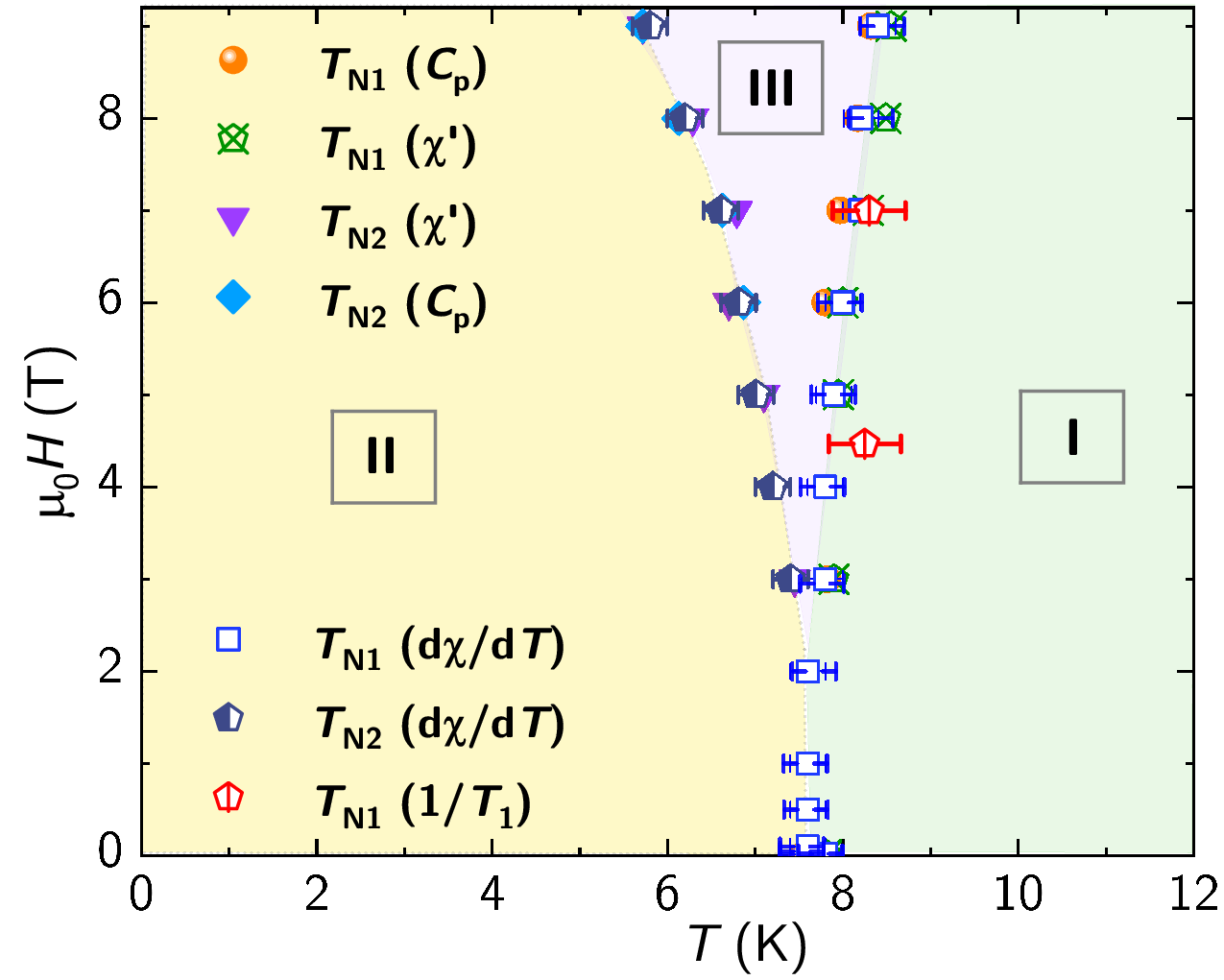}	
\caption{\label{Fig13} The $H-T$ phase diagram highlighting the evolution of $T_{\rm N1}$ and $T_{\rm N2}$ with $H$, obtained from the dc and ac susceptibility, heat capacity, and NMR measurements. For dc susceptibility, the transition temperatures are determined by plotting the temperature derivative. Three distinct phases are marked as (I), (II), and (III).}
\end{figure}

%For an ideal 2D Heisenberg TLAFM, a magnetic order is significantly influenced by frustration and quantum fluctuations. In a molecular field approximation, the ground state in a 2D triangular lattice is achieved with the magnetization sublattices for the planar structure arrange in a non-collinear $120\degree$ spin-structure in zero magnetic field ~\cite{Collins605}. The ground state of 2D TLAFM is highly degenerate in the presence of an externally applied magnetic field. However, the thermal effects and quantum fluctuations along with interplanar interactions lift this ground state degeneracy. The quantum fluctuations not only stabilize but also alter the various magnetic structures, selecting the phase with planar spin structure for a purely 2D Heisenberg TLAFM ~\cite{Golosov69}. Further, a collinear magnetic phase is stabilized by the fluctuations and easy-axis anistotropy in which a characteristic feature of magnetization plateau i.e., one-third of saturation magnetization is observed in $M-H$ curve. This magnetic structure is referred to as up-up-down ($uud$) state corresponding to two spins up and one spin down. This $uud$ phase has been experimentally discovered in several low and high spin TLAFMs for due to easy axis and easy-plane anisotropies, featuring the similar magnetic phase diagram ~\cite{Svistov094434,Ishii17001,Stefanie7982,Zvereva1550}.

The $H - T$ phase diagram of RFHPO presented in Fig.~\ref{Fig13} has been constructed using transition temperatures from $\chi(T)$, $\chi^{\prime}(T)$, $C_{\rm p}(T)$, and $1/T_1(T)$. Three regimes are identified as: (I) high-temperature paramagnetic state, (II) an AFM ordered state, and (III) a field-induced phase. In zero-field, the system undergoes a single transition at $T_{\rm N1}\simeq 7.8$~K into a commensurate AFM state, as confirmed by the rectangular NMR lineshape. Remarkably, $T_{\rm N1}$ shifts to higher temperatures for $\mu_0H > 3$~T, a contrasting behaviour expected for a conventional AFM. Typically, this unusual increase can possibly be attributed to the interplay of AFM LRO and quantum fluctuations. In low-dimensional and frustrated magnets, quantum fluctuations usually suppress the magnetic LRO. In RFHPO, when magnetic field is applied, a small field partially quenches/weakens these fluctuations; as a result, $T_{\rm N1}$ is shifted towards high temperatures. In higher fields, it is expected to shift back to low temperatures. A similar field-driven shift of the transition temperature has been observed in other high-spin frustrated TLAFs, such as Ba$_3$MnNb$_2$O$_9$, NH$_4$Fe(PO$_3$F)$_2$, and Rb$_4$Mn(MoO$_4$)$_3$~\cite{Lee224402,Ishii17001,Mohanty184435}. For $\mu_0H > 3$~T, an additional transition $T_{\rm N2}$ emerges, giving rise to the intermediate regime (III). Unlike $T_{\rm N1}$, $T_{\rm N2}$ shifts to low temperatures with increasing field, consistent with a conventional AFM nature. Clearly, there are at least two magnetic phases (II and III) that possess different spin reorientations. Therefore, we presume that this unusual phase boundary separating II and III could be due to magnetic anisotropy, which induces spin canting and hence, the complex phase diagram, akin to the behavior reported in several TLAFs~\cite{Lee224402,Garlea011038,Seabra214418}.
%Such field induced complexity is a hallmark of TLAFs~\cite{Seabra214418,Mitsuda513,Susuki267201}. 

In an ideal 2D Heisenberg TLAF, the ground state in zero-field is a non-collinear $120^{\circ}$ structure~\cite{Collins605}. Under applied fields, quantum fluctuations and interlayer couplings lift the degeneracy, stabilizing distinct spin configurations. A particularly interesting example is the stabilization of a collinear up–up–down ($uud$) state by the magnetic field, which is manifested as a $1/3$ magnetization plateau~\cite{Golosov69}. Indeed, such a plateau has been experimentally observed in several TLAFs ranging from low-spin ($S=1/2$) to high-spin cases~\cite{Shirata057205,Sakhratov014431}.
%Zvereva1550 For RFHPO, based on the value of average exchange coupling, the saturation field is estimated to be about $\mu_0H_{\rm sat} \simeq 50$~T, which lies beyond the field range of our present experiments.
In principle, the large spin ($S=5/2$) and weak anisotropy of Fe$^{3+}$ ions should still allow for the stabilization of a field-induced $1/3$ plateau in RFHPO. However, since the expected saturation field ($\mu_0H_{\rm sat} \simeq 50$~T) of RFHPO is very high, probing this $1/3$ plateau would require high magnetic fields, which is beyond the field range of our present experiments.

%Overall, the $H$–$T$ phase diagram demonstrates that RbFe(HPO$_3$)$_2$ belongs to the broader class of spin-$5/2$ TLAFs where moderate frustration ($f \approx 7$) and weak anisotropy combine to produce unconventional field-dependent transitions.

In summary, we have presented a comprehensive study of thermodynamic, static, and dynamic properties of RbFe(HPO$_3$)$_2$. The compound features equilateral triangular layers of Fe$^{3+}$ ($S=5/2$) ions stacked along the $c$-axis. Our analysis of $\chi(T)$ and $K_{\rm iso}(T)$ validates the Heisenberg TLAF model with an average exchange coupling of $J/k_{\rm B}\simeq 2.8$~K. It undergoes a magnetic ordering at $T_{\rm N1} \simeq 7.8$~K in zero-field, characterized by a distinct anomaly in $\chi(T)$, $C_{\rm p}(T)$, and $1/T_1(T)$. The $^{31}$P NMR spectra reveal two inequivalent phosphorus sites and the $K$–$\chi$ analysis yields isotropic hyperfine couplings of $A_{\rm hf}^{\rm iso} = 0.55(2)$~T/$\mu_{\rm B}$ and $0.25(3)$~T/$\mu_{\rm B}$ for the P1 and P2 sites, respectively. The NMR spectra further confirm a commensurate antiferromagnetic ordering below $T_{\rm N1}$. While the thermodynamic measurements establish a field-induced transition at $T_{\rm N2}$, this feature is absent in the $1/T_1(T)$ data. The resulting $H–-T$ phase diagram highlights the delicate competition between frustration, anisotropy, and dimensionality in stabilizing multiple distinct magnetic phases. Further studies on high-quality single crystals are essential to resolve the ambiguity about the field-induced transition at $T_{\rm N2}$, spin structures, and possible magnetization plateaus in this compound.

\acknowledgments
VN, SJS, SPP, and RN acknowledge SERB, India, for financial support bearing sanction Grant No.~CRG/2022/000997 and DST-FIST with Grant No.~SR/FST/PS-II/2018/54(C). SJS acknowledges the Fulbright-Nehru Doctoral Research Fellowship Award No.~2997/FNDR/2024-2025 and the Prime Minister's Research Fellowship (PMRF) scheme, Government of India, to be a visiting research scholar at the Ames National Laboratory. The research was supported by the US Department of Energy, Office of Basic Energy Sciences, Division of Materials Sciences and Engineering. Ames National Laboratory is operated for the US Department of Energy by Iowa State University under Contract No.~DEAC02-07CH11358.

%\bibliography{RFHPO}

\begin{thebibliography}{67}%
	\makeatletter
	\providecommand \@ifxundefined [1]{%
		\@ifx{#1\undefined}
	}%
	\providecommand \@ifnum [1]{%
		\ifnum #1\expandafter \@firstoftwo
		\else \expandafter \@secondoftwo
		\fi
	}%
	\providecommand \@ifx [1]{%
		\ifx #1\expandafter \@firstoftwo
		\else \expandafter \@secondoftwo
		\fi
	}%
	\providecommand \natexlab [1]{#1}%
	\providecommand \enquote  [1]{``#1''}%
	\providecommand \bibnamefont  [1]{#1}%
	\providecommand \bibfnamefont [1]{#1}%
	\providecommand \citenamefont [1]{#1}%
	\providecommand \href@noop [0]{\@secondoftwo}%
	\providecommand \href [0]{\begingroup \@sanitize@url \@href}%
	\providecommand \@href[1]{\@@startlink{#1}\@@href}%
	\providecommand \@@href[1]{\endgroup#1\@@endlink}%
	\providecommand \@sanitize@url [0]{\catcode `\\12\catcode `\$12\catcode
		`\&12\catcode `\#12\catcode `\^12\catcode `\_12\catcode `\%12\relax}%
	\providecommand \@@startlink[1]{}%
	\providecommand \@@endlink[0]{}%
	\providecommand \url  [0]{\begingroup\@sanitize@url \@url }%
	\providecommand \@url [1]{\endgroup\@href {#1}{\urlprefix }}%
	\providecommand \urlprefix  [0]{URL }%
	\providecommand \Eprint [0]{\href }%
	\providecommand \doibase [0]{https://doi.org/}%
	\providecommand \selectlanguage [0]{\@gobble}%
	\providecommand \bibinfo  [0]{\@secondoftwo}%
	\providecommand \bibfield  [0]{\@secondoftwo}%
	\providecommand \translation [1]{[#1]}%
	\providecommand \BibitemOpen [0]{}%
	\providecommand \bibitemStop [0]{}%
	\providecommand \bibitemNoStop [0]{.\EOS\space}%
	\providecommand \EOS [0]{\spacefactor3000\relax}%
	\providecommand \BibitemShut  [1]{\csname bibitem#1\endcsname}%
	\let\auto@bib@innerbib\@empty
	%</preamble>
	\bibitem [{\citenamefont {Diep}(2013)}]{Diep2013}%
	\BibitemOpen
	\bibfield  {author} {\bibinfo {author} {\bibfnamefont {H.~T.}\ \bibnamefont
			{Diep}},\ }\href {https://doi.org/10.1142/8676} {\emph {\bibinfo {title}
			{Frustrated Spin Systems}}},\ \bibinfo {edition} {2nd}\ ed.\ (\bibinfo
	{publisher} {WORLD SCIENTIFIC},\ \bibinfo {year} {2013})\BibitemShut
	{NoStop}%
	\bibitem [{\citenamefont {Anderson}(1973)}]{Anderson153}%
	\BibitemOpen
	\bibfield  {author} {\bibinfo {author} {\bibfnamefont {P.}~\bibnamefont
			{Anderson}},\ }\bibfield  {title} {\bibinfo {title} {Resonating valence
			bonds: A new kind of insulator?},\ }\href
	{https://doi.org/https://doi.org/10.1016/0025-5408(73)90167-0} {\bibfield
		{journal} {\bibinfo  {journal} {Mater. Res. Bull.}\ }\textbf {\bibinfo
			{volume} {8}},\ \bibinfo {pages} {153} (\bibinfo {year} {1973})}\BibitemShut
	{NoStop}%
	\bibitem [{\citenamefont {Balents}(2010)}]{Balents199}%
	\BibitemOpen
	\bibfield  {author} {\bibinfo {author} {\bibfnamefont {L.}~\bibnamefont
			{Balents}},\ }\bibfield  {title} {\bibinfo {title} {Spin liquids in
			frustrated magnets},\ }\href {https://doi.org/10.1038/nature08917} {\bibfield
		{journal} {\bibinfo  {journal} {Nature}\ }\textbf {\bibinfo {volume}
			{464}},\ \bibinfo {pages} {199} (\bibinfo {year} {2010})}\BibitemShut
	{NoStop}%
	\bibitem [{\citenamefont {Jolicoeur}\ and\ \citenamefont
		{Le~Guillou}(1989)}]{Jolicoeur2727}%
	\BibitemOpen
	\bibfield  {author} {\bibinfo {author} {\bibfnamefont {T.}~\bibnamefont
			{Jolicoeur}}\ and\ \bibinfo {author} {\bibfnamefont {J.~C.}\ \bibnamefont
			{Le~Guillou}},\ }\bibfield  {title} {\bibinfo {title} {{Spin-wave results for
				the triangular Heisenberg antiferromagnet}},\ }\href
	{https://doi.org/10.1103/PhysRevB.40.2727} {\bibfield  {journal} {\bibinfo
			{journal} {Phys. Rev. B}\ }\textbf {\bibinfo {volume} {40}},\ \bibinfo
		{pages} {2727} (\bibinfo {year} {1989})}\BibitemShut {NoStop}%
	\bibitem [{\citenamefont {Chubukov}\ \emph {et~al.}(1994)\citenamefont
		{Chubukov}, \citenamefont {Sachdev},\ and\ \citenamefont
		{Senthil}}]{Chubukov8891}%
	\BibitemOpen
	\bibfield  {author} {\bibinfo {author} {\bibfnamefont {A.~V.}\ \bibnamefont
			{Chubukov}}, \bibinfo {author} {\bibfnamefont {S.}~\bibnamefont {Sachdev}},\
		and\ \bibinfo {author} {\bibfnamefont {T.}~\bibnamefont {Senthil}},\
	}\bibfield  {title} {\bibinfo {title} {{Large-S expansion for quantum
				antiferromagnets on a triangular lattice}},\ }\href
	{https://doi.org/10.1088/0953-8984/6/42/019} {\bibfield  {journal} {\bibinfo
			{journal} {J. Phys. Condens. Matter.}\ }\textbf {\bibinfo {volume} {6}},\
		\bibinfo {pages} {8891} (\bibinfo {year} {1994})}\BibitemShut {NoStop}%
	\bibitem [{\citenamefont {Smirnov}\ \emph {et~al.}(2009)\citenamefont
		{Smirnov}, \citenamefont {Svistov}, \citenamefont {Prozorova}, \citenamefont
		{Zheludev}, \citenamefont {Lumsden}, \citenamefont {Ressouche}, \citenamefont
		{Petrenko}, \citenamefont {Nishikawa}, \citenamefont {Kimura}, \citenamefont
		{Hagiwara}, \citenamefont {Kindo}, \citenamefont {Shapiro},\ and\
		\citenamefont {Demianets}}]{Smirnov037202}%
	\BibitemOpen
	\bibfield  {author} {\bibinfo {author} {\bibfnamefont {A.~I.}\ \bibnamefont
			{Smirnov}}, \bibinfo {author} {\bibfnamefont {L.~E.}\ \bibnamefont
			{Svistov}}, \bibinfo {author} {\bibfnamefont {L.~A.}\ \bibnamefont
			{Prozorova}}, \bibinfo {author} {\bibfnamefont {A.}~\bibnamefont {Zheludev}},
		\bibinfo {author} {\bibfnamefont {M.~D.}\ \bibnamefont {Lumsden}}, \bibinfo
		{author} {\bibfnamefont {E.}~\bibnamefont {Ressouche}}, \bibinfo {author}
		{\bibfnamefont {O.~A.}\ \bibnamefont {Petrenko}}, \bibinfo {author}
		{\bibfnamefont {K.}~\bibnamefont {Nishikawa}}, \bibinfo {author}
		{\bibfnamefont {S.}~\bibnamefont {Kimura}}, \bibinfo {author} {\bibfnamefont
			{M.}~\bibnamefont {Hagiwara}}, \bibinfo {author} {\bibfnamefont
			{K.}~\bibnamefont {Kindo}}, \bibinfo {author} {\bibfnamefont {A.~Y.}\
			\bibnamefont {Shapiro}},\ and\ \bibinfo {author} {\bibfnamefont {L.~N.}\
			\bibnamefont {Demianets}},\ }\bibfield  {title} {\bibinfo {title} {Chiral and
			collinear ordering in a distorted triangular antiferromagnet},\ }\href
	{https://doi.org/10.1103/PhysRevLett.102.037202} {\bibfield  {journal}
		{\bibinfo  {journal} {Phys. Rev. Lett.}\ }\textbf {\bibinfo {volume} {102}},\
		\bibinfo {pages} {037202} (\bibinfo {year} {2009})}\BibitemShut {NoStop}%
	\bibitem [{\citenamefont {Gallegos}\ \emph {et~al.}(2025)\citenamefont
		{Gallegos}, \citenamefont {Jiang}, \citenamefont {White},\ and\ \citenamefont
		{Chernyshev}}]{Gallegos196702}%
	\BibitemOpen
	\bibfield  {author} {\bibinfo {author} {\bibfnamefont {C.~A.}\ \bibnamefont
			{Gallegos}}, \bibinfo {author} {\bibfnamefont {S.}~\bibnamefont {Jiang}},
		\bibinfo {author} {\bibfnamefont {S.~R.}\ \bibnamefont {White}},\ and\
		\bibinfo {author} {\bibfnamefont {A.~L.}\ \bibnamefont {Chernyshev}},\
	}\bibfield  {title} {\bibinfo {title} {{Phase Diagram of the Easy-Axis
				Triangular-Lattice ${J}_{1}\text{\ensuremath{-}}{J}_{2}$ Model}},\ }\href
	{https://doi.org/10.1103/PhysRevLett.134.196702} {\bibfield  {journal}
		{\bibinfo  {journal} {Phys. Rev. Lett.}\ }\textbf {\bibinfo {volume} {134}},\
		\bibinfo {pages} {196702} (\bibinfo {year} {2025})}\BibitemShut {NoStop}%
	\bibitem [{\citenamefont {Melchy}\ and\ \citenamefont
		{Zhitomirsky}(2009)}]{Melchy064411}%
	\BibitemOpen
	\bibfield  {author} {\bibinfo {author} {\bibfnamefont {P.-E.}\ \bibnamefont
			{Melchy}}\ and\ \bibinfo {author} {\bibfnamefont {M.~E.}\ \bibnamefont
			{Zhitomirsky}},\ }\bibfield  {title} {\bibinfo {title} {Interplay of
			anisotropy and frustration: Triple transitions in a triangular-lattice
			antiferromagnet},\ }\href {https://doi.org/10.1103/PhysRevB.80.064411}
	{\bibfield  {journal} {\bibinfo  {journal} {Phys. Rev. B}\ }\textbf {\bibinfo
			{volume} {80}},\ \bibinfo {pages} {064411} (\bibinfo {year}
		{2009})}\BibitemShut {NoStop}%
	\bibitem [{\citenamefont {Seabra}\ \emph {et~al.}(2011)\citenamefont {Seabra},
		\citenamefont {Momoi}, \citenamefont {Sindzingre},\ and\ \citenamefont
		{Shannon}}]{Seabra214418}%
	\BibitemOpen
	\bibfield  {author} {\bibinfo {author} {\bibfnamefont {L.}~\bibnamefont
			{Seabra}}, \bibinfo {author} {\bibfnamefont {T.}~\bibnamefont {Momoi}},
		\bibinfo {author} {\bibfnamefont {P.}~\bibnamefont {Sindzingre}},\ and\
		\bibinfo {author} {\bibfnamefont {N.}~\bibnamefont {Shannon}},\ }\bibfield
	{title} {\bibinfo {title} {Phase diagram of the classical heisenberg
			antiferromagnet on a triangular lattice in an applied magnetic field},\
	}\href {https://doi.org/10.1103/PhysRevB.84.214418} {\bibfield  {journal}
		{\bibinfo  {journal} {Phys. Rev. B}\ }\textbf {\bibinfo {volume} {84}},\
		\bibinfo {pages} {214418} (\bibinfo {year} {2011})}\BibitemShut {NoStop}%
	\bibitem [{\citenamefont {Collins}\ and\ \citenamefont
		{Petrenko}(1997)}]{Collins605}%
	\BibitemOpen
	\bibfield  {author} {\bibinfo {author} {\bibfnamefont {M.~F.}\ \bibnamefont
			{Collins}}\ and\ \bibinfo {author} {\bibfnamefont {O.~A.}\ \bibnamefont
			{Petrenko}},\ }\bibfield  {title} {\bibinfo {title} {Review/synthèse:
			Triangular antiferromagnets},\ }\href {https://doi.org/{10.1139/p97-007}}
	{\bibfield  {journal} {\bibinfo  {journal} {Can. J. Phys.}\ }\textbf
		{\bibinfo {volume} {75}},\ \bibinfo {pages} {605} (\bibinfo {year}
		{1997})}\BibitemShut {NoStop}%
	\bibitem [{\citenamefont {Kadowaki}\ \emph {et~al.}(1987)\citenamefont
		{Kadowaki}, \citenamefont {Ubukoshi},\ and\ \citenamefont
		{Hirakawa}}]{Kadowaki751}%
	\BibitemOpen
	\bibfield  {author} {\bibinfo {author} {\bibfnamefont {H.}~\bibnamefont
			{Kadowaki}}, \bibinfo {author} {\bibfnamefont {K.}~\bibnamefont {Ubukoshi}},\
		and\ \bibinfo {author} {\bibfnamefont {K.}~\bibnamefont {Hirakawa}},\
	}\bibfield  {title} {\bibinfo {title} {{Neutron Scattering Study of
				Successive Phase Transitions in Triangular Lattice Antiferromagnet
				CsNi${\mathrm{Cl}}_{3}$ }},\ }\href {https://doi.org/10.1143/JPSJ.56.751}
	{\bibfield  {journal} {\bibinfo  {journal} {J. Phys. Soc. Jpn}\ }\textbf
		{\bibinfo {volume} {56}},\ \bibinfo {pages} {751} (\bibinfo {year}
		{1987})}\BibitemShut {NoStop}%
	\bibitem [{\citenamefont {Doi}\ \emph {et~al.}(2004)\citenamefont {Doi},
		\citenamefont {Hinatsu},\ and\ \citenamefont {Ohoyama}}]{Doi8923}%
	\BibitemOpen
	\bibfield  {author} {\bibinfo {author} {\bibfnamefont {Y.}~\bibnamefont
			{Doi}}, \bibinfo {author} {\bibfnamefont {Y.}~\bibnamefont {Hinatsu}},\ and\
		\bibinfo {author} {\bibfnamefont {K.}~\bibnamefont {Ohoyama}},\ }\bibfield
	{title} {\bibinfo {title} {{Structural and magnetic properties of
				pseudo-two-dimensional triangular antiferromagnets
				${\mathrm{Ba}_{3}}{\mathrm{MSb}}_{2}{\mathrm{O}}_{9}$ (M = Mn, Co, and
				Ni)}},\ }\href {https://doi.org/10.1088/0953-8984/16/49/009} {\bibfield
		{journal} {\bibinfo  {journal} {J. Phys. Condens. Matter.}\ }\textbf
		{\bibinfo {volume} {16}},\ \bibinfo {pages} {8923} (\bibinfo {year}
		{2004})}\BibitemShut {NoStop}%
	\bibitem [{\citenamefont {Lee}\ \emph {et~al.}(2014)\citenamefont {Lee},
		\citenamefont {Choi}, \citenamefont {Huang}, \citenamefont {Ma},
		\citenamefont {Dela~Cruz}, \citenamefont {Matsuda}, \citenamefont {Tian},
		\citenamefont {Dun}, \citenamefont {Dong},\ and\ \citenamefont
		{Zhou}}]{Lee224402}%
	\BibitemOpen
	\bibfield  {author} {\bibinfo {author} {\bibfnamefont {M.}~\bibnamefont
			{Lee}}, \bibinfo {author} {\bibfnamefont {E.~S.}\ \bibnamefont {Choi}},
		\bibinfo {author} {\bibfnamefont {X.}~\bibnamefont {Huang}}, \bibinfo
		{author} {\bibfnamefont {J.}~\bibnamefont {Ma}}, \bibinfo {author}
		{\bibfnamefont {C.~R.}\ \bibnamefont {Dela~Cruz}}, \bibinfo {author}
		{\bibfnamefont {M.}~\bibnamefont {Matsuda}}, \bibinfo {author} {\bibfnamefont
			{W.}~\bibnamefont {Tian}}, \bibinfo {author} {\bibfnamefont {Z.~L.}\
			\bibnamefont {Dun}}, \bibinfo {author} {\bibfnamefont {S.}~\bibnamefont
			{Dong}},\ and\ \bibinfo {author} {\bibfnamefont {H.~D.}\ \bibnamefont
			{Zhou}},\ }\bibfield  {title} {\bibinfo {title} {{Magnetic phase diagram and
				multiferroicity of ${\mathrm{Ba}}_{3}{\mathrm{MnNb}}_{2}{\mathrm{O}}_{9}$: A
				spin-$\frac{5}{2}$ triangular lattice antiferromagnet with weak easy-axis
				anisotropy}},\ }\href {https://doi.org/10.1103/PhysRevB.90.224402} {\bibfield
		{journal} {\bibinfo  {journal} {Phys. Rev. B}\ }\textbf {\bibinfo {volume}
			{90}},\ \bibinfo {pages} {224402} (\bibinfo {year} {2014})}\BibitemShut
	{NoStop}%
	\bibitem [{\citenamefont {Ranjith}\ \emph {et~al.}(2017)\citenamefont
		{Ranjith}, \citenamefont {Brinda}, \citenamefont {Arjun}, \citenamefont
		{Hegde},\ and\ \citenamefont {Nath}}]{Ranjith115804}%
	\BibitemOpen
	\bibfield  {author} {\bibinfo {author} {\bibfnamefont {K.~M.}\ \bibnamefont
			{Ranjith}}, \bibinfo {author} {\bibfnamefont {K.}~\bibnamefont {Brinda}},
		\bibinfo {author} {\bibfnamefont {U.}~\bibnamefont {Arjun}}, \bibinfo
		{author} {\bibfnamefont {N.~G.}\ \bibnamefont {Hegde}},\ and\ \bibinfo
		{author} {\bibfnamefont {R.}~\bibnamefont {Nath}},\ }\bibfield  {title}
	{\bibinfo {title} {{Double phase transition in the triangular antiferromagnet
				${\mathrm{Ba}_{3}}{\mathrm{CoTa}_{2}}{\mathrm{O}_{9}}$}},\ }\href
	{https://doi.org/10.1088/1361-648X/aa57be} {\bibfield  {journal} {\bibinfo
			{journal} {J. Phys.: Condens. Matter}\ }\textbf {\bibinfo {volume} {29}},\
		\bibinfo {pages} {115804} (\bibinfo {year} {2017})}\BibitemShut {NoStop}%
	\bibitem [{\citenamefont {Lal}\ \emph {et~al.}(2023)\citenamefont {Lal},
		\citenamefont {Sebastian}, \citenamefont {Islam}, \citenamefont {Saravanan},
		\citenamefont {Uhlarz}, \citenamefont {Skourski},\ and\ \citenamefont
		{Nath}}]{Lal014429}%
	\BibitemOpen
	\bibfield  {author} {\bibinfo {author} {\bibfnamefont {S.}~\bibnamefont
			{Lal}}, \bibinfo {author} {\bibfnamefont {S.~J.}\ \bibnamefont {Sebastian}},
		\bibinfo {author} {\bibfnamefont {S.~S.}\ \bibnamefont {Islam}}, \bibinfo
		{author} {\bibfnamefont {M.~P.}\ \bibnamefont {Saravanan}}, \bibinfo {author}
		{\bibfnamefont {M.}~\bibnamefont {Uhlarz}}, \bibinfo {author} {\bibfnamefont
			{Y.}~\bibnamefont {Skourski}},\ and\ \bibinfo {author} {\bibfnamefont
			{R.}~\bibnamefont {Nath}},\ }\bibfield  {title} {\bibinfo {title} {{Double
				magnetic transitions and exotic field-induced phase in the triangular lattice
				antiferromagnets
				${\mathrm{Sr}}_{3}\mathrm{Co}{(\mathrm{Nb},\mathrm{Ta})}_{2}{\mathrm{O}}_{9}$}},\
	}\href {https://doi.org/10.1103/PhysRevB.108.014429} {\bibfield  {journal}
		{\bibinfo  {journal} {Phys. Rev. B}\ }\textbf {\bibinfo {volume} {108}},\
		\bibinfo {pages} {014429} (\bibinfo {year} {2023})}\BibitemShut {NoStop}%
	\bibitem [{\citenamefont {Sebastian}\ \emph {et~al.}(2022)\citenamefont
		{Sebastian}, \citenamefont {Islam}, \citenamefont {Jain}, \citenamefont
		{Yusuf}, \citenamefont {Uhlarz},\ and\ \citenamefont
		{Nath}}]{Sebastian104425}%
	\BibitemOpen
	\bibfield  {author} {\bibinfo {author} {\bibfnamefont {S.~J.}\ \bibnamefont
			{Sebastian}}, \bibinfo {author} {\bibfnamefont {S.~S.}\ \bibnamefont
			{Islam}}, \bibinfo {author} {\bibfnamefont {A.}~\bibnamefont {Jain}},
		\bibinfo {author} {\bibfnamefont {S.~M.}\ \bibnamefont {Yusuf}}, \bibinfo
		{author} {\bibfnamefont {M.}~\bibnamefont {Uhlarz}},\ and\ \bibinfo {author}
		{\bibfnamefont {R.}~\bibnamefont {Nath}},\ }\bibfield  {title} {\bibinfo
		{title} {{Collinear order in the spin-$\frac{5}{2}$ triangular-lattice
				antiferromagnet ${\mathrm{Na}}_{3}\mathrm{Fe}{({\mathrm{PO}}_{4})}_{2}$}},\
	}\href {https://doi.org/10.1103/PhysRevB.105.104425} {\bibfield  {journal}
		{\bibinfo  {journal} {Phys. Rev. B}\ }\textbf {\bibinfo {volume} {105}},\
		\bibinfo {pages} {104425} (\bibinfo {year} {2022})}\BibitemShut {NoStop}%
	\bibitem [{\citenamefont {Bhattacharya}\ \emph {et~al.}(2024)\citenamefont
		{Bhattacharya}, \citenamefont {Mohanty}, \citenamefont {Hillier},
		\citenamefont {Telling}, \citenamefont {Nath},\ and\ \citenamefont
		{Majumder}}]{MajumderL060403}%
	\BibitemOpen
	\bibfield  {author} {\bibinfo {author} {\bibfnamefont {K.}~\bibnamefont
			{Bhattacharya}}, \bibinfo {author} {\bibfnamefont {S.}~\bibnamefont
			{Mohanty}}, \bibinfo {author} {\bibfnamefont {A.~D.}\ \bibnamefont
			{Hillier}}, \bibinfo {author} {\bibfnamefont {M.~T.~F.}\ \bibnamefont
			{Telling}}, \bibinfo {author} {\bibfnamefont {R.}~\bibnamefont {Nath}},\ and\
		\bibinfo {author} {\bibfnamefont {M.}~\bibnamefont {Majumder}},\ }\bibfield
	{title} {\bibinfo {title} {{Evidence of quantum spin liquid state in a
				${\mathrm{Cu}}^{2+}$-based $S=\frac{1}{2}$ triangular lattice
				antiferromagnet}},\ }\href {https://doi.org/10.1103/PhysRevB.110.L060403}
	{\bibfield  {journal} {\bibinfo  {journal} {Phys. Rev. B}\ }\textbf {\bibinfo
			{volume} {110}},\ \bibinfo {pages} {L060403} (\bibinfo {year}
		{2024})}\BibitemShut {NoStop}%
	\bibitem [{\citenamefont {Somesh}\ \emph {et~al.}(2021)\citenamefont {Somesh},
		\citenamefont {Furukawa}, \citenamefont {Simutis}, \citenamefont {Bert},
		\citenamefont {Prinz-Zwick}, \citenamefont {B\"uttgen}, \citenamefont
		{Zorko}, \citenamefont {Tsirlin}, \citenamefont {Mendels},\ and\
		\citenamefont {Nath}}]{Somesh104422}%
	\BibitemOpen
	\bibfield  {author} {\bibinfo {author} {\bibfnamefont {K.}~\bibnamefont
			{Somesh}}, \bibinfo {author} {\bibfnamefont {Y.}~\bibnamefont {Furukawa}},
		\bibinfo {author} {\bibfnamefont {G.}~\bibnamefont {Simutis}}, \bibinfo
		{author} {\bibfnamefont {F.}~\bibnamefont {Bert}}, \bibinfo {author}
		{\bibfnamefont {M.}~\bibnamefont {Prinz-Zwick}}, \bibinfo {author}
		{\bibfnamefont {N.}~\bibnamefont {B\"uttgen}}, \bibinfo {author}
		{\bibfnamefont {A.}~\bibnamefont {Zorko}}, \bibinfo {author} {\bibfnamefont
			{A.~A.}\ \bibnamefont {Tsirlin}}, \bibinfo {author} {\bibfnamefont
			{P.}~\bibnamefont {Mendels}},\ and\ \bibinfo {author} {\bibfnamefont
			{R.}~\bibnamefont {Nath}},\ }\bibfield  {title} {\bibinfo {title} {Universal
			fluctuating regime in triangular chromate antiferromagnets},\ }\href
	{https://doi.org/10.1103/PhysRevB.104.104422} {\bibfield  {journal} {\bibinfo
			{journal} {Phys. Rev. B}\ }\textbf {\bibinfo {volume} {104}},\ \bibinfo
		{pages} {104422} (\bibinfo {year} {2021})}\BibitemShut {NoStop}%
	\bibitem [{\citenamefont {Serrano-González}\ \emph {et~al.}(1998)\citenamefont
		{Serrano-González}, \citenamefont {Bramwell}, \citenamefont {Harris},
		\citenamefont {Kariuki}, \citenamefont {Nixon}, \citenamefont {Parkin},\ and\
		\citenamefont {Ritter}}]{Serrano6314}%
	\BibitemOpen
	\bibfield  {author} {\bibinfo {author} {\bibfnamefont {H.}~\bibnamefont
			{Serrano-González}}, \bibinfo {author} {\bibfnamefont {S.~T.}\ \bibnamefont
			{Bramwell}}, \bibinfo {author} {\bibfnamefont {K.~D.~M.}\ \bibnamefont
			{Harris}}, \bibinfo {author} {\bibfnamefont {B.~M.}\ \bibnamefont {Kariuki}},
		\bibinfo {author} {\bibfnamefont {L.}~\bibnamefont {Nixon}}, \bibinfo
		{author} {\bibfnamefont {I.~P.}\ \bibnamefont {Parkin}},\ and\ \bibinfo
		{author} {\bibfnamefont {C.}~\bibnamefont {Ritter}},\ }\bibfield  {title}
	{\bibinfo {title} {{Magnetic structures of the triangular lattice magnets
				$A{\mathrm{Fe}(\mathrm{SO}_{4})_{2}(A=\mathrm{K, Rb,Cs})}$ }},\ }\href
	{https://doi.org/10.1063/1.367915} {\bibfield  {journal} {\bibinfo  {journal}
			{J. Appl. Phys.}\ }\textbf {\bibinfo {volume} {83}},\ \bibinfo {pages} {6314}
		(\bibinfo {year} {1998})}\BibitemShut {NoStop}%
	\bibitem [{\citenamefont {Inami}\ \emph {et~al.}(1996)\citenamefont {Inami},
		\citenamefont {Ajiro},\ and\ \citenamefont {Goto}}]{Inami2374}%
	\BibitemOpen
	\bibfield  {author} {\bibinfo {author} {\bibfnamefont {T.}~\bibnamefont
			{Inami}}, \bibinfo {author} {\bibfnamefont {Y.}~\bibnamefont {Ajiro}},\ and\
		\bibinfo {author} {\bibfnamefont {T.}~\bibnamefont {Goto}},\ }\bibfield
	{title} {\bibinfo {title} {{Magnetization Process of the Triangular Lattice
				Antiferromagnets, $\mathrm{RbFe}(\mathrm{MoO}_{4})_{2}$ and
				$\mathrm{CsFe}(\mathrm{SO}_{4})_{2}$}},\ }\href
	{https://doi.org/10.1143/JPSJ.65.2374} {\bibfield  {journal} {\bibinfo
			{journal} {J. Phys. Soc.}\ }\textbf {\bibinfo {volume} {65}},\ \bibinfo
		{pages} {2374} (\bibinfo {year} {1996})}\BibitemShut {NoStop}%
	\bibitem [{\citenamefont {Nilsen}\ \emph {et~al.}(2015)\citenamefont {Nilsen},
		\citenamefont {Raja}, \citenamefont {Tsirlin}, \citenamefont {Mutka},
		\citenamefont {Kasinathan}, \citenamefont {Ritter},\ and\ \citenamefont
		{Rønnow}}]{Nilsen113035}%
	\BibitemOpen
	\bibfield  {author} {\bibinfo {author} {\bibfnamefont {G.}~\bibnamefont
			{Nilsen}}, \bibinfo {author} {\bibfnamefont {A.}~\bibnamefont {Raja}},
		\bibinfo {author} {\bibfnamefont {A.}~\bibnamefont {Tsirlin}}, \bibinfo
		{author} {\bibfnamefont {H.}~\bibnamefont {Mutka}}, \bibinfo {author}
		{\bibfnamefont {D.}~\bibnamefont {Kasinathan}}, \bibinfo {author}
		{\bibfnamefont {C.}~\bibnamefont {Ritter}},\ and\ \bibinfo {author}
		{\bibfnamefont {H.}~\bibnamefont {Rønnow}},\ }\bibfield  {title} {\bibinfo
		{title} {{One-dimensional quantum magnetism in the anhydrous alum
				$\mathrm{KTi}$(${\mathrm{SO}}_{4})_{2}$}},\ }\href
	{https://doi.org/10.1088/1367-2630/17/11/113035} {\bibfield  {journal}
		{\bibinfo  {journal} {New J. Phys.}\ }\textbf {\bibinfo {volume} {17}},\
		\bibinfo {pages} {113035} (\bibinfo {year} {2015})}\BibitemShut {NoStop}%
	\bibitem [{\citenamefont {Svistov}\ \emph {et~al.}(2004)\citenamefont
		{Svistov}, \citenamefont {Smirnov}, \citenamefont {Prozorova}, \citenamefont
		{Petrenko}, \citenamefont {Shapiro},\ and\ \citenamefont
		{Dem’yanets}}]{Svistov204}%
	\BibitemOpen
	\bibfield  {author} {\bibinfo {author} {\bibfnamefont {L.~E.}\ \bibnamefont
			{Svistov}}, \bibinfo {author} {\bibfnamefont {A.~I.}\ \bibnamefont
			{Smirnov}}, \bibinfo {author} {\bibfnamefont {L.~A.}\ \bibnamefont
			{Prozorova}}, \bibinfo {author} {\bibfnamefont {O.~A.}\ \bibnamefont
			{Petrenko}}, \bibinfo {author} {\bibfnamefont {A.~Y.}\ \bibnamefont
			{Shapiro}},\ and\ \bibinfo {author} {\bibfnamefont {L.~N.}\ \bibnamefont
			{Dem’yanets}},\ }\bibfield  {title} {\bibinfo {title} {{On the possible
				coexistence of spiral and collinear structures in antiferromagnetic
				$\mathrm{KFe}(\mathrm{MoO}_{4})_{2}$}},\ }\href
	{https://doi.org/10.1134/1.1808851} {\bibfield  {journal} {\bibinfo
			{journal} {JETP Lett.}\ }\textbf {\bibinfo {volume} {80}},\ \bibinfo {pages}
		{204} (\bibinfo {year} {2004})}\BibitemShut {NoStop}%
	\bibitem [{\citenamefont {Svistov}\ \emph {et~al.}(2003)\citenamefont
		{Svistov}, \citenamefont {Smirnov}, \citenamefont {Prozorova}, \citenamefont
		{Petrenko}, \citenamefont {Demianets},\ and\ \citenamefont
		{Shapiro}}]{Svistov094434}%
	\BibitemOpen
	\bibfield  {author} {\bibinfo {author} {\bibfnamefont {L.~E.}\ \bibnamefont
			{Svistov}}, \bibinfo {author} {\bibfnamefont {A.~I.}\ \bibnamefont
			{Smirnov}}, \bibinfo {author} {\bibfnamefont {L.~A.}\ \bibnamefont
			{Prozorova}}, \bibinfo {author} {\bibfnamefont {O.~A.}\ \bibnamefont
			{Petrenko}}, \bibinfo {author} {\bibfnamefont {L.~N.}\ \bibnamefont
			{Demianets}},\ and\ \bibinfo {author} {\bibfnamefont {A.~Y.}\ \bibnamefont
			{Shapiro}},\ }\bibfield  {title} {\bibinfo {title} {{Quasi-two-dimensional
				antiferromagnet on a triangular lattice
				$\mathrm{RbFe}(\mathrm{MoO}_{4})_{2}$}},\ }\href
	{https://doi.org/10.1103/PhysRevB.67.094434} {\bibfield  {journal} {\bibinfo
			{journal} {Phys. Rev. B}\ }\textbf {\bibinfo {volume} {67}},\ \bibinfo
		{pages} {094434} (\bibinfo {year} {2003})}\BibitemShut {NoStop}%
	\bibitem [{\citenamefont {Mitamura}\ \emph {et~al.}(2014)\citenamefont
		{Mitamura}, \citenamefont {Watanuki}, \citenamefont {Kaneko}, \citenamefont
		{Onozaki}, \citenamefont {Amou}, \citenamefont {Kittaka}, \citenamefont
		{Kobayashi}, \citenamefont {Shimura}, \citenamefont {Yamamoto}, \citenamefont
		{Suzuki}, \citenamefont {Chi},\ and\ \citenamefont
		{Sakakibara}}]{Mitamura147202}%
	\BibitemOpen
	\bibfield  {author} {\bibinfo {author} {\bibfnamefont {H.}~\bibnamefont
			{Mitamura}}, \bibinfo {author} {\bibfnamefont {R.}~\bibnamefont {Watanuki}},
		\bibinfo {author} {\bibfnamefont {K.}~\bibnamefont {Kaneko}}, \bibinfo
		{author} {\bibfnamefont {N.}~\bibnamefont {Onozaki}}, \bibinfo {author}
		{\bibfnamefont {Y.}~\bibnamefont {Amou}}, \bibinfo {author} {\bibfnamefont
			{S.}~\bibnamefont {Kittaka}}, \bibinfo {author} {\bibfnamefont
			{R.}~\bibnamefont {Kobayashi}}, \bibinfo {author} {\bibfnamefont
			{Y.}~\bibnamefont {Shimura}}, \bibinfo {author} {\bibfnamefont
			{I.}~\bibnamefont {Yamamoto}}, \bibinfo {author} {\bibfnamefont
			{K.}~\bibnamefont {Suzuki}}, \bibinfo {author} {\bibfnamefont
			{S.}~\bibnamefont {Chi}},\ and\ \bibinfo {author} {\bibfnamefont
			{T.}~\bibnamefont {Sakakibara}},\ }\bibfield  {title} {\bibinfo {title}
		{{Spin-Chirality-Driven Ferroelectricity on a Perfect Triangular Lattice
				Antiferromagnet}},\ }\href {https://doi.org/10.1103/PhysRevLett.113.147202}
	{\bibfield  {journal} {\bibinfo  {journal} {Phys. Rev. Lett.}\ }\textbf
		{\bibinfo {volume} {113}},\ \bibinfo {pages} {147202} (\bibinfo {year}
		{2014})}\BibitemShut {NoStop}%
	\bibitem [{\citenamefont {Abdeldaim}\ \emph {et~al.}(2019)\citenamefont
		{Abdeldaim}, \citenamefont {Badrtdinov}, \citenamefont {Gibbs}, \citenamefont
		{Manuel}, \citenamefont {Walker}, \citenamefont {Le}, \citenamefont {Wu},
		\citenamefont {Wardecki}, \citenamefont {Eriksson}, \citenamefont {Kvashnin},
		\citenamefont {Tsirlin},\ and\ \citenamefont {Nilsen}}]{Abdel214427}%
	\BibitemOpen
	\bibfield  {author} {\bibinfo {author} {\bibfnamefont {A.~H.}\ \bibnamefont
			{Abdeldaim}}, \bibinfo {author} {\bibfnamefont {D.~I.}\ \bibnamefont
			{Badrtdinov}}, \bibinfo {author} {\bibfnamefont {A.~S.}\ \bibnamefont
			{Gibbs}}, \bibinfo {author} {\bibfnamefont {P.}~\bibnamefont {Manuel}},
		\bibinfo {author} {\bibfnamefont {H.~C.}\ \bibnamefont {Walker}}, \bibinfo
		{author} {\bibfnamefont {M.~D.}\ \bibnamefont {Le}}, \bibinfo {author}
		{\bibfnamefont {C.~H.}\ \bibnamefont {Wu}}, \bibinfo {author} {\bibfnamefont
			{D.}~\bibnamefont {Wardecki}}, \bibinfo {author} {\bibfnamefont {S.-G.}\
			\bibnamefont {Eriksson}}, \bibinfo {author} {\bibfnamefont {Y.~O.}\
			\bibnamefont {Kvashnin}}, \bibinfo {author} {\bibfnamefont {A.~A.}\
			\bibnamefont {Tsirlin}},\ and\ \bibinfo {author} {\bibfnamefont {G.~J.}\
			\bibnamefont {Nilsen}},\ }\bibfield  {title} {\bibinfo {title} {{Large
				easy-axis anisotropy in the one-dimensional magnet
				$\mathrm{BaMo}{({\mathrm{PO}}_{4})}_{2}$}},\ }\href
	{https://doi.org/10.1103/PhysRevB.100.214427} {\bibfield  {journal} {\bibinfo
			{journal} {Phys. Rev. B}\ }\textbf {\bibinfo {volume} {100}},\ \bibinfo
		{pages} {214427} (\bibinfo {year} {2019})}\BibitemShut {NoStop}%
	\bibitem [{\citenamefont {Svistov}\ \emph {et~al.}(2006)\citenamefont
		{Svistov}, \citenamefont {Smirnov}, \citenamefont {Prozorova}, \citenamefont
		{Petrenko}, \citenamefont {Micheler}, \citenamefont {B\"uttgen},
		\citenamefont {Shapiro},\ and\ \citenamefont {Demianets}}]{Svistov024412}%
	\BibitemOpen
	\bibfield  {author} {\bibinfo {author} {\bibfnamefont {L.~E.}\ \bibnamefont
			{Svistov}}, \bibinfo {author} {\bibfnamefont {A.~I.}\ \bibnamefont
			{Smirnov}}, \bibinfo {author} {\bibfnamefont {L.~A.}\ \bibnamefont
			{Prozorova}}, \bibinfo {author} {\bibfnamefont {O.~A.}\ \bibnamefont
			{Petrenko}}, \bibinfo {author} {\bibfnamefont {A.}~\bibnamefont {Micheler}},
		\bibinfo {author} {\bibfnamefont {N.}~\bibnamefont {B\"uttgen}}, \bibinfo
		{author} {\bibfnamefont {A.~Y.}\ \bibnamefont {Shapiro}},\ and\ \bibinfo
		{author} {\bibfnamefont {L.~N.}\ \bibnamefont {Demianets}},\ }\bibfield
	{title} {\bibinfo {title} {{Magnetic phase diagram, critical behavior, and
				two-dimensional to three-dimensional crossover in the triangular lattice
				antiferromagnet $\mathrm{RbFe}(\mathrm{MoO}_{4})_{2}$}},\ }\href
	{https://doi.org/10.1103/PhysRevB.74.024412} {\bibfield  {journal} {\bibinfo
			{journal} {Phys. Rev. B}\ }\textbf {\bibinfo {volume} {74}},\ \bibinfo
		{pages} {024412} (\bibinfo {year} {2006})}\BibitemShut {NoStop}%
	\bibitem [{\citenamefont {Siebeneichler}\ \emph {et~al.}(2022)\citenamefont
		{Siebeneichler}, \citenamefont {Dorn}, \citenamefont {Ovchinnikov},
		\citenamefont {Papawassiliou}, \citenamefont {da~Silva}, \citenamefont
		{Smetana}, \citenamefont {Pell},\ and\ \citenamefont
		{Mudring}}]{Stefanie7982}%
	\BibitemOpen
	\bibfield  {author} {\bibinfo {author} {\bibfnamefont {S.}~\bibnamefont
			{Siebeneichler}}, \bibinfo {author} {\bibfnamefont {K.~V.}\ \bibnamefont
			{Dorn}}, \bibinfo {author} {\bibfnamefont {A.}~\bibnamefont {Ovchinnikov}},
		\bibinfo {author} {\bibfnamefont {W.}~\bibnamefont {Papawassiliou}}, \bibinfo
		{author} {\bibfnamefont {I.}~\bibnamefont {da~Silva}}, \bibinfo {author}
		{\bibfnamefont {V.}~\bibnamefont {Smetana}}, \bibinfo {author} {\bibfnamefont
			{A.~J.}\ \bibnamefont {Pell}},\ and\ \bibinfo {author} {\bibfnamefont
			{A.-V.}\ \bibnamefont {Mudring}},\ }\bibfield  {title} {\bibinfo {title}
		{{Frustration and 120° Magnetic Ordering in the Layered Triangular
				Antiferromagnets ${\mathrm{AFe}}({\mathrm{PO}_{3}}{\mathrm{F}})_{2}$ (A = K,
				(${\mathrm{NH}_{4}})_{2}$Cl, N${\mathrm{H}_{4}}$, Rb, and Cs)}},\ }\href
	{https://doi.org/{10.1021/acs.chemmater.2c01916}} {\bibfield  {journal}
		{\bibinfo  {journal} {Chem. Mater.}\ }\textbf {\bibinfo {volume} {34}},\
		\bibinfo {pages} {7982} (\bibinfo {year} {2022})}\BibitemShut {NoStop}%
	\bibitem [{\citenamefont {Mohanty}\ \emph {et~al.}(2025)\citenamefont
		{Mohanty}, \citenamefont {Ranjith}, \citenamefont {Saramgi}, \citenamefont
		{Skourski}, \citenamefont {B\"uchner}, \citenamefont {Grafe},\ and\
		\citenamefont {Nath}}]{Mohanty184435}%
	\BibitemOpen
	\bibfield  {author} {\bibinfo {author} {\bibfnamefont {S.}~\bibnamefont
			{Mohanty}}, \bibinfo {author} {\bibfnamefont {K.~M.}\ \bibnamefont
			{Ranjith}}, \bibinfo {author} {\bibfnamefont {C.~S.}\ \bibnamefont
			{Saramgi}}, \bibinfo {author} {\bibfnamefont {Y.}~\bibnamefont {Skourski}},
		\bibinfo {author} {\bibfnamefont {B.}~\bibnamefont {B\"uchner}}, \bibinfo
		{author} {\bibfnamefont {H.-J.}\ \bibnamefont {Grafe}},\ and\ \bibinfo
		{author} {\bibfnamefont {R.}~\bibnamefont {Nath}},\ }\bibfield  {title}
	{\bibinfo {title} {{Ground-state properties of the spin-$\frac{5}{2}$
				frustrated triangular lattice antiferromagnet
				${\mathrm{NH}}_{4}{\mathrm{Fe}(\mathrm{PO}}_{3}{\mathrm{F})}_{2}$}},\ }\href
	{https://doi.org/10.1103/PhysRevB.111.184435} {\bibfield  {journal} {\bibinfo
			{journal} {Phys. Rev. B}\ }\textbf {\bibinfo {volume} {111}},\ \bibinfo
		{pages} {184435} (\bibinfo {year} {2025})}\BibitemShut {NoStop}%
	\bibitem [{\citenamefont {Hamchaoui}\ \emph {et~al.}(2013)\citenamefont
		{Hamchaoui}, \citenamefont {Alonzo}, \citenamefont {Venegas-Yazigi},
		\citenamefont {Rebbah},\ and\ \citenamefont {{Le Fur}}}]{Hamchaoui295}%
	\BibitemOpen
	\bibfield  {author} {\bibinfo {author} {\bibfnamefont {F.}~\bibnamefont
			{Hamchaoui}}, \bibinfo {author} {\bibfnamefont {V.}~\bibnamefont {Alonzo}},
		\bibinfo {author} {\bibfnamefont {D.}~\bibnamefont {Venegas-Yazigi}},
		\bibinfo {author} {\bibfnamefont {H.}~\bibnamefont {Rebbah}},\ and\ \bibinfo
		{author} {\bibfnamefont {E.}~\bibnamefont {{Le Fur}}},\ }\bibfield  {title}
	{\bibinfo {title} {{Six novel transition-metal phosphite compounds, with
				structure related to yavapaiite: Crystal structures and magnetic and thermal
				properties of ${A^{I}}[{M^{III}(\mathrm{HPO}_{3})_{2}}]$ ($A \mathrm{=K,
					NH_{4}, Rb}$ and $M = {\mathrm{V, Fe}}$)}},\ }\href
	{https://doi.org/https://doi.org/10.1016/j.jssc.2012.10.007} {\bibfield
		{journal} {\bibinfo  {journal} {J. Sol. State Chem.}\ }\textbf {\bibinfo
			{volume} {198}},\ \bibinfo {pages} {295} (\bibinfo {year}
		{2013})}\BibitemShut {NoStop}%
	\bibitem [{\citenamefont {Todo}\ and\ \citenamefont {Kato}(2001)}]{Todo047203}%
	\BibitemOpen
	\bibfield  {author} {\bibinfo {author} {\bibfnamefont {S.}~\bibnamefont
			{Todo}}\ and\ \bibinfo {author} {\bibfnamefont {K.}~\bibnamefont {Kato}},\
	}\bibfield  {title} {\bibinfo {title} {{Cluster Algorithms for General-
				$\mathit{S}$ Quantum Spin Systems}},\ }\href
	{https://doi.org/10.1103/PhysRevLett.87.047203} {\bibfield  {journal}
		{\bibinfo  {journal} {Phys. Rev. Lett.}\ }\textbf {\bibinfo {volume} {87}},\
		\bibinfo {pages} {047203} (\bibinfo {year} {2001})}\BibitemShut {NoStop}%
	\bibitem [{\citenamefont {Albuquerque}\ \emph {et~al.}(2007)\citenamefont
		{Albuquerque}, \citenamefont {Alet}, \citenamefont {Corboz}, \citenamefont
		{Dayal}, \citenamefont {Feiguin}, \citenamefont {Fuchs}, \citenamefont
		{Gamper}, \citenamefont {Gull}, \citenamefont {Gürtler}, \citenamefont
		{Honecker}, \citenamefont {Igarashi}, \citenamefont {Körner}, \citenamefont
		{Kozhevnikov}, \citenamefont {Läuchli}, \citenamefont {Manmana},
		\citenamefont {Matsumoto}, \citenamefont {McCulloch}, \citenamefont {Michel},
		\citenamefont {Noack}, \citenamefont {Pawłowski}, \citenamefont {Pollet},
		\citenamefont {Pruschke}, \citenamefont {Schollwöck}, \citenamefont {Todo},
		\citenamefont {Trebst}, \citenamefont {Troyer}, \citenamefont {Werner},\ and\
		\citenamefont {Wessel}}]{Albuquerque1187}%
	\BibitemOpen
	\bibfield  {author} {\bibinfo {author} {\bibfnamefont {A.}~\bibnamefont
			{Albuquerque}}, \bibinfo {author} {\bibfnamefont {F.}~\bibnamefont {Alet}},
		\bibinfo {author} {\bibfnamefont {P.}~\bibnamefont {Corboz}}, \bibinfo
		{author} {\bibfnamefont {P.}~\bibnamefont {Dayal}}, \bibinfo {author}
		{\bibfnamefont {A.}~\bibnamefont {Feiguin}}, \bibinfo {author} {\bibfnamefont
			{S.}~\bibnamefont {Fuchs}}, \bibinfo {author} {\bibfnamefont
			{L.}~\bibnamefont {Gamper}}, \bibinfo {author} {\bibfnamefont
			{E.}~\bibnamefont {Gull}}, \bibinfo {author} {\bibfnamefont {S.}~\bibnamefont
			{Gürtler}}, \bibinfo {author} {\bibfnamefont {A.}~\bibnamefont {Honecker}},
		\bibinfo {author} {\bibfnamefont {R.}~\bibnamefont {Igarashi}}, \bibinfo
		{author} {\bibfnamefont {M.}~\bibnamefont {Körner}}, \bibinfo {author}
		{\bibfnamefont {A.}~\bibnamefont {Kozhevnikov}}, \bibinfo {author}
		{\bibfnamefont {A.}~\bibnamefont {Läuchli}}, \bibinfo {author}
		{\bibfnamefont {S.}~\bibnamefont {Manmana}}, \bibinfo {author} {\bibfnamefont
			{M.}~\bibnamefont {Matsumoto}}, \bibinfo {author} {\bibfnamefont
			{I.}~\bibnamefont {McCulloch}}, \bibinfo {author} {\bibfnamefont
			{F.}~\bibnamefont {Michel}}, \bibinfo {author} {\bibfnamefont
			{R.}~\bibnamefont {Noack}}, \bibinfo {author} {\bibfnamefont
			{G.}~\bibnamefont {Pawłowski}}, \bibinfo {author} {\bibfnamefont
			{L.}~\bibnamefont {Pollet}}, \bibinfo {author} {\bibfnamefont
			{T.}~\bibnamefont {Pruschke}}, \bibinfo {author} {\bibfnamefont
			{U.}~\bibnamefont {Schollwöck}}, \bibinfo {author} {\bibfnamefont
			{S.}~\bibnamefont {Todo}}, \bibinfo {author} {\bibfnamefont {S.}~\bibnamefont
			{Trebst}}, \bibinfo {author} {\bibfnamefont {M.}~\bibnamefont {Troyer}},
		\bibinfo {author} {\bibfnamefont {P.}~\bibnamefont {Werner}},\ and\ \bibinfo
		{author} {\bibfnamefont {S.}~\bibnamefont {Wessel}},\ }\bibfield  {title}
	{\bibinfo {title} {{The ALPS project release 1.3: Open-source software for
				strongly correlated systems}},\ }\href
	{https://doi.org/https://doi.org/10.1016/j.jmmm.2006.10.304} {\bibfield
		{journal} {\bibinfo  {journal} {J. Magn. Magn. Mater.}\ }\textbf {\bibinfo
			{volume} {310}},\ \bibinfo {pages} {1187} (\bibinfo {year}
		{2007})}\BibitemShut {NoStop}%
	\bibitem [{\citenamefont {Bain}\ and\ \citenamefont {Berry}(2008)}]{Bain532}%
	\BibitemOpen
	\bibfield  {author} {\bibinfo {author} {\bibfnamefont {G.~A.}\ \bibnamefont
			{Bain}}\ and\ \bibinfo {author} {\bibfnamefont {J.~F.}\ \bibnamefont
			{Berry}},\ }\bibfield  {title} {\bibinfo {title} {Diamagnetic corrections and
			pascal's constants},\ }\href {https://doi.org/10.1021/ed085p532} {\bibfield
		{journal} {\bibinfo  {journal} {J. Chem. Edu.}\ }\textbf {\bibinfo {volume}
			{85}},\ \bibinfo {pages} {532} (\bibinfo {year} {2008})}\BibitemShut
	{NoStop}%
	\bibitem [{\citenamefont {Domb}\ and\ \citenamefont {Miedema}(1964)}]{Domb296}%
	\BibitemOpen
	\bibfield  {author} {\bibinfo {author} {\bibfnamefont {C.}~\bibnamefont
			{Domb}}\ and\ \bibinfo {author} {\bibfnamefont {A.}~\bibnamefont {Miedema}},\
	}\bibfield  {title} {\bibinfo {title} {{Chapter VI Magnetic Transitions}},\
	}\href {https://doi.org/https://doi.org/10.1016/S0079-6417(08)60154-7}
	{\bibfield  {journal} {\bibinfo  {journal} {Prog. Low Temp. Phys.}\ }\textbf
		{\bibinfo {volume} {4}},\ \bibinfo {pages} {296} (\bibinfo {year}
		{1964})}\BibitemShut {NoStop}%
	\bibitem [{\citenamefont {Kawamura}\ and\ \citenamefont
		{Miyashita}(1985)}]{Kawamura4530}%
	\BibitemOpen
	\bibfield  {author} {\bibinfo {author} {\bibfnamefont {H.}~\bibnamefont
			{Kawamura}}\ and\ \bibinfo {author} {\bibfnamefont {S.}~\bibnamefont
			{Miyashita}},\ }\bibfield  {title} {\bibinfo {title} {{Phase Transition of
				the Heisenberg Antiferromagnet on the Triangular Lattice in a Magnetic
				Field}},\ }\href {https://doi.org/10.1143/JPSJ.54.4530} {\bibfield  {journal}
		{\bibinfo  {journal} {J. Phys. Soc. Jpn.}\ }\textbf {\bibinfo {volume}
			{54}},\ \bibinfo {pages} {4530} (\bibinfo {year} {1985})}\BibitemShut
	{NoStop}%
	\bibitem [{\citenamefont {Delmas}\ \emph {et~al.}(1978)\citenamefont {Delmas},
		\citenamefont {{Le Flem}}, \citenamefont {Fouassier},\ and\ \citenamefont
		{Hagenmuller}}]{Delmas55}%
	\BibitemOpen
	\bibfield  {author} {\bibinfo {author} {\bibfnamefont {C.}~\bibnamefont
			{Delmas}}, \bibinfo {author} {\bibfnamefont {G.}~\bibnamefont {{Le Flem}}},
		\bibinfo {author} {\bibfnamefont {C.}~\bibnamefont {Fouassier}},\ and\
		\bibinfo {author} {\bibfnamefont {P.}~\bibnamefont {Hagenmuller}},\
	}\bibfield  {title} {\bibinfo {title} {{Etude comparative des proprietes
				magnetiques des oxydes lamellaires $A{\mathrm{CrO}_{2}} {(A \mathrm{= Li, Na,
						K)—II}}$: Calcul des integrales d'echange}},\ }\href
	{https://doi.org/https://doi.org/10.1016/0022-3697(78)90200-7} {\bibfield
		{journal} {\bibinfo  {journal} {J. Phys. Chem. Sol.}\ }\textbf {\bibinfo
			{volume} {39}},\ \bibinfo {pages} {55} (\bibinfo {year} {1978})}\BibitemShut
	{NoStop}%
	\bibitem [{\citenamefont {Schmidt}\ \emph {et~al.}(2011)\citenamefont
		{Schmidt}, \citenamefont {Lohmann},\ and\ \citenamefont
		{Richter}}]{Schmidt104443}%
	\BibitemOpen
	\bibfield  {author} {\bibinfo {author} {\bibfnamefont {H.-J.}\ \bibnamefont
			{Schmidt}}, \bibinfo {author} {\bibfnamefont {A.}~\bibnamefont {Lohmann}},\
		and\ \bibinfo {author} {\bibfnamefont {J.}~\bibnamefont {Richter}},\
	}\bibfield  {title} {\bibinfo {title} {{Eighth-order high-temperature
				expansion for general Heisenberg Hamiltonians}},\ }\href
	{https://doi.org/10.1103/PhysRevB.84.104443} {\bibfield  {journal} {\bibinfo
			{journal} {Phys. Rev. B}\ }\textbf {\bibinfo {volume} {84}},\ \bibinfo
		{pages} {104443} (\bibinfo {year} {2011})}\BibitemShut {NoStop}%
	\bibitem [{\citenamefont {Nalbandyan}\ \emph {et~al.}(2015)\citenamefont
		{Nalbandyan}, \citenamefont {Zvereva}, \citenamefont {Nikulin}, \citenamefont
		{Shukaev}, \citenamefont {Whangbo}, \citenamefont {Koo}, \citenamefont
		{Abdel-Hafiez}, \citenamefont {Chen}, \citenamefont {Koo}, \citenamefont
		{Vasiliev},\ and\ \citenamefont {Klingeler}}]{Nalbandyan1705}%
	\BibitemOpen
	\bibfield  {author} {\bibinfo {author} {\bibfnamefont {V.~B.}\ \bibnamefont
			{Nalbandyan}}, \bibinfo {author} {\bibfnamefont {E.~A.}\ \bibnamefont
			{Zvereva}}, \bibinfo {author} {\bibfnamefont {A.~Y.}\ \bibnamefont
			{Nikulin}}, \bibinfo {author} {\bibfnamefont {I.~L.}\ \bibnamefont
			{Shukaev}}, \bibinfo {author} {\bibfnamefont {M.-H.}\ \bibnamefont
			{Whangbo}}, \bibinfo {author} {\bibfnamefont {H.-J.}\ \bibnamefont {Koo}},
		\bibinfo {author} {\bibfnamefont {M.}~\bibnamefont {Abdel-Hafiez}}, \bibinfo
		{author} {\bibfnamefont {X.-J.}\ \bibnamefont {Chen}}, \bibinfo {author}
		{\bibfnamefont {C.}~\bibnamefont {Koo}}, \bibinfo {author} {\bibfnamefont
			{A.~N.}\ \bibnamefont {Vasiliev}},\ and\ \bibinfo {author} {\bibfnamefont
			{R.}~\bibnamefont {Klingeler}},\ }\bibfield  {title} {\bibinfo {title} {{New
				Phase of ${\mathrm{MnSb}_{2}{\mathrm{O}_{6}}}$ Prepared by Ion Exchange:
				Structural, Magnetic, and Thermodynamic Properties}},\ }\href
	{https://doi.org/{10.1021/ic502666c}} {\bibfield  {journal} {\bibinfo
			{journal} {Inorg. Chem.}\ }\textbf {\bibinfo {volume} {54}},\ \bibinfo
		{pages} {1705} (\bibinfo {year} {2015})}\BibitemShut {NoStop}%
	\bibitem [{\citenamefont {Ghosh}\ \emph {et~al.}(2018)\citenamefont {Ghosh},
		\citenamefont {Chen}, \citenamefont {Qiu}, \citenamefont {Hsieh},
		\citenamefont {Shao}, \citenamefont {Du}, \citenamefont {Wang}, \citenamefont
		{Chin}, \citenamefont {Chiou}, \citenamefont {Ray}, \citenamefont {Tsai},
		\citenamefont {Pao}, \citenamefont {Lin}, \citenamefont {Lee}, \citenamefont
		{Sankar}, \citenamefont {Chou},\ and\ \citenamefont {Pong}}]{Ghosh15779}%
	\BibitemOpen
	\bibfield  {author} {\bibinfo {author} {\bibfnamefont {A.}~\bibnamefont
			{Ghosh}}, \bibinfo {author} {\bibfnamefont {K.-H.}\ \bibnamefont {Chen}},
		\bibinfo {author} {\bibfnamefont {X.-S.}\ \bibnamefont {Qiu}}, \bibinfo
		{author} {\bibfnamefont {S.~H.}\ \bibnamefont {Hsieh}}, \bibinfo {author}
		{\bibfnamefont {Y.~C.}\ \bibnamefont {Shao}}, \bibinfo {author}
		{\bibfnamefont {C.~H.}\ \bibnamefont {Du}}, \bibinfo {author} {\bibfnamefont
			{H.~T.}\ \bibnamefont {Wang}}, \bibinfo {author} {\bibfnamefont {Y.~Y.}\
			\bibnamefont {Chin}}, \bibinfo {author} {\bibfnamefont {J.~W.}\ \bibnamefont
			{Chiou}}, \bibinfo {author} {\bibfnamefont {S.~C.}\ \bibnamefont {Ray}},
		\bibinfo {author} {\bibfnamefont {H.~M.}\ \bibnamefont {Tsai}}, \bibinfo
		{author} {\bibfnamefont {C.~W.}\ \bibnamefont {Pao}}, \bibinfo {author}
		{\bibfnamefont {H.~J.}\ \bibnamefont {Lin}}, \bibinfo {author} {\bibfnamefont
			{J.~F.}\ \bibnamefont {Lee}}, \bibinfo {author} {\bibfnamefont
			{R.}~\bibnamefont {Sankar}}, \bibinfo {author} {\bibfnamefont {F.~C.}\
			\bibnamefont {Chou}},\ and\ \bibinfo {author} {\bibfnamefont {W.~F.}\
			\bibnamefont {Pong}},\ }\bibfield  {title} {\bibinfo {title} {{Anisotropy in
				the magnetic interaction and lattice-orbital coupling of single crystal
				Ni$_3$TeO$_6$}},\ }\href {https://doi.org/10.1038/s41598-018-33976-w}
	{\bibfield  {journal} {\bibinfo  {journal} {Sci. Rep.}\ }\textbf {\bibinfo
			{volume} {8}},\ \bibinfo {pages} {15779} (\bibinfo {year}
		{2018})}\BibitemShut {NoStop}%
	\bibitem [{\citenamefont {Nath}\ \emph
		{et~al.}(2014{\natexlab{a}})\citenamefont {Nath}, \citenamefont {Ranjith},
		\citenamefont {Sichelschmidt}, \citenamefont {Baenitz}, \citenamefont
		{Skourski}, \citenamefont {Alet}, \citenamefont {Rousochatzakis},\ and\
		\citenamefont {Tsirlin}}]{Nath014407}%
	\BibitemOpen
	\bibfield  {author} {\bibinfo {author} {\bibfnamefont {R.}~\bibnamefont
			{Nath}}, \bibinfo {author} {\bibfnamefont {K.~M.}\ \bibnamefont {Ranjith}},
		\bibinfo {author} {\bibfnamefont {J.}~\bibnamefont {Sichelschmidt}}, \bibinfo
		{author} {\bibfnamefont {M.}~\bibnamefont {Baenitz}}, \bibinfo {author}
		{\bibfnamefont {Y.}~\bibnamefont {Skourski}}, \bibinfo {author}
		{\bibfnamefont {F.}~\bibnamefont {Alet}}, \bibinfo {author} {\bibfnamefont
			{I.}~\bibnamefont {Rousochatzakis}},\ and\ \bibinfo {author} {\bibfnamefont
			{A.~A.}\ \bibnamefont {Tsirlin}},\ }\bibfield  {title} {\bibinfo {title}
		{{Hindered magnetic order from mixed dimensionalities in
				${\text{CuP}}_{2}{\text{O}}_{6}$}},\ }\href
	{https://doi.org/10.1103/PhysRevB.89.014407} {\bibfield  {journal} {\bibinfo
			{journal} {Phys. Rev. B}\ }\textbf {\bibinfo {volume} {89}},\ \bibinfo
		{pages} {014407} (\bibinfo {year} {2014}{\natexlab{a}})}\BibitemShut
	{NoStop}%
	\bibitem [{\citenamefont {Ishii}\ \emph {et~al.}(2011)\citenamefont {Ishii},
		\citenamefont {Tanaka}, \citenamefont {Onuma}, \citenamefont {Nambu},
		\citenamefont {Tokunaga}, \citenamefont {Sakakibara}, \citenamefont
		{Kawashima}, \citenamefont {Maeno}, \citenamefont {Broholm}, \citenamefont
		{P.~Gautreaux}, \citenamefont {Chan},\ and\ \citenamefont
		{Nakatsuji}}]{Ishii17001}%
	\BibitemOpen
	\bibfield  {author} {\bibinfo {author} {\bibfnamefont {R.}~\bibnamefont
			{Ishii}}, \bibinfo {author} {\bibfnamefont {S.}~\bibnamefont {Tanaka}},
		\bibinfo {author} {\bibfnamefont {K.}~\bibnamefont {Onuma}}, \bibinfo
		{author} {\bibfnamefont {Y.}~\bibnamefont {Nambu}}, \bibinfo {author}
		{\bibfnamefont {M.}~\bibnamefont {Tokunaga}}, \bibinfo {author}
		{\bibfnamefont {T.}~\bibnamefont {Sakakibara}}, \bibinfo {author}
		{\bibfnamefont {N.}~\bibnamefont {Kawashima}}, \bibinfo {author}
		{\bibfnamefont {Y.}~\bibnamefont {Maeno}}, \bibinfo {author} {\bibfnamefont
			{C.}~\bibnamefont {Broholm}}, \bibinfo {author} {\bibfnamefont
			{D.}~\bibnamefont {P.~Gautreaux}}, \bibinfo {author} {\bibfnamefont {J.~Y.}\
			\bibnamefont {Chan}},\ and\ \bibinfo {author} {\bibfnamefont
			{S.}~\bibnamefont {Nakatsuji}},\ }\bibfield  {title} {\bibinfo {title}
		{{Successive phase transitions and phase diagrams for the
				quasi-two-dimensional easy-axis triangular antiferromagnet
				${\mathrm{Rb}_{4}}{\mathrm{Mn(MoO_4)_{3}}}$}},\ }\href
	{https://doi.org/10.1209/0295-5075/94/17001} {\bibfield  {journal} {\bibinfo
			{journal} {Europhys. Lett.}\ }\textbf {\bibinfo {volume} {94}},\ \bibinfo
		{pages} {17001} (\bibinfo {year} {2011})}\BibitemShut {NoStop}%
	\bibitem [{\citenamefont {Yue}\ \emph {et~al.}(2024)\citenamefont {Yue},
		\citenamefont {Lu}, \citenamefont {Yan}, \citenamefont {Wang}, \citenamefont
		{Guo}, \citenamefont {Guo}, \citenamefont {Chen}, \citenamefont {Chen},\ and\
		\citenamefont {Mei}}]{Yue214430}%
	\BibitemOpen
	\bibfield  {author} {\bibinfo {author} {\bibfnamefont {L.}~\bibnamefont
			{Yue}}, \bibinfo {author} {\bibfnamefont {Z.}~\bibnamefont {Lu}}, \bibinfo
		{author} {\bibfnamefont {K.}~\bibnamefont {Yan}}, \bibinfo {author}
		{\bibfnamefont {L.}~\bibnamefont {Wang}}, \bibinfo {author} {\bibfnamefont
			{S.}~\bibnamefont {Guo}}, \bibinfo {author} {\bibfnamefont {R.}~\bibnamefont
			{Guo}}, \bibinfo {author} {\bibfnamefont {P.}~\bibnamefont {Chen}}, \bibinfo
		{author} {\bibfnamefont {X.}~\bibnamefont {Chen}},\ and\ \bibinfo {author}
		{\bibfnamefont {J.-W.}\ \bibnamefont {Mei}},\ }\bibfield  {title} {\bibinfo
		{title} {{Magnetic phase transitions in the triangular-lattice spin-1 dimer
				compound ${\mathrm{K}}_{2}{\mathrm{Ni}}_{2}{({\mathrm{SeO}}_{3})}_{3}$}},\
	}\href {https://doi.org/10.1103/PhysRevB.109.214430} {\bibfield  {journal}
		{\bibinfo  {journal} {Phys. Rev. B}\ }\textbf {\bibinfo {volume} {109}},\
		\bibinfo {pages} {214430} (\bibinfo {year} {2024})}\BibitemShut {NoStop}%
	\bibitem [{\citenamefont {Gopal}(2012)}]{Gopal2012}%
	\BibitemOpen
	\bibfield  {author} {\bibinfo {author} {\bibfnamefont {E.}~\bibnamefont
			{Gopal}},\ }\href {https://books.google.co.in/books?id=Rj3jBwAAQBAJ} {\emph
		{\bibinfo {title} {Specific Heats at Low Temperatures}}},\ The International
	Cryogenics Monograph Series\ (\bibinfo  {publisher} {Springer US},\ \bibinfo
	{year} {2012})\BibitemShut {NoStop}%
	\bibitem [{\citenamefont {Sebastian}\ \emph {et~al.}(2021)\citenamefont
		{Sebastian}, \citenamefont {Somesh}, \citenamefont {Nandi}, \citenamefont
		{Ahmed}, \citenamefont {Bag}, \citenamefont {Baenitz}, \citenamefont {Koo},
		\citenamefont {Sichelschmidt}, \citenamefont {Tsirlin}, \citenamefont
		{Furukawa},\ and\ \citenamefont {Nath}}]{Sebastian064413}%
	\BibitemOpen
	\bibfield  {author} {\bibinfo {author} {\bibfnamefont {S.~J.}\ \bibnamefont
			{Sebastian}}, \bibinfo {author} {\bibfnamefont {K.}~\bibnamefont {Somesh}},
		\bibinfo {author} {\bibfnamefont {M.}~\bibnamefont {Nandi}}, \bibinfo
		{author} {\bibfnamefont {N.}~\bibnamefont {Ahmed}}, \bibinfo {author}
		{\bibfnamefont {P.}~\bibnamefont {Bag}}, \bibinfo {author} {\bibfnamefont
			{M.}~\bibnamefont {Baenitz}}, \bibinfo {author} {\bibfnamefont
			{B.}~\bibnamefont {Koo}}, \bibinfo {author} {\bibfnamefont {J.}~\bibnamefont
			{Sichelschmidt}}, \bibinfo {author} {\bibfnamefont {A.~A.}\ \bibnamefont
			{Tsirlin}}, \bibinfo {author} {\bibfnamefont {Y.}~\bibnamefont {Furukawa}},\
		and\ \bibinfo {author} {\bibfnamefont {R.}~\bibnamefont {Nath}},\ }\bibfield
	{title} {\bibinfo {title} {{Quasi-one-dimensional magnetism in the
				spin-$\frac{1}{2}$ antiferromagnet
				${\mathrm{BaNa}}_{2}\mathrm{Cu}{({\mathrm{VO}}_{4})}_{2}$}},\ }\href
	{https://doi.org/10.1103/PhysRevB.103.064413} {\bibfield  {journal} {\bibinfo
			{journal} {Phys. Rev. B}\ }\textbf {\bibinfo {volume} {103}},\ \bibinfo
		{pages} {064413} (\bibinfo {year} {2021})}\BibitemShut {NoStop}%
	\bibitem [{\citenamefont {Fitzgerel}\ and\ \citenamefont
		{Verhoek}(1960)}]{Fitzgerel545}%
	\BibitemOpen
	\bibfield  {author} {\bibinfo {author} {\bibfnamefont {R.~K.}\ \bibnamefont
			{Fitzgerel}}\ and\ \bibinfo {author} {\bibfnamefont {F.~H.}\ \bibnamefont
			{Verhoek}},\ }\bibfield  {title} {\bibinfo {title} {{The law of Dulong and
				Petit}},\ }\href {https://doi.org/10.1021/ed037p545} {\bibfield  {journal}
		{\bibinfo  {journal} {J. Chem. Edu.}\ }\textbf {\bibinfo {volume} {37}},\
		\bibinfo {pages} {545} (\bibinfo {year} {1960})}\BibitemShut {NoStop}%
	\bibitem [{\citenamefont {Nath}\ \emph
		{et~al.}(2014{\natexlab{b}})\citenamefont {Nath}, \citenamefont {Ranjith},
		\citenamefont {Roy}, \citenamefont {Johnston}, \citenamefont {Furukawa},\
		and\ \citenamefont {Tsirlin}}]{Nath024431}%
	\BibitemOpen
	\bibfield  {author} {\bibinfo {author} {\bibfnamefont {R.}~\bibnamefont
			{Nath}}, \bibinfo {author} {\bibfnamefont {K.~M.}\ \bibnamefont {Ranjith}},
		\bibinfo {author} {\bibfnamefont {B.}~\bibnamefont {Roy}}, \bibinfo {author}
		{\bibfnamefont {D.~C.}\ \bibnamefont {Johnston}}, \bibinfo {author}
		{\bibfnamefont {Y.}~\bibnamefont {Furukawa}},\ and\ \bibinfo {author}
		{\bibfnamefont {A.~A.}\ \bibnamefont {Tsirlin}},\ }\bibfield  {title}
	{\bibinfo {title} {{Magnetic transitions in the spin-$\frac{5}{2}$ frustrated
				magnet $\mathrm{Bi}{\mathrm{Mn}}_{2}{\mathrm{PO}}_{6}$ and strong lattice
				softening in $\mathrm{Bi}{\mathrm{Mn}}_{2}{\mathrm{PO}}_{6}$ and
				$\mathrm{Bi}{\mathrm{Zn}}_{2}{\mathrm{PO}}_{6}$ below 200 K}},\ }\href
	{https://doi.org/10.1103/PhysRevB.90.024431} {\bibfield  {journal} {\bibinfo
			{journal} {Phys. Rev. B}\ }\textbf {\bibinfo {volume} {90}},\ \bibinfo
		{pages} {024431} (\bibinfo {year} {2014}{\natexlab{b}})}\BibitemShut
	{NoStop}%
	\bibitem [{\citenamefont {Takatsu}\ \emph {et~al.}(2009)\citenamefont
		{Takatsu}, \citenamefont {Yoshizawa}, \citenamefont {Yonezawa},\ and\
		\citenamefont {Maeno}}]{Takatsu104424}%
	\BibitemOpen
	\bibfield  {author} {\bibinfo {author} {\bibfnamefont {H.}~\bibnamefont
			{Takatsu}}, \bibinfo {author} {\bibfnamefont {H.}~\bibnamefont {Yoshizawa}},
		\bibinfo {author} {\bibfnamefont {S.}~\bibnamefont {Yonezawa}},\ and\
		\bibinfo {author} {\bibfnamefont {Y.}~\bibnamefont {Maeno}},\ }\bibfield
	{title} {\bibinfo {title} {{Critical behavior of the metallic
				triangular-lattice Heisenberg antiferromagnet ${\text{PdCrO}}_{2}$}},\ }\href
	{https://doi.org/10.1103/PhysRevB.79.104424} {\bibfield  {journal} {\bibinfo
			{journal} {Phys. Rev. B}\ }\textbf {\bibinfo {volume} {79}},\ \bibinfo
		{pages} {104424} (\bibinfo {year} {2009})}\BibitemShut {NoStop}%
	\bibitem [{\citenamefont {Islam}\ \emph {et~al.}(2018)\citenamefont {Islam},
		\citenamefont {Ranjith}, \citenamefont {Baenitz}, \citenamefont {Skourski},
		\citenamefont {Tsirlin},\ and\ \citenamefont {Nath}}]{Islam174432}%
	\BibitemOpen
	\bibfield  {author} {\bibinfo {author} {\bibfnamefont {S.~S.}\ \bibnamefont
			{Islam}}, \bibinfo {author} {\bibfnamefont {K.~M.}\ \bibnamefont {Ranjith}},
		\bibinfo {author} {\bibfnamefont {M.}~\bibnamefont {Baenitz}}, \bibinfo
		{author} {\bibfnamefont {Y.}~\bibnamefont {Skourski}}, \bibinfo {author}
		{\bibfnamefont {A.~A.}\ \bibnamefont {Tsirlin}},\ and\ \bibinfo {author}
		{\bibfnamefont {R.}~\bibnamefont {Nath}},\ }\bibfield  {title} {\bibinfo
		{title} {{Frustration of square cupola in
				Sr(TiO)${\mathrm{Cu}}_{4}({\mathrm{PO}}_{4}{)}_{4}$}},\ }\href
	{https://doi.org/10.1103/PhysRevB.97.174432} {\bibfield  {journal} {\bibinfo
			{journal} {Phys. Rev. B}\ }\textbf {\bibinfo {volume} {97}},\ \bibinfo
		{pages} {174432} (\bibinfo {year} {2018})}\BibitemShut {NoStop}%
	\bibitem [{\citenamefont {Yogi}\ \emph {et~al.}(2015)\citenamefont {Yogi},
		\citenamefont {Ahmed}, \citenamefont {Nath}, \citenamefont {Tsirlin},
		\citenamefont {Kundu}, \citenamefont {Mahajan}, \citenamefont
		{Sichelschmidt}, \citenamefont {Roy},\ and\ \citenamefont
		{Furukawa}}]{Yogi024413}%
	\BibitemOpen
	\bibfield  {author} {\bibinfo {author} {\bibfnamefont {A.}~\bibnamefont
			{Yogi}}, \bibinfo {author} {\bibfnamefont {N.}~\bibnamefont {Ahmed}},
		\bibinfo {author} {\bibfnamefont {R.}~\bibnamefont {Nath}}, \bibinfo {author}
		{\bibfnamefont {A.~A.}\ \bibnamefont {Tsirlin}}, \bibinfo {author}
		{\bibfnamefont {S.}~\bibnamefont {Kundu}}, \bibinfo {author} {\bibfnamefont
			{A.~V.}\ \bibnamefont {Mahajan}}, \bibinfo {author} {\bibfnamefont
			{J.}~\bibnamefont {Sichelschmidt}}, \bibinfo {author} {\bibfnamefont
			{B.}~\bibnamefont {Roy}},\ and\ \bibinfo {author} {\bibfnamefont
			{Y.}~\bibnamefont {Furukawa}},\ }\bibfield  {title} {\bibinfo {title}
		{{Antiferromagnetism of
				${\mathrm{Zn}}_{2}\mathrm{VO}{{(\mathrm{PO}}_{4})}_{2}$ and the dilution with
				${\mathrm{Ti}}^{4+}$}},\ }\href {https://doi.org/10.1103/PhysRevB.91.024413}
	{\bibfield  {journal} {\bibinfo  {journal} {Phys. Rev. B}\ }\textbf {\bibinfo
			{volume} {91}},\ \bibinfo {pages} {024413} (\bibinfo {year}
		{2015})}\BibitemShut {NoStop}%
	\bibitem [{\citenamefont {Slichter}(2013)}]{slichter2013}%
	\BibitemOpen
	\bibfield  {author} {\bibinfo {author} {\bibfnamefont {C.~P.}\ \bibnamefont
			{Slichter}},\ }\href@noop {} {\emph {\bibinfo {title} {{Principles of
					magnetic resonance}}}},\ Vol.~\bibinfo {volume} {1}\ (\bibinfo  {publisher}
	{Springer Science \& Business Media},\ \bibinfo {year} {2013})\BibitemShut
	{NoStop}%
	\bibitem [{\citenamefont {Sebastian}\ \emph {et~al.}(2025)\citenamefont
		{Sebastian}, \citenamefont {Kolay}, \citenamefont {B}, \citenamefont {Ding},
		\citenamefont {Furukawa},\ and\ \citenamefont {Nath}}]{Sebastian104428}%
	\BibitemOpen
	\bibfield  {author} {\bibinfo {author} {\bibfnamefont {S.~J.}\ \bibnamefont
			{Sebastian}}, \bibinfo {author} {\bibfnamefont {R.}~\bibnamefont {Kolay}},
		\bibinfo {author} {\bibfnamefont {A.}~\bibnamefont {B}}, \bibinfo {author}
		{\bibfnamefont {Q.-P.}\ \bibnamefont {Ding}}, \bibinfo {author}
		{\bibfnamefont {Y.}~\bibnamefont {Furukawa}},\ and\ \bibinfo {author}
		{\bibfnamefont {R.}~\bibnamefont {Nath}},\ }\bibfield  {title} {\bibinfo
		{title} {{Spin fluctuations, absence of magnetic order, and crystal electric
				field studies in the ${\mathrm{Yb}}^{3+}$-based triangular lattice
				antiferromagnet ${\mathrm{Rb}}_{3}\mathrm{Yb}{({\mathrm{VO}}_{4})}_{2}$}},\
	}\href {https://doi.org/10.1103/ydpn-8wgf} {\bibfield  {journal} {\bibinfo
			{journal} {Phys. Rev. B}\ }\textbf {\bibinfo {volume} {112}},\ \bibinfo
		{pages} {104428} (\bibinfo {year} {2025})}\BibitemShut {NoStop}%
	\bibitem [{\citenamefont {Ambika}\ \emph {et~al.}(2022)\citenamefont {Ambika},
		\citenamefont {Ding}, \citenamefont {Sebastian}, \citenamefont {Nath},\ and\
		\citenamefont {Furukawa}}]{Ambika015803}%
	\BibitemOpen
	\bibfield  {author} {\bibinfo {author} {\bibfnamefont {D.~V.}\ \bibnamefont
			{Ambika}}, \bibinfo {author} {\bibfnamefont {Q.-P.}\ \bibnamefont {Ding}},
		\bibinfo {author} {\bibfnamefont {S.~J.}\ \bibnamefont {Sebastian}}, \bibinfo
		{author} {\bibfnamefont {R.}~\bibnamefont {Nath}},\ and\ \bibinfo {author}
		{\bibfnamefont {Y.}~\bibnamefont {Furukawa}},\ }\bibfield  {title} {\bibinfo
		{title} {{Static and dynamic magnetic properties of the spin-$\frac52$
				triangle lattice antiferromagnet
				$\mathrm{Na}_3\mathrm{Fe}\mathrm{{(PO_{4})_{2}}}$ studied by
				$^{31}\mathrm{P}$ NMR}},\ }\href {https://doi.org/10.1088/1361-648X/ac9e37}
	{\bibfield  {journal} {\bibinfo  {journal} {J. Phys. Condens. Matter.}\
		}\textbf {\bibinfo {volume} {35}},\ \bibinfo {pages} {015803} (\bibinfo
		{year} {2022})}\BibitemShut {NoStop}%
	\bibitem [{\citenamefont {Yamada}\ and\ \citenamefont
		{Sakata}(1986)}]{Yamada1751}%
	\BibitemOpen
	\bibfield  {author} {\bibinfo {author} {\bibfnamefont {Y.}~\bibnamefont
			{Yamada}}\ and\ \bibinfo {author} {\bibfnamefont {A.}~\bibnamefont
			{Sakata}},\ }\bibfield  {title} {\bibinfo {title} {{An Analysis Method of
				Antiferromagnetic Powder Patterns in Spin-Echo NMR under External Fields}},\
	}\href {https://doi.org/10.1143/JPSJ.55.1751} {\bibfield  {journal} {\bibinfo
			{journal} {J. Phys. Soc. Jpn.}\ }\textbf {\bibinfo {volume} {55}},\ \bibinfo
		{pages} {1751} (\bibinfo {year} {1986})}\BibitemShut {NoStop}%
	\bibitem [{\citenamefont {Ranjith}\ \emph {et~al.}(2016)\citenamefont
		{Ranjith}, \citenamefont {Nath}, \citenamefont {Majumder}, \citenamefont
		{Kasinathan}, \citenamefont {Skoulatos}, \citenamefont {Keller},
		\citenamefont {Skourski}, \citenamefont {Baenitz},\ and\ \citenamefont
		{Tsirlin}}]{Ranjith014415}%
	\BibitemOpen
	\bibfield  {author} {\bibinfo {author} {\bibfnamefont {K.~M.}\ \bibnamefont
			{Ranjith}}, \bibinfo {author} {\bibfnamefont {R.}~\bibnamefont {Nath}},
		\bibinfo {author} {\bibfnamefont {M.}~\bibnamefont {Majumder}}, \bibinfo
		{author} {\bibfnamefont {D.}~\bibnamefont {Kasinathan}}, \bibinfo {author}
		{\bibfnamefont {M.}~\bibnamefont {Skoulatos}}, \bibinfo {author}
		{\bibfnamefont {L.}~\bibnamefont {Keller}}, \bibinfo {author} {\bibfnamefont
			{Y.}~\bibnamefont {Skourski}}, \bibinfo {author} {\bibfnamefont
			{M.}~\bibnamefont {Baenitz}},\ and\ \bibinfo {author} {\bibfnamefont {A.~A.}\
			\bibnamefont {Tsirlin}},\ }\bibfield  {title} {\bibinfo {title}
		{{Commensurate and incommensurate magnetic order in spin-1 chains stacked on
				the triangular lattice in
				${\mathrm{Li}}_{2}{\mathrm{NiW}}_{2}{\mathrm{O}}_{8}$}},\ }\href
	{https://doi.org/10.1103/PhysRevB.94.014415} {\bibfield  {journal} {\bibinfo
			{journal} {Phys. Rev. B}\ }\textbf {\bibinfo {volume} {94}},\ \bibinfo
		{pages} {014415} (\bibinfo {year} {2016})}\BibitemShut {NoStop}%
	\bibitem [{\citenamefont {Ranjith}\ \emph {et~al.}(2015)\citenamefont
		{Ranjith}, \citenamefont {Majumder}, \citenamefont {Baenitz}, \citenamefont
		{Tsirlin},\ and\ \citenamefont {Nath}}]{Ranjith024422}%
	\BibitemOpen
	\bibfield  {author} {\bibinfo {author} {\bibfnamefont {K.~M.}\ \bibnamefont
			{Ranjith}}, \bibinfo {author} {\bibfnamefont {M.}~\bibnamefont {Majumder}},
		\bibinfo {author} {\bibfnamefont {M.}~\bibnamefont {Baenitz}}, \bibinfo
		{author} {\bibfnamefont {A.~A.}\ \bibnamefont {Tsirlin}},\ and\ \bibinfo
		{author} {\bibfnamefont {R.}~\bibnamefont {Nath}},\ }\bibfield  {title}
	{\bibinfo {title} {{Frustrated three-dimensional antiferromagnet
				${\text{Li}}_{2}{\text{CuW}}_{2}{\text{O}}_{8}$: $^{7}\mathrm{Li}$ NMR and
				the effect of nonmagnetic dilution}},\ }\href
	{https://doi.org/10.1103/PhysRevB.92.024422} {\bibfield  {journal} {\bibinfo
			{journal} {Phys. Rev. B}\ }\textbf {\bibinfo {volume} {92}},\ \bibinfo
		{pages} {024422} (\bibinfo {year} {2015})}\BibitemShut {NoStop}%
	\bibitem [{\citenamefont {Nath}\ \emph {et~al.}(2009)\citenamefont {Nath},
		\citenamefont {Furukawa}, \citenamefont {Borsa}, \citenamefont {Kaul},
		\citenamefont {Baenitz}, \citenamefont {Geibel},\ and\ \citenamefont
		{Johnston}}]{Nath214430}%
	\BibitemOpen
	\bibfield  {author} {\bibinfo {author} {\bibfnamefont {R.}~\bibnamefont
			{Nath}}, \bibinfo {author} {\bibfnamefont {Y.}~\bibnamefont {Furukawa}},
		\bibinfo {author} {\bibfnamefont {F.}~\bibnamefont {Borsa}}, \bibinfo
		{author} {\bibfnamefont {E.~E.}\ \bibnamefont {Kaul}}, \bibinfo {author}
		{\bibfnamefont {M.}~\bibnamefont {Baenitz}}, \bibinfo {author} {\bibfnamefont
			{C.}~\bibnamefont {Geibel}},\ and\ \bibinfo {author} {\bibfnamefont {D.~C.}\
			\bibnamefont {Johnston}},\ }\bibfield  {title} {\bibinfo {title}
		{{Single-crystal $^{31}\text{P}$ NMR studies of the frustrated square-lattice
				compound ${\text{Pb}}_{2}(\text{VO}){({\text{PO}}_{4})}_{2}$}},\ }\href
	{https://doi.org/10.1103/PhysRevB.80.214430} {\bibfield  {journal} {\bibinfo
			{journal} {Phys. Rev. B}\ }\textbf {\bibinfo {volume} {80}},\ \bibinfo
		{pages} {214430} (\bibinfo {year} {2009})}\BibitemShut {NoStop}%
	\bibitem [{\citenamefont {Collins}(1989)}]{Collins1989}%
	\BibitemOpen
	\bibfield  {author} {\bibinfo {author} {\bibfnamefont {M.~F.}\ \bibnamefont
			{Collins}},\ }\href@noop {} {\emph {\bibinfo {title} {{Magnetic Critical
					Scattering}}}}\ (\bibinfo  {publisher} {Oxford University Press},\ \bibinfo
	{year} {1989})\BibitemShut {NoStop}%
	\bibitem [{\citenamefont {Ozeki}\ and\ \citenamefont {Ito}(2007)}]{Ozeki2007}%
	\BibitemOpen
	\bibfield  {author} {\bibinfo {author} {\bibfnamefont {Y.}~\bibnamefont
			{Ozeki}}\ and\ \bibinfo {author} {\bibfnamefont {N.}~\bibnamefont {Ito}},\
	}\bibfield  {title} {\bibinfo {title} {{Nonequilibrium relaxation method}},\
	}\href {https://doi.org/10.1088/1751-8113/40/31/R01} {\bibfield  {journal}
		{\bibinfo  {journal} {J. Phys. A: Math. Theor.}\ }\textbf {\bibinfo {volume}
			{40}},\ \bibinfo {pages} {R149} (\bibinfo {year} {2007})}\BibitemShut
	{NoStop}%
	\bibitem [{\citenamefont {Bramwell}\ and\ \citenamefont
		{Holdsworth}(1994)}]{Bramwell1994}%
	\BibitemOpen
	\bibfield  {author} {\bibinfo {author} {\bibfnamefont {S.~T.}\ \bibnamefont
			{Bramwell}}\ and\ \bibinfo {author} {\bibfnamefont {P.~C.~W.}\ \bibnamefont
			{Holdsworth}},\ }\bibfield  {title} {\bibinfo {title} {Magnetization: A
			characteristic of the kosterlitz-thouless-berezinskii transition},\ }\href
	{https://doi.org/10.1103/PhysRevB.49.8811} {\bibfield  {journal} {\bibinfo
			{journal} {Phys. Rev. B}\ }\textbf {\bibinfo {volume} {49}},\ \bibinfo
		{pages} {8811} (\bibinfo {year} {1994})}\BibitemShut {NoStop}%
	\bibitem [{\citenamefont {Benner}\ and\ \citenamefont
		{Boucher}(1990)}]{Benner1990}%
	\BibitemOpen
	\bibfield  {author} {\bibinfo {author} {\bibfnamefont {H.}~\bibnamefont
			{Benner}}\ and\ \bibinfo {author} {\bibfnamefont {J.}~\bibnamefont
			{Boucher}},\ }\href@noop {} {\emph {\bibinfo {title} {{Spin Dynamics in the
					Paramagnetic Regime: NMR and EPR in Two-Dimensional Magnets in Magnetic
					Properties of Layered Transition Metal Compounds}}}}\ (\bibinfo  {publisher}
	{Editor LJ de Jongh, Kluwer Academic Publishers, Dordrecht, Boston, London},\
	\bibinfo {year} {1990})\BibitemShut {NoStop}%
	\bibitem [{\citenamefont {Cui}\ \emph {et~al.}(2017)\citenamefont {Cui},
		\citenamefont {Ding}, \citenamefont {Meier}, \citenamefont {B\"ohmer},
		\citenamefont {Kong}, \citenamefont {Borisov}, \citenamefont {Lee},
		\citenamefont {Bud'ko}, \citenamefont {Valent\'{\i}}, \citenamefont
		{Canfield},\ and\ \citenamefont {Furukawa}}]{Cui104512}%
	\BibitemOpen
	\bibfield  {author} {\bibinfo {author} {\bibfnamefont {J.}~\bibnamefont
			{Cui}}, \bibinfo {author} {\bibfnamefont {Q.-P.}\ \bibnamefont {Ding}},
		\bibinfo {author} {\bibfnamefont {W.~R.}\ \bibnamefont {Meier}}, \bibinfo
		{author} {\bibfnamefont {A.~E.}\ \bibnamefont {B\"ohmer}}, \bibinfo {author}
		{\bibfnamefont {T.}~\bibnamefont {Kong}}, \bibinfo {author} {\bibfnamefont
			{V.}~\bibnamefont {Borisov}}, \bibinfo {author} {\bibfnamefont
			{Y.}~\bibnamefont {Lee}}, \bibinfo {author} {\bibfnamefont {S.~L.}\
			\bibnamefont {Bud'ko}}, \bibinfo {author} {\bibfnamefont {R.}~\bibnamefont
			{Valent\'{\i}}}, \bibinfo {author} {\bibfnamefont {P.~C.}\ \bibnamefont
			{Canfield}},\ and\ \bibinfo {author} {\bibfnamefont {Y.}~\bibnamefont
			{Furukawa}},\ }\bibfield  {title} {\bibinfo {title} {{Magnetic fluctuations
				and superconducting properties of ${\mathrm{CaKFe}}_{4}{\mathrm{As}}_{4}$
				studied by $^{75}\mathrm{As}$ NMR}},\ }\href
	{https://doi.org/10.1103/PhysRevB.96.104512} {\bibfield  {journal} {\bibinfo
			{journal} {Phys. Rev. B}\ }\textbf {\bibinfo {volume} {96}},\ \bibinfo
		{pages} {104512} (\bibinfo {year} {2017})}\BibitemShut {NoStop}%
	\bibitem [{\citenamefont {Ding}\ \emph {et~al.}(2017)\citenamefont {Ding},
		\citenamefont {Meier}, \citenamefont {B\"ohmer}, \citenamefont {Bud'ko},
		\citenamefont {Canfield},\ and\ \citenamefont {Furukawa}}]{Ding220510}%
	\BibitemOpen
	\bibfield  {author} {\bibinfo {author} {\bibfnamefont {Q.-P.}\ \bibnamefont
			{Ding}}, \bibinfo {author} {\bibfnamefont {W.~R.}\ \bibnamefont {Meier}},
		\bibinfo {author} {\bibfnamefont {A.~E.}\ \bibnamefont {B\"ohmer}}, \bibinfo
		{author} {\bibfnamefont {S.~L.}\ \bibnamefont {Bud'ko}}, \bibinfo {author}
		{\bibfnamefont {P.~C.}\ \bibnamefont {Canfield}},\ and\ \bibinfo {author}
		{\bibfnamefont {Y.}~\bibnamefont {Furukawa}},\ }\bibfield  {title} {\bibinfo
		{title} {{NMR study of the new magnetic superconductor
				$\mathrm{CaK}{({\mathrm{Fe}}_{0.951}{\mathrm{Ni}}_{0.049})}_{4}{\mathrm{As}}_{4}$:
				Microscopic coexistence of the hedgehog spin-vortex crystal and
				superconductivity}},\ }\href {https://doi.org/10.1103/PhysRevB.96.220510}
	{\bibfield  {journal} {\bibinfo  {journal} {Phys. Rev. B}\ }\textbf {\bibinfo
			{volume} {96}},\ \bibinfo {pages} {220510} (\bibinfo {year}
		{2017})}\BibitemShut {NoStop}%
	\bibitem [{\citenamefont {Ding}\ \emph {et~al.}(2018)\citenamefont {Ding},
		\citenamefont {Meier}, \citenamefont {Cui}, \citenamefont {Xu}, \citenamefont
		{B\"ohmer}, \citenamefont {Bud'ko}, \citenamefont {Canfield},\ and\
		\citenamefont {Furukawa}}]{Ding137204}%
	\BibitemOpen
	\bibfield  {author} {\bibinfo {author} {\bibfnamefont {Q.-P.}\ \bibnamefont
			{Ding}}, \bibinfo {author} {\bibfnamefont {W.~R.}\ \bibnamefont {Meier}},
		\bibinfo {author} {\bibfnamefont {J.}~\bibnamefont {Cui}}, \bibinfo {author}
		{\bibfnamefont {M.}~\bibnamefont {Xu}}, \bibinfo {author} {\bibfnamefont
			{A.~E.}\ \bibnamefont {B\"ohmer}}, \bibinfo {author} {\bibfnamefont {S.~L.}\
			\bibnamefont {Bud'ko}}, \bibinfo {author} {\bibfnamefont {P.~C.}\
			\bibnamefont {Canfield}},\ and\ \bibinfo {author} {\bibfnamefont
			{Y.}~\bibnamefont {Furukawa}},\ }\bibfield  {title} {\bibinfo {title}
		{{Hedgehog Spin-Vortex Crystal Antiferromagnetic Quantum Criticality in
				$\mathrm{CaK}({\mathrm{Fe}}_{1\ensuremath{-}x}{\mathrm{Ni}}_{x}{)}_{4}{\mathrm{As}}_{4}$
				Revealed by NMR}},\ }\href {https://doi.org/10.1103/PhysRevLett.121.137204}
	{\bibfield  {journal} {\bibinfo  {journal} {Phys. Rev. Lett.}\ }\textbf
		{\bibinfo {volume} {121}},\ \bibinfo {pages} {137204} (\bibinfo {year}
		{2018})}\BibitemShut {NoStop}%
	\bibitem [{\citenamefont {Moriya}(1963)}]{Moriya516}%
	\BibitemOpen
	\bibfield  {author} {\bibinfo {author} {\bibfnamefont {T.}~\bibnamefont
			{Moriya}},\ }\bibfield  {title} {\bibinfo {title} {{The Effect of
				Electron-Electron Interaction on the Nuclear Spin Relaxation in Metals}},\
	}\href {https://doi.org/10.1143/JPSJ.18.516} {\bibfield  {journal} {\bibinfo
			{journal} {J. Phys. Soc. Jpn.}\ }\textbf {\bibinfo {volume} {18}},\ \bibinfo
		{pages} {516} (\bibinfo {year} {1963})}\BibitemShut {NoStop}%
	\bibitem [{\citenamefont {Garlea}\ \emph {et~al.}(2019)\citenamefont {Garlea},
		\citenamefont {Sanjeewa}, \citenamefont {McGuire}, \citenamefont {Batista},
		\citenamefont {Samarakoon}, \citenamefont {Graf}, \citenamefont {Winn},
		\citenamefont {Ye}, \citenamefont {Hoffmann},\ and\ \citenamefont
		{Kolis}}]{Garlea011038}%
	\BibitemOpen
	\bibfield  {author} {\bibinfo {author} {\bibfnamefont {V.~O.}\ \bibnamefont
			{Garlea}}, \bibinfo {author} {\bibfnamefont {L.~D.}\ \bibnamefont
			{Sanjeewa}}, \bibinfo {author} {\bibfnamefont {M.~A.}\ \bibnamefont
			{McGuire}}, \bibinfo {author} {\bibfnamefont {C.~D.}\ \bibnamefont
			{Batista}}, \bibinfo {author} {\bibfnamefont {A.~M.}\ \bibnamefont
			{Samarakoon}}, \bibinfo {author} {\bibfnamefont {D.}~\bibnamefont {Graf}},
		\bibinfo {author} {\bibfnamefont {B.}~\bibnamefont {Winn}}, \bibinfo {author}
		{\bibfnamefont {F.}~\bibnamefont {Ye}}, \bibinfo {author} {\bibfnamefont
			{C.}~\bibnamefont {Hoffmann}},\ and\ \bibinfo {author} {\bibfnamefont
			{J.~W.}\ \bibnamefont {Kolis}},\ }\bibfield  {title} {\bibinfo {title}
		{{Exotic Magnetic Field-Induced Spin-Superstructures in a Mixed
				Honeycomb-Triangular Lattice System}},\ }\href
	{https://doi.org/10.1103/PhysRevX.9.011038} {\bibfield  {journal} {\bibinfo
			{journal} {Phys. Rev. X}\ }\textbf {\bibinfo {volume} {9}},\ \bibinfo {pages}
		{011038} (\bibinfo {year} {2019})}\BibitemShut {NoStop}%
	\bibitem [{\citenamefont {Chubukov}\ and\ \citenamefont
		{Golosov}(1991)}]{Golosov69}%
	\BibitemOpen
	\bibfield  {author} {\bibinfo {author} {\bibfnamefont {A.~V.}\ \bibnamefont
			{Chubukov}}\ and\ \bibinfo {author} {\bibfnamefont {D.~I.}\ \bibnamefont
			{Golosov}},\ }\bibfield  {title} {\bibinfo {title} {Quantum theory of an
			antiferromagnet on a triangular lattice in a magnetic field},\ }\href
	{https://doi.org/10.1088/0953-8984/3/1/005} {\bibfield  {journal} {\bibinfo
			{journal} {J. Phys. Condens. Matter.}\ }\textbf {\bibinfo {volume} {3}},\
		\bibinfo {pages} {69} (\bibinfo {year} {1991})}\BibitemShut {NoStop}%
	\bibitem [{\citenamefont {Shirata}\ \emph {et~al.}(2012)\citenamefont
		{Shirata}, \citenamefont {Tanaka}, \citenamefont {Matsuo},\ and\
		\citenamefont {Kindo}}]{Shirata057205}%
	\BibitemOpen
	\bibfield  {author} {\bibinfo {author} {\bibfnamefont {Y.}~\bibnamefont
			{Shirata}}, \bibinfo {author} {\bibfnamefont {H.}~\bibnamefont {Tanaka}},
		\bibinfo {author} {\bibfnamefont {A.}~\bibnamefont {Matsuo}},\ and\ \bibinfo
		{author} {\bibfnamefont {K.}~\bibnamefont {Kindo}},\ }\bibfield  {title}
	{\bibinfo {title} {{Experimental Realization of a Spin-$1/2$
				Triangular-Lattice Heisenberg Antiferromagnet}},\ }\href
	{https://doi.org/10.1103/PhysRevLett.108.057205} {\bibfield  {journal}
		{\bibinfo  {journal} {Phys. Rev. Lett.}\ }\textbf {\bibinfo {volume} {108}},\
		\bibinfo {pages} {057205} (\bibinfo {year} {2012})}\BibitemShut {NoStop}%
	\bibitem [{\citenamefont {Sakhratov}\ \emph {et~al.}(2022)\citenamefont
		{Sakhratov}, \citenamefont {Prokhnenko}, \citenamefont {Shapiro},
		\citenamefont {Zhou}, \citenamefont {Svistov}, \citenamefont {Reyes},\ and\
		\citenamefont {Petrenko}}]{Sakhratov014431}%
	\BibitemOpen
	\bibfield  {author} {\bibinfo {author} {\bibfnamefont {Y.~A.}\ \bibnamefont
			{Sakhratov}}, \bibinfo {author} {\bibfnamefont {O.}~\bibnamefont
			{Prokhnenko}}, \bibinfo {author} {\bibfnamefont {A.~Y.}\ \bibnamefont
			{Shapiro}}, \bibinfo {author} {\bibfnamefont {H.~D.}\ \bibnamefont {Zhou}},
		\bibinfo {author} {\bibfnamefont {L.~E.}\ \bibnamefont {Svistov}}, \bibinfo
		{author} {\bibfnamefont {A.~P.}\ \bibnamefont {Reyes}},\ and\ \bibinfo
		{author} {\bibfnamefont {O.~A.}\ \bibnamefont {Petrenko}},\ }\bibfield
	{title} {\bibinfo {title} {{High-field magnetic structure of the triangular
				antiferromagnet $\mathrm{RbFe}({\mathrm{MoO}}_{4}{)}_{2}$}},\ }\href
	{https://doi.org/10.1103/PhysRevB.105.014431} {\bibfield  {journal} {\bibinfo
			{journal} {Phys. Rev. B}\ }\textbf {\bibinfo {volume} {105}},\ \bibinfo
		{pages} {014431} (\bibinfo {year} {2022})}\BibitemShut {NoStop}%
\end{thebibliography}

\end{document}